\documentclass[prx, reprint, amsmath,amssymb,showpacs,floatfix,longbibliography, aps, twocolumn, superscriptaddress, letterpaper, accepted=2021-12-15]{quantumarticle}
\pdfoutput=1
\usepackage{graphicx}
\usepackage{dcolumn}
\usepackage{bm}
\usepackage{color}
\usepackage{xcolor}
\usepackage{hyperref}
\usepackage{enumitem}
\usepackage{cancel}
\hypersetup{colorlinks=true,citecolor=blue,linkcolor=blue, urlcolor=blue}
\hypersetup{linktocpage}
\usepackage{amsfonts}
\usepackage{environ}
\usepackage{soul} %highlights using the command \hl{}
\usepackage[numbers,sort&compress]{natbib}

\newtheorem{definition}{Definition}
\newtheorem{proposition}{Proposition}
\newtheorem{lemma}{Lemma}

\newcommand{\RR}{\mathbb{R}}
\newcommand{\ZZ}{{\mathbb{Z}}}
\newcommand{\spt}{\text{SPT}}
\newcommand{\stab}{\mathcal{S}}
\newcommand{\loc}{\text{loc}}
\newcommand{\cs}{{\text{CS}}}
\newcommand{\sym}{{\text{sym}}}
\newcommand{\Gone}{H}
\newcommand{\Gtwo}{K}
\newcommand{\simp}{\Delta}

\NewEnviron{eqs}{%
\begin{equation}\begin{split}
    \BODY
\end{split}\end{equation}
}

\definecolor{KKgreen}{RGB}{0,100,0}

\newtheorem{innercustomthm}{Proposition}
\newenvironment{customthm}[1]
  {\renewcommand\theinnercustomthm{#1}\innercustomthm}
  {\endinnercustomthm}

\begin{document}

\title{Symmetry-protected sign problem and magic in quantum phases of matter}

\author{Tyler D. Ellison}
\email[E-mail: ]{ellisont@uw.edu}
\affiliation {University of Washington, Seattle, WA 98195, USA}
\affiliation {Perimeter Institute for Theoretical Physics, Waterloo, Ontario N2L 2Y5, Canada}

\author{Kohtaro Kato}
\affiliation {Institute for Quantum Information and Matter,
California Institute of Technology, Pasadena, CA 91125, USA}
\affiliation{Center for Quantum Information and Quantum Biology,\\
Institute for Open and Transdisciplinary Research Initiatives,\\ Osaka University, Osaka 560-8531, Japan}

\author{Zi-Wen Liu}
\affiliation {Perimeter Institute for Theoretical Physics, Waterloo, Ontario N2L 2Y5, Canada}

\author{Timothy H. Hsieh}
\email[E-mail: ]{thsieh@pitp.ca}
\affiliation {Perimeter Institute for Theoretical Physics, Waterloo, Ontario N2L 2Y5, Canada}

% \date{\today}

\begin{abstract}

We introduce the concepts of a symmetry-protected sign problem and symmetry-protected magic to study the complexity of symmetry-protected topological (SPT) phases of matter. In particular, we say a state has a symmetry-protected sign problem or symmetry-protected magic, if finite-depth quantum circuits composed of symmetric gates are unable to transform the state into a non-negative real wave function or stabilizer state, respectively.  We prove that states belonging to certain SPT phases have these properties, as a result of their anomalous symmetry action at a boundary. For example, we find that one-dimensional $\ZZ_2 \times \ZZ_2$ SPT states (e.g. cluster state) have a symmetry-protected sign problem, and {two-dimensional $\ZZ_2$ SPT states (e.g. Levin-Gu state) have symmetry-protected magic. Furthermore, we comment on the relation between a symmetry-protected sign problem and the computational wire property of one-dimensional SPT states. In an appendix, we also introduce explicit decorated domain wall models of SPT phases, which may be of independent interest. }

\end{abstract}

\maketitle

% \onecolumngrid

\tableofcontents

\section{Introduction} \label{sec: introduction}

The concept of entanglement is an important tool for diagnosing the complexity of quantum states and has led to a deeper understanding of quantum phases of matter and quantum phase transitions. However, entanglement by itself does not fully capture the quantum complexity of a state -- some quantum states can be efficiently simulated by classical systems, despite the presence of entanglement.
This motivates using diagnostics beyond entanglement to assess the quantum complexity of many-body states and to further inform us of the quantum information structures intrinsic to phases of matter. 
In this work, we focus on two means for evaluating the complexity of a state: (i) its `magic' and (ii) its sign structure. 

Magic is an assessment of the extent to which a state can be expressed as a stabilizer state \cite{VMGE14}. Since stabilizer states can be efficiently stored and manipulated on classical computers \cite{G99}, magic can be regarded as a measure of the complexity of a state.
The sign structure of a state, on the other hand, relates to the difficulty in expressing a state as a non-negative state -- i.e., a state with real non-negative probability amplitudes in a local basis \cite{GF15,H16}. Complex probability amplitudes are responsible for inherently non-classical phenomena, such as quantum interference, so the sign structure can be used to characterize the quantum nature of a state. 

The sign structure of a state is, of course, basis dependent, so to make a meaningful assessment of the complexity of the state, we consider the sign structure modulo local basis changes. Following Ref.~\cite{H16}, we say the wave function has a sign problem if the amplitudes cannot be made non-negative by any local basis transformation. This notion of a sign problem implies that any gapped parent Hamiltonian has a sign problem in the stoquastic sense \cite{H16}. Therefore, the sign problem at the level of the wave function also implies that there is an obstacle to efficiently simulating the system using Monte Carlo methods.

While the magic in a many-body state and the notion of a sign problem are promising metrics for the quantum complexity of states, they are notoriously challenging to study analytically and numerically, although substantial progress has been made \cite{H16,TCFMF20,PhysRevLett.94.170201,Marvian2019,klassen2020hardness,Klassen2019twolocalqubit,levy2019mitigating,WZY20, Hangleitereabb8341,VMGE14,WCS20,SMB20,LW20,RK17,GSR20,SGR20,DGS20,DGS20-2}. We therefore propose a simplification by imposing symmetry constraints.  In particular, we introduce symmetry-protected magic and a symmetry-protected sign problem. These simplified diagnostics of the complexity of a state allow us to make analytical statements about the structure of quantum information in quantum phases of matter.

More specifically, we consider
% celebrated 
symmetry-protected topological (SPT) phases of matter, whose properties can be characterized by short-range entangled (SRE) states. 
Despite the short-range entanglement, SPT phases are responsible for a rich set of quantum phenomena including the helical edge modes at the boundary of topological insulators \cite{KM05,KM05-2} and symmetry-protected degeneracies useful for measurement-based quantum computing \cite{RBH01,RWPWS17,ROWSN19,SNBER19,DAM20,DW18,M10}. 
It is therefore valuable to have a complete understanding of the quantum information structures of SPT phases to be able to both simulate their novel behaviors and harness their resources for quantum computing.

In this work, we contribute to the understanding of the quantum complexity of SPT states, by showing that certain SPT states have symmetry-protected magic and that some possess a symmetry-protected sign problem. The symmetry-protected magic implies that the SPT states have magic that cannot be removed by making local symmetry-preserving changes to the state. This builds on the work of Refs.~\cite{MM16,MM18,Y17}, in which particular finely tuned SPT states are shown to have magic.
The symmetry-protected sign problem, in contrast, informs us about the sign structure of SPT states and poses an obstruction to finding a non-negative representation through local symmetry-respecting basis changes. To the best of our knowledge, this constitutes the first analytic proof of a (symmetry-protected) sign problem at the level of the wave function. We speculate that our methods for evaluating symmetry-protected sign problems may also be valuable for diagnosing sign problems in the absence of symmetry.

\vspace{1.5mm}
\noindent \begin{center}\emph{Structure of the paper:}\end{center}
\vspace{1.5mm}

Our main application of symmetry-protected magic and a symmetry-protected sign problem are to SPT states. Therefore, we begin by defining SPT states and SPT phases in Section~\ref{sec: definition of SPT phases}. For convenience, we work with a definition of SPT phases phrased in terms of finite-depth quantum circuits. Then, in Section~\ref{sec: anomalous boundary symmetry action}, we describe a characteristic feature of SPT phases -- the symmetry acts anomalously near a boundary.
In the following section, Section~\ref{sec: strange correlators}, we discuss how the effects of the anomalous symmetry action can be detected using a strange correlator. To illustrate these concepts on a concrete example, we apply them to the $1$D cluster state in Section~\ref{sec: example: cluster state}. 

We then move on to assess the complexity of SPT states, starting with symmetry-protected magic in Section~\ref{sec: symmetry protected magic}. We first review the stabilizer formalism in Section~\ref{sec: review of the stabilizer formalism} before defining symmetry-protected magic in Section~\ref{sec: definition of symmetry protected magic}. Subsequently, in Section~\ref{sec: symmetry protected magic in SPT states}, we use the anomalous boundary symmetry action to show that a subset of SPT states (belonging to group cohomology SPT phases in spatial dimensions $D \geq 2$) have symmetry-protected magic.

Next, we turn to the symmetry-protected sign problem in Section~\ref{sec: symmetry protected sign problem}. In Section~\ref{sec: definition of symmetry protected sign problem}, we give a precise definition for a symmetry-protected sign problem, and then, in Section~\ref{sec: symmetry protected sign problem for SPT states}, we argue that SPT states in dimensions {$D = 1$} have a symmetry-protected sign problem relative to local bases where the symmetry is diagonal. The argument relies on the expected ``strange correlations" in SPT states. We also provide a second argument in Section~\ref{sec: symmetry protected sign problem for SPT states} based on the incompatibility between the computational wire property of one-dimensional SPT phases and bounds on measurement-induced entanglement in non-negative wave functions \cite{H16}.

We conclude by commenting on relations to previous work and by proposing future directions for studying the quantum complexity of topological phases of matter. We also state a number of conjectures, and in particular, we conjecture that states defined on qubits and belonging to the double semion phase have magic that is robust to arbitrary unitary local operations.

\section{Primer on SPT phases} \label{sec: primer on SPT phases}

To begin, we 
define SPT phases in terms of the circuit complexity of states, following Ref.~\cite{CGW10}. 
 We then describe a characteristic property of (nontrivial) SPT phases in Section~\ref{sec: anomalous boundary symmetry action}: the symmetry acts on the system in an anomalous fashion in the presence of a boundary. In certain cases, the effects of the anomalous symmetry action can be detected using strange correlators, which we define in Section~\ref{sec: strange correlators}. In Section~\ref{sec: example: cluster state}, we illustrate the concepts of the anomalous boundary symmetry action and strange correlators with an example of a well-known SPT state - the $1$D cluster state.

\subsection{Definition of SPT phases} \label{sec: definition of SPT phases}

In this section, we define SPT states and SPT phases using finite-depth quantum circuits (FDQCs). Recall that a FDQC is any unitary operator that can be written in the form:
\begin{align}
    \mathcal{U} = \prod_{\ell = 1}^d \bigg ( \prod_{j_\ell} U_{j_\ell} \bigg).
\end{align}
Here, the first product runs over layers, up to a depth $d$, and $j_\ell$ indexes unitary operators $U_{j_\ell}$ in the layer $\ell$. The unitary operators $U_{j_\ell}$, referred to as gates, are taken to be local\footnote{Throughout the text, by local, we mean that the support of the operator can be contained in a ball of fixed finite diameter.}  and to have non-overlapping supports within a given layer.
We note that the circuit is ``finite-depth'', if the depth $d$ is both finite and constant in the system size.

To define SPT states in $D$ dimensions, we consider Hilbert spaces of the form:
\begin{align}
{\mathcal{H}= \bigotimes_{i \in \Lambda} \mathcal{H}_i},
\end{align}
where $i$ labels sites on a lattice $\Lambda$ embedded in a $D$ dimensional manifold without boundary. Each site $i$ hosts a finite-dimensional Hilbert space $\mathcal{H}_i$. For SPT phases protected by a $G$ symmetry, we assume the $G$ symmetry is represented by an \textit{onsite} representation.\footnote{Note that, unless otherwise stated, we take the symmetry to be a unitary finite Abelian $0$-form symmetry.} That is, every $g$ in $G$ is represented by an 
operator:
\begin{align}
   u(g)=\prod_{i \in \Lambda}u_i(g),
\end{align}
with each $u_i(g)$ forming a linear representation
of $G$ on $\mathcal{H}_i$.
With this, an SPT state is any state that satisfies the following three conditions:
\begin{itemize}
    \item \textbf{Short-range entangled:} It can be prepared from a product state 
by a finite-depth quantum circuit.
    \item \textbf{Symmetric:} It is invariant under the onsite representation of the $G$ symmetry. 
    \item \textbf{SPT parent Hamiltonian:} It is the unique ground state of a symmetric local gapped Hamiltonian.
\end{itemize}

The SPT states are then organized into SPT phases by imposing an equivalence relation. Two SPT states are equivalent, or belong to the same phase, if one can be constructed from the other by a FDQC composed of symmetric gates -- with the possible use of ancillary lower-dimensional SPT states.
We say an SPT state is trivial if it belongs to the same equivalence class as a product state, whereas a nontrivial SPT state has entanglement that cannot be removed by making symmetry preserving local changes to the state. In other words, a nontrivial SPT state cannot be disentangled by applying a FDQC with symmetric gates.

\subsection{Anomalous symmetry action at a boundary} \label{sec: anomalous boundary symmetry action}

Having defined SPT phases, an important question is: what properties characterize an SPT phase? 
For nontrivial SPT phases, the symmetry action near a boundary is anomalous -- i.e., there is an obstruction to finding an \textit{effective} boundary symmetry action that is onsite.\footnote{Moreover, the effective boundary symmetry action cannot be made onsite through a combination of taking the tensor product with the effective boundary symmetry action of lower-dimensional SPT phases and conjugation by a FDQC.}
In what follows, we give a heuristic description of the effective boundary symmetry action, and we refer to Ref.~\cite{EN14} for more details. 
In Appendix~\ref{app: universality of the anomalous symmetry action}, we outline an argument that the obstruction gives a well-defined quantized invariant of the SPT phase.

\begin{figure}[t]
\centering
\includegraphics[width=.52\textwidth,trim={10cm 3.5cm 3cm 6.6cm},clip]{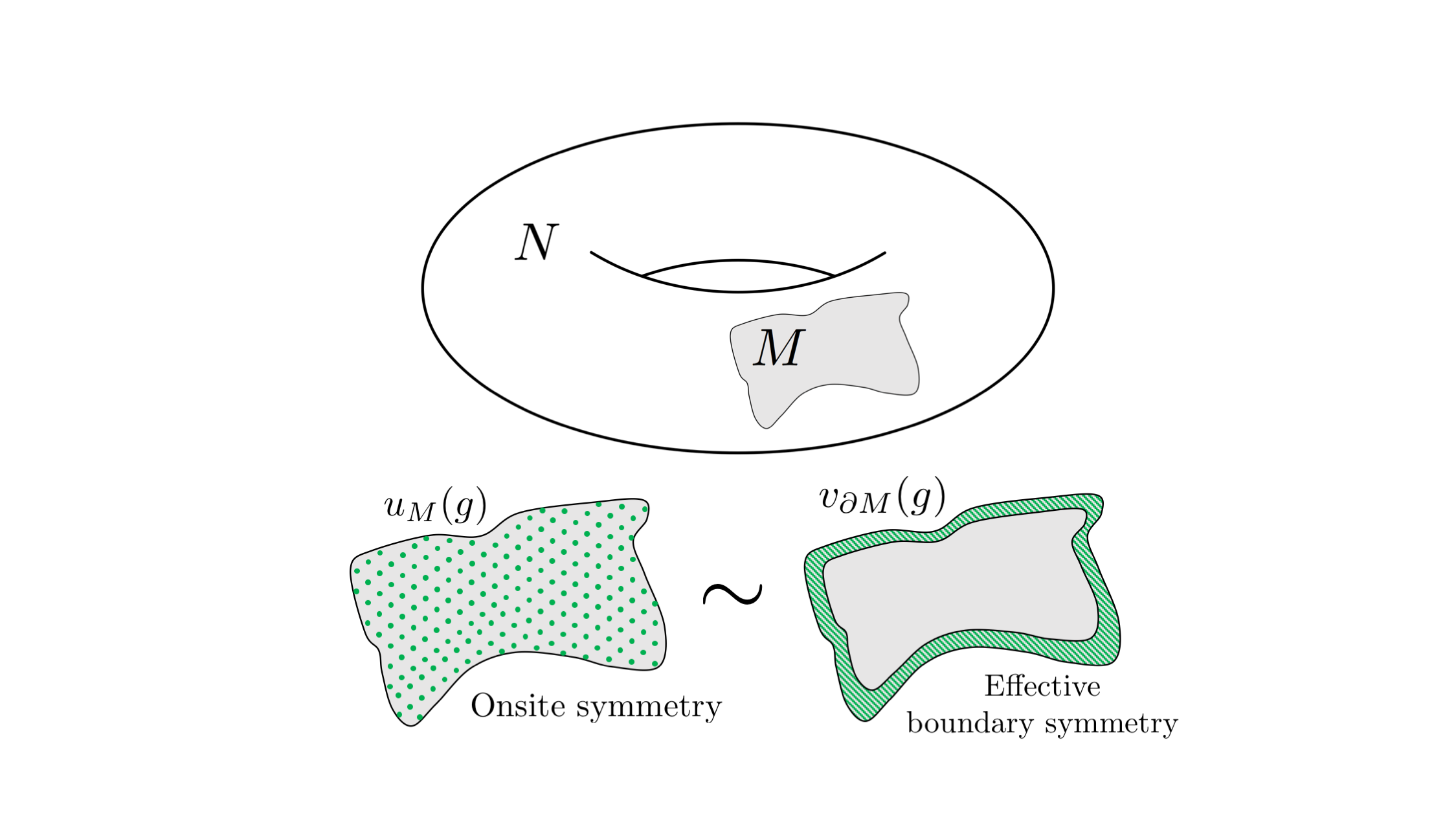}
\caption{To determine the SPT phase associated to a given SPT state, we compute an effective boundary symmetry action. This is done by truncating a corresponding SPT Hamiltonian defined on a closed manifold $N$ to a submanifold $M$ with boundary. In the low-energy Hilbert space of the truncated Hamiltonian, the onsite symmetry action $u_M(g)$ (green dots) on $M$ is equivalent to an effective boundary symmetry action $v_{\partial M}(g)$ (striped green) supported near the boundary of $M$. The symbol ``$\sim$'' denotes that $u_M(g)$ and $v_{\partial M}(g)$ are only required to be equivalent in the low-energy Hilbert space. We use the effective boundary symmetry action to show that certain SPT states have symmetry-protected magic in Section~\ref{sec: symmetry protected magic in SPT states}.}
\label{fig:M_restriction}
\end{figure}

To describe the effective boundary symmetry action, we first define the boundary Hilbert space. We consider a choice of SPT state along with a parent SPT Hamiltonian
on a manifold $N$ {without} boundary and call the energy gap between the ground state and the first excited state $\Delta$. We then imagine truncating the Hamiltonian to a submanifold $M$ {with} boundary by removing any terms whose support includes sites outside of $M$ (Fig.~\ref{fig:M_restriction}).\footnote{We assume $M$ is large compared with the Lieb-Robinson length of a FDQC that prepares the ground state of the Hamiltonian on $N$.} Furthermore, we use the tensor product structure to restrict the Hilbert space and onsite symmetry to $M$. 

After restricting to $M$, we expect the spectrum of the truncated Hamiltonian to look qualitatively different -- states now possibly lie within the energy window $\Delta$. The boundary Hilbert space is defined as the Hilbert space spanned by the states within the bulk gap $\Delta$. We assume that these low-energy states are similar to the ground state of the un-truncated Hamiltonian in regions far from the boundary.\footnote{More precisely, the reduced density matrices agree on regions sufficiently far from the boundary. This is the ${\text{TQO-}2}$ assumption in Refs.~\cite{BHM10} and \cite{BH11}.}
Hence, the low-energy states correspond to excitations localized near the boundary or degenerate ground states.

With this, the effective boundary symmetry action is any unitary linear representation of the $G$ symmetry, such that (i) its support is localized\footnote{In particular, we assume the effective boundary symmetry action is supported on $M$ and within a fixed distance from the boundary of $M$.} near the boundary of $M$ and (ii) its action agrees with the global symmetry on states within the boundary Hilbert space (Fig.~\ref{fig:M_restriction}).
While the symmetry on $M$ is onsite, the effective boundary symmetry action may be non-onsite.

Ref.~\cite{EN14} showed that certain SPT phases, known as group cohomology phases \cite{CGLW13}, exhibit an obstruction to an onsite effective boundary symmetry action captured by group cohomology. In particular, in $D$-dimensions with a $G$ symmetry, the obstruction corresponds to an element of $\mathcal{H}^{D+1}[G,U(1)]$, i.e., the $(D+1)^\text{th}$ group cohomology of $G$ with coefficients in $U(1)$. It is believed that $\mathcal{H}^{D+1}[G,U(1)]$ gives a complete classification of (bosonic) SPT phases protected by unitary symmetries in dimensions $D<4$ \cite{Gj19,K14,Y19}. We refer to SPT phases characterized by a nontrivial element of $\mathcal{H}^{D+1}[G,U(1)]$ as nontrivial group cohomology phases. 

We would like to point out that, according to the K\"{u}nneth theorem \cite{HW13}, a partial classification of SPT phases protected by a product group $\Gone \times \Gtwo$ is given by:
\begin{align} \label{eq: Kunneth part}
    \mathcal{H}^1[\Gone,\mathcal{H}^D[\Gtwo,U(1)]].
\end{align}
The SPT phases characterized by the group cohomology in Eq.~\eqref{eq: Kunneth part} are the focus of Proposition~\ref{prop: symmetry-protected magic SPT type iii} in Section~\ref{sec: symmetry protected magic}. These SPT phases can be described by decorated domain wall models \cite{CLV14}, where the ground state is a superposition of $\Gone$ domain configurations with $(D-1)$-dimensional $\Gtwo$ SPT states hosted on the domain walls. In Appendix~\ref{app: ddw models}, we argue that, for some element of $\Gone$, the corresponding effective boundary symmetry action is implemented by a FDQC that prepares a $(D-1)$-dimensional $\Gtwo$ SPT state from a product state. The effective boundary symmetry is notably not onsite because the $(D-1)$-dimensional $\Gtwo$ SPT state is entangled (for $D > 1$).

\subsection{Strange correlator} \label{sec: strange correlators}

The anomalous symmetry action at a boundary, in the previous section, enforces long-range entanglement in states describing nontrivial SPT phases on a manifold with boundary.\footnote{We note that, by an SPT phase on a manifold with boundary, we have in mind a collection of ground states, where the parent Hamiltonians are truncated SPT Hamiltonians with an arbitrary local symmetric Hamiltonian supported near the boundary.} This has been shown carefully in one and two spatial dimensions using a tensor network approach \cite{CLW11} and is believed to hold in higher dimensions. We emphasize that on a manifold \textit{without} boundary, the states in an SPT phase are short-range entangled by definition -- the long-range entanglement only appears explicitly when a boundary to a trivial SPT phase is exposed.

\begin{figure*}[t]
\centering
\includegraphics[width=1\textwidth, trim={0cm 13cm 0cm 15cm},clip]{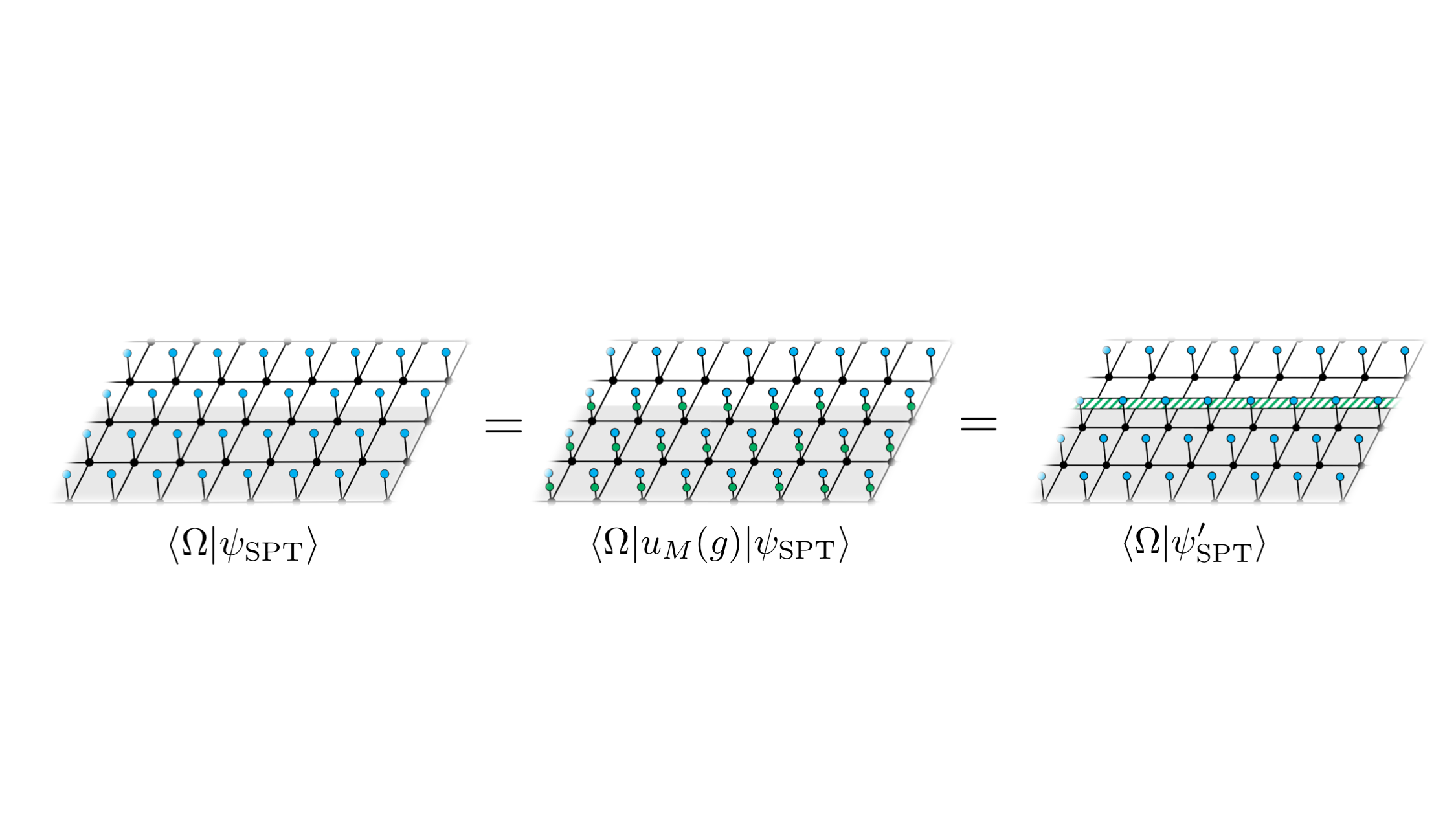}
\caption{The overlap between $\langle \Omega |$ (light blue tensors) and $|\psi_\spt \rangle$ (black tensors) takes the form of a Euclidean partition function for a ($D-1$)-dimensional system, for which, a spacetime configuration corresponds to a set of fixed indices on the virtual bonds. The ($D-1$)-dimensional system defined by $\langle \Omega | \psi_\spt \rangle$ is invariant under an anomalous symmetry if $|\psi_\spt \rangle$ belongs to a nontrivial SPT phase. This can be argued by first using the symmetry of $\langle \Omega |$ to insert the symmetry action $u_M(g)$ (green circles) restricted to a region $M$ (shaded gray). By the arguments in Ref.~\cite{WBMSHV16}, $u_M(g)$ applied to $|\psi_\spt \rangle$ is equivalent to inserting a certain tensor network operator (striped green) along the virtual bonds on the boundary of $M$. If $|\psi_\spt \rangle$ is a nontrivial SPT state, then the effective symmetry action on the virtual bonds is anomalous, and the ($D-1$)-dimensional system has an anomalous symmetry. $|\psi_\spt'\rangle$ denotes the state with the tensor network operator applied on the virtual bonds. This motivates the use of strange correlators to prove a symmetry-protected sign problem in certain SPT states (see Section~\ref{sec: symmetry protected sign problem for SPT states}).}
\label{fig: strangecorrelatormotivation}
\end{figure*}

One tool that has been developed to probe the long-range entanglement of SPT phases in the presence of a boundary is the strange correlator \cite{SV11,YBRSX14,VBWBHV18, BVHV18}. The strange correlator takes the general form:
\begin{align} \label{eq: strange correlator definition}
    \frac{\langle \Omega | \mathcal{O}_i\mathcal{O}_j |\psi_\spt \rangle}{\langle \Omega | \psi_\spt \rangle},
\end{align}
where $\langle\Omega|$ is a symmetric product state, $|\psi_\spt \rangle$ is an SPT state on a manifold \textit{without} boundary, and $\mathcal{O}_i$ and $\mathcal{O}_j$ are operators localized near the sites $i$ and $j$. {More specifically, $\mathcal{O}_i$ and $\mathcal{O}_j$ are elements of a strange order parameter, given by a set $\{\mathcal{O}_i,\mathcal{O}_j\}$ consisting of a pair of operators $\mathcal{O}_i,\mathcal{O}_j$ for each pair of sites $i,j$. Furthermore, the pairs of operators $\mathcal{O}_i,\mathcal{O}_j$ of a strange order parameter must satisfy the following three properties:
\begin{itemize}
    \item \textbf{Local:} There is a constant $r$, independent of $i,j$, such that the support of $\mathcal{O}_i$ and $\mathcal{O}_j$ can be separately contained within a ball of radius $r$.
    \item \textbf{Bounded norm:} $\mathcal{O}_i$ and $\mathcal{O}_j$ have bounded norm, i.e., $||\mathcal{O}_i||,||\mathcal{O}_j|| \leq 1$ in the operator norm. 
    \item \textbf{Charged:} $\mathcal{O}_i$ and $\mathcal{O}_j$ have nontrivial definite charge under the symmetry. That is, for a finite Abelian symmetry $G$ and any $g \in G$, $\mathcal{O}_k$ satisfies: 
\begin{align}
    u(g) \mathcal{O}_k u(g)^\dagger = e^{i \kappa(g)} \mathcal{O}_k,
\end{align}
where $e^{i \kappa(g)}$ forms a nontrivial one dimensional representation of $G$.
\end{itemize}}

The general expectation is that, for a nontrivial SPT state in either one or two dimensions, there exists a strange order parameter such that the strange correlator in Eq.~\eqref{eq: strange correlator definition} has a power law decay or is constant as the separation between $i$ and $j$ goes to infinity. This is based on numerous examples as well as physical intuition from a tensor network representation of $\langle \Omega|\psi_\spt \rangle$. 

Given a tensor network representation of the $D$-dimensional SPT state $|\psi_\spt \rangle$, we can interpret the overlap $\langle \Omega |\psi_\spt \rangle$ as a partition function for a $(D-1)$-dimensional system, as pictured in Fig.~\ref{fig: strangecorrelatormotivation}. The $(D-1)$-dimensional system is invariant under an anomalous symmetry, similar to the anomalous boundary symmetry action of an SPT phase. This can be seen by acting with the symmetry restricted to a subregion $M$ with a boundary. Ref.~\cite{WBMSHV16} argued that the symmetry action on $M$ can be replaced with an effective symmetry action on the virtual bonds of the tensor network along the boundary of $M$ (see Fig.~\ref{fig: strangecorrelatormotivation}). 

For a nontrivial SPT state, the effective symmetry action on the virtual bonds is anomalous, and hence, the ($D-1$)-dimensional partition function $\langle \Omega| \psi_\spt \rangle$ is invariant under an anomalous symmetry action. This implies that $\langle \Omega|\psi_\spt \rangle$ should be thought of as a partition function for a long-range entangled state, and the strange correlator probes the correlations in this state. Therefore, the strange correlator measures correlations similar to those that arise on the boundary of a state in an SPT phase. (See also Refs.~\cite{SV11} and \cite{YBRSX14} for a physical interpretation of the strange correlator.)

The use of strange correlators can be rigorously justified for $1$D SPT states using the notion of string-order parameters. We illustrate this for the cluster state assuming the working definition of SPT phases, given in Section~\ref{sec: definition of SPT phases}. We claim that the argument can be generalized straightforwardly to other $1$D SPT states following Refs.~\cite{EN14,PT12}. 

\subsection{Example: cluster state} \label{sec: example: cluster state}

To make the discussion more concrete, we describe the cluster state, an example of a nontrivial $1$D SPT state with a $\ZZ_2 \times \ZZ_2$ symmetry. The cluster state is defined on a $1$D lattice with $2N$ qubits and periodic boundary conditions.
% (the site $2N+1$ is the same as the site $1$). 
We denote the Pauli X and Pauli Z operator at the site $i$ by $X_i$ and $Z_i$, respectively. The onsite $\ZZ_2 \times \ZZ_2$ symmetry is then generated by the operators:
\begin{align}
u\boldsymbol{(}(g,1)\boldsymbol{)}\equiv \prod_j X_{2j}, \quad u\boldsymbol{(}(g,1)\boldsymbol{)}\equiv \prod_j X_{2j+1},
\end{align}
where we have labeled the elements of $\ZZ_2 \times \ZZ_2$ as:
\begin{align}
    \ZZ_2 \times \ZZ_2 = \{1,(g,1),(1,g),(g,g)\}.
\end{align}

The cluster state can be prepared from a product state by the FDQC $\mathcal{U}_\cs$ given as:
\begin{align} \label{eq: cluster state FDQC}
    \mathcal{U}_\cs \equiv \prod_{\langle i, i+1 \rangle} CZ_{i(i+1)}.
\end{align}
Here, the product is over pairs of neighboring sites, and the control-Z operator $CZ_{i(i+1)}$ is the two qubit operator whose action on an arbitrary computational basis state $|a\rangle_i |b \rangle_{i+1}$ is:
\begin{align}
    CZ_{i(i+1)}|a\rangle_i |b \rangle_{i+1} = (-1)^{ab} |a\rangle_i |b \rangle_{i+1}, \quad a,b \in \{0,1\}.
\end{align}
Explicitly, the cluster state is:
\begin{align}
    |\psi_\cs\rangle \equiv \mathcal{U}_\cs| \! + \ldots +\rangle,
\end{align}
where $|\! + \ldots + \rangle$ is the simultaneous $+1$ eigenstate of all Pauli X operators.

A parent Hamiltonian for the cluster state is: 
\begin{align} \label{eq: cluster state Hamiltonian}
    H_\cs \equiv \mathcal{U}_\cs \left( -\sum_i X_i \right) \mathcal{U}_\cs^\dagger = - \sum_{i=1}^{2N} Z_{i-1} X_i Z_{i+1}.
\end{align}
$H_\cs$ is gapped and has a unique ground state given that it has the same spectrum as the paramagnet Hamiltonian: $ -\sum_i X_i $. Further, it can be checked that each term of $H_\cs$ is symmetric. Therefore, $H_\cs$ is an SPT Hamiltonian.

To see that the ground state is in a nontrivial SPT phase, we introduce a boundary and study the effective symmetry action near the boundary, as described below.

\vspace{4mm}
\noindent \begin{center}\emph{Anomalous boundary symmetry action:}\end{center}
\vspace{1.5mm}

In dimension $D=1$, SPT phases with a $G$ symmetry are classified by $\mathcal{H}^{2}[G,U(1)]$, where the elements of $\mathcal{H}^{2}[G,U(1)]$ correspond to projective representations of $G$ \cite{CGW11,FK11}. We compute an effective boundary symmetry action for the cluster state model and show that it forms a projective representation of $\ZZ_2 \times \ZZ_2$. This is the nontrivial element of $\mathcal{H}^2[\ZZ_2 \times \ZZ_2, U(1)]=\ZZ_2$.

To start, we truncate the Hamiltonian $H_\cs$ in Eq.~\eqref{eq: cluster state Hamiltonian} to a lattice with $2M$ sites and open boundary conditions. This gives us the Hamiltonian $H^M_\cs$:
\begin{align}
    H^M_\cs \equiv -\sum_{i=2}^{2M-1} Z_{i-1} X_i Z_{i+1}.
\end{align}
$H^M_\cs$ has a $4$-fold degenerate ground state subspace, which follows from the fact that we have removed the terms associated to the sites $i=1$ and $i=2M$. The degenerate ground state subspace of $H^M_\cs$ defines the boundary Hilbert space.

We now derive an effective boundary symmetry action. The states in the boundary Hilbert space are $+1$ eigenstates of the terms in $H^M_\cs$, since the terms are mutually commuting and un-frustrated. Therefore, in the boundary Hilbert space, we have:
\begin{align} \label{eq: CS term in boundary Hilbert space}
    Z_{i-1} X_i Z_{i+1} \sim 1, \quad \forall i \in \{2, \ldots, 2M-1\},
\end{align}
where $\sim$ emphasizes that this holds in the boundary Hilbert space. {Note that products of the Hamiltonian terms in Eq.~\eqref{eq: CS term in boundary Hilbert space} resemble the symmetry action away from the endpoints:
\begin{align} 
    \prod_{j=1}^{M-1}Z_{2j-1} X_{2j} Z_{2j+1} &=   Z_1Z_{2M-1} \prod_{j=1}^{M-1} X_{2j} , \\
    \prod_{j=1}^{M-1}Z_{2j} X_{2j+1} Z_{2j+2} &=   Z_2Z_{2M} \prod_{j=1}^{M-1} X_{2j} .
\end{align}}
Consequently, using the relation in Eq.~\eqref{eq: CS term in boundary Hilbert space}, the generators of the $\ZZ_2 \times \ZZ_2$ symmetry can be written in the boundary Hilbert space as:
\begin{align} \label{eq: CS effective boundary symmetry action}
    u\boldsymbol{(}(g,1)\boldsymbol{)} \sim  Z_1 (Z_{2M-1}X_{2M}), \quad u\boldsymbol{(}(1,g)\boldsymbol{)} \sim (X_1 Z_2)  Z_{2M}.
\end{align} 
We define the right-hand side of the equations in Eq.~\eqref{eq: CS effective boundary symmetry action} as the operators:
\begin{align} \label{eq: define CS effective boundary symmetry action}
    v\boldsymbol{(}(g,1)\boldsymbol{)} \equiv Z_1 (Z_{2M-1}X_{2M}), \quad v\boldsymbol{(}(1,g)\boldsymbol{)} \equiv (X_1 Z_2)  Z_{2M}.
\end{align}
These 
define a $\ZZ_2 \times \ZZ_2$ effective boundary symmetry action, since they form a unitary linear representation of $\ZZ_2 \times \ZZ_2$, are localized near the boundary, and, by definition, agree with the global symmetry action in the boundary Hilbert space. 

The effective boundary symmetry action generated by the operators in Eq.~\eqref{eq: define CS effective boundary symmetry action} is not onsite -- i.e., it is not in the form of a tensor product of linear representations at each site (as defined in Section~\ref{sec: definition of SPT phases}). Instead, the action at endpoints $i=1$ and $i=2M$ are independently projective representations of $\ZZ_2 \times \ZZ_2$. For example, at the endpoint $i=1$, we have:
\begin{align}
    v_{i=1}\boldsymbol{(}(g,1)\boldsymbol{)} \equiv Z_1, \quad v_{i=1}\boldsymbol{(}(1,g)\boldsymbol{)} \equiv X_1 Z_2.
\end{align}
These give a projective representation, as can be seen by the commutation relations between $v_{i=1}\boldsymbol{(}(g,1)\boldsymbol{)}$ and $v_{i=1}\boldsymbol{(}(1,g)\boldsymbol{)}$:
\begin{align} \label{eq: CS projective representation}
    v_{i=1}\boldsymbol{(}(g,1)\boldsymbol{)}v_{i=1}\boldsymbol{(}(1,g)\boldsymbol{)}=-v_{i=1}\boldsymbol{(}(1,g)\boldsymbol{)}v_{i=1}\boldsymbol{(}(g,1)\boldsymbol{)}.
\end{align}
Thus, the effective boundary symmetry action is anomalous.\footnote{Importantly, the projective representations at the endpoints cannot be made into linear representations by conjugating by a FDQC.} In Appendix~\ref{app: universality of the anomalous symmetry action}, we show that the anomalous symmetry action implies that the cluster state cannot be disentangled by a FDQC composed of symmetric gates.

\vspace{1.5mm}
\noindent \begin{center}\emph{Strange correlator:}\end{center}
\vspace{1.5mm}

For the cluster state, we can use the exactly solvable Hamiltonian $H_\cs$ to identify a suitable strange order parameter. To see this, we consider the product of Hamiltonian terms:
\begin{eqs} \label{eq: cluster state product of Hamiltonian terms}
   \prod_{k=i}^{j-1}Z_{2k}X_{2k+1}Z_{2k+2} = ( \prod_{k=i}^{j-1} X_{2k+1} ) Z_{2i}Z_{2j}.
\end{eqs}
This gives us the identity:
\begin{eqs} \label{eq: cluster state string order}
    \langle + \ldots +| \psi_\cs \rangle &= \langle + \ldots +| ( \prod_{k=i}^{j-1} X_{2k+1} ) Z_{2i}Z_{2j} |\psi_\cs \rangle \\ &= \langle +\ldots +| Z_{2i}Z_{2j} |\psi_\cs \rangle,
\end{eqs}
where the first equality comes from the fact that $|\psi_\cs \rangle$ is a $+1$ eigenstate of each term of $H_\cs$, and the second equality uses that $\langle + \ldots +|$ is a $+1$ eigenstate of every Pauli X operator. From Eq.~\eqref{eq: cluster state string order}, we have:
\begin{align} \label{eq: cluster state strange order parameter}
    \frac{\langle +\ldots +| Z_{2i}Z_{2j} |\psi_\cs \rangle}{\langle +\ldots + |\psi_\cs \rangle} = 1,
\end{align}
for any choice of $i$ and $j$. {We see that the set $\{Z_{2i},Z_{2j}\}$ can be used as a strange order parameter. {Note that the operators $\{Z_{2i},Z_{2j}\}$ are charged under the $\ZZ_2 \times \ZZ_2$ symmetry, as required for a strange order parameter.} }
Moreover, we have found a strange order parameter for $|\psi_\cs \rangle$ such that the strange correlator is constant in the separation of $2i$ and $2j$, according to Eq.~\eqref{eq: cluster state strange order parameter}.

As for other states in the same phase as $|\psi_\cs \rangle$, we can use the operator in Eq.~\eqref{eq: cluster state product of Hamiltonian terms} to identify a strange order parameter for which the strange correlator is constant in the limit $|i-j| \to \infty$. For example, let $|\psi_\cs '\rangle$ be the state prepared from $|\psi_\cs \rangle$ by the FDQC $\mathcal{U}_\sym$ composed of symmetric gates:
\begin{align}
    |\psi_\cs'\rangle \equiv \mathcal{U}_\sym |\psi_\cs \rangle.
\end{align}
$|\psi_\cs' \rangle$ is invariant under the operator:
\begin{align} \label{eq: cluster state prime string order}
    \mathcal{U}_\sym \left[( \prod_{k=i}^{j-1} X_{2k+1} ) Z_{2i}Z_{2j} \right] \mathcal{U}_\sym^\dagger.
\end{align}
%\zw{Big brackets, same below}
Since $\mathcal{U}_\sym$ is built from symmetric gates, the operator in Eq.~\eqref{eq: cluster state prime string order} is equal to:
\begin{align}
   (\prod_{k=i}^{j-1} X_{2k+1}) \mathcal{O}_i \mathcal{O}_j,
\end{align}
for some unitary local charged operators $\mathcal{O}_i$ and $\mathcal{O}_j$. Following Eqs.~\eqref{eq: cluster state string order} and \eqref{eq: cluster state strange order parameter}, we can define a strange order parameter from the collection of $\mathcal{O}_i$ and $\mathcal{O}_j$ for varying endpoints $i$ and $j$. Thus, every state in the SPT phase (as defined in Section~\ref{sec: definition of SPT phases}) admits a strange order parameter with a constant strange correlator.

We note that the operators in Eqs.~\eqref{eq: cluster state string order} and \eqref{eq: cluster state prime string order} are the more familiar string-order parameters that characterize $1$D SPT phases \cite{PT12}. These naturally lead to strange order parameters with a constant strange correlation.

\section{Symmetry-protected magic} \label{sec: symmetry protected magic}

In this section, we introduce symmetry-protected magic and demonstrate that it is a feature of a large class of SPT states. To start, we review the stabilizer formalism and describe how it can be simulated efficiently on a classical computer. The stabilizer formalism is insufficient for universal quantum computing, but leads to the concept of magic -- a resource that can be used to help overcome the limitations of the stabilizer formalism. We then define the notion of symmetry-protected magic and use it to assess the magic in SPT states. In particular, we show that SPT states belonging to group cohomology phases in $D\geq 2$ dimensions have symmetry-protected magic.

\subsection{Review of the stabilizer formalism} \label{sec: review of the stabilizer formalism}

The stabilizer formalism has been instrumental to our understanding of the complexity of quantum phases of matter and often provides simple, workable examples, such as the cluster state model in Section~\ref{sec: example: cluster state}. In this section, we summarize the key concepts of the stabilizer formalism to ensure the text is self-contained. We refer to Refs.~\cite{NC10, G97, GMC14, VFGE12,VMGE14} for more thorough reviews. 

To keep the discussion general, we describe the stabilizer formalism on systems of $q$-dimensional qudits (i.e. $q$-dimensional Hilbert spaces), where $q$ is not necessarily prime. We begin by generalizing the usual Pauli Z and Pauli X operators to: 
\begin{align} \label{eq: Z and X definitions}
   Z \equiv \sum_{j \in \ZZ_q} e^{\frac{2 \pi i}{q}j} |j\rangle \langle j |, \quad X \equiv \sum_{j \in \ZZ_q} |j+1\rangle \langle j |,
\end{align}
where the computational basis states for a $q$-dimensional qudit are labeled by $j \in \ZZ_q$.
If $q$ is odd, the Pauli operators are generated by products of $Z$ and $X$, and if $q$ is even, the Pauli operators are generated by products of $Z$, $X$, and the phase $i$.\footnote{The group generated by the Pauli operators is commonly called the Heisenberg-Weyl group.} For systems of more than one qudit, we refer to a tensor product of Pauli operators as a Pauli string. We say a Pauli string is Z-type or X-type if, up to a phase, it consists of only products of $Z$ operators or $X$ operators, respectively. 

{With this, we can introduce stabilizer states. Stabilizer states are defined by the property that for any stabilizer state $|\psi_\stab\rangle$, there exists a group $\mathcal{G}$ of mutually commuting Pauli strings such that $|\psi_\stab \rangle$ is the unique state satisfying:
\begin{align}
    S|\psi_\stab \rangle = |\psi_\stab \rangle, \quad \forall S \in \mathcal{G}.
\end{align}
Importantly, the stabilizer state is \textit{uniquely} specified by the Abelian group $\mathcal{G}$ of Pauli strings. We refer to the elements in $\mathcal{G}$ as stabilizers and call the group $\mathcal{G}$ a stabilizer group. We say the stabilizer group $\mathcal{G}$ ``stabilizes'' or ``fixes'' $|\psi_\stab \rangle$ to mean that $|\psi_\stab \rangle$ is in the simultaneous $+1$ eigenspace of all of the stabilizers. More generally, any group of mutually commuting Pauli strings (which does not include $-1$) is a stabilizer group, although it may not fix a unique stabilizer state.} 

{As a first example, the product state $|\! +\ldots + \rangle$ is a stabilizer state. 
$|\! +\ldots +\rangle$ is uniquely stabilized by the stabilizer group $\mathcal{G}_0$ generated by a Pauli X operator for each site:
\begin{align}
   \mathcal{G}_0 \equiv \langle X_i : i \in \text{sites} \rangle. 
\end{align}
Similarly, the cluster state $|\psi_\cs\rangle$ presented in Section~\ref{sec: example: cluster state} is a stabilizer state. The corresponding stabilizer group $\mathcal{G}_\cs$ is generated by the terms of the cluster state Hamiltonian $H_\cs$:
\begin{align}
    \mathcal{G}_\cs \equiv \langle Z_{i-1}X_i Z_{i+1} : i \in \text{sites} \rangle.
\end{align}}

{A useful class of unitary operators in the context of stabilizer states are the Clifford unitaries. A Clifford unitary is any unitary operator that maps Pauli strings to Pauli strings by conjugation. Explicitly, for any Pauli string $P$, a Clifford unitary $U$ satisfies:
\begin{align}
    UPU^\dagger = Q,
\end{align}
for some Pauli string $Q$. Consequently, Clifford unitaries also map stabilizer states to stabilizer states. For example, the FDQC $\mathcal{U}_\cs$, introduced in Eq.~\eqref{eq: cluster state FDQC}, is a Clifford unitary. It maps $\mathcal{G}_0$ to $\mathcal{G}_\cs$ by conjugation and maps $|\! + \ldots + \rangle$ to $|\psi_\cs \rangle$.}

At this point, one can define a computational scheme based on applying Clifford unitaries to stabilizer states and making measurements of Pauli strings. However, this restricted set of operations -- the stabilizer operations -- can be efficiently simulated by a classical computer. This is the statement of the Gottesman--Knill theorem \cite{G99} and a consequence of the fact that a stabilizer state can be fully characterized by a stabilizer group. Indeed, stabilizer groups that uniquely fix a stabilizer state are generated by a number of Pauli strings that grows linearly in the number of sites \cite{G14}. A stabilizer state can therefore be efficiently specified by a stabilizer group, and moreover, the effects of evolution by a Clifford unitary and measurements of Pauli strings can be determined by appropriately modifying the stabilizer group. We see that the stabilizer operations are no more powerful than a classical computer, and additional ingredients are needed to promote it to a universal set of operations.

Before describing how the stabilizer formalism can be supplemented to achieve universal quantum computation, we remark that the generators of a stabilizer group can be used to build a stabilizer Hamiltonian. More specifically, given a stabilizer group $\mathcal{G}$ that uniquely stabilizes a state $|\psi_\stab \rangle$, we can construct a Hamiltonian:
\begin{align}
    H_\stab \equiv - \sum_{S \in \mathcal{S}}S+\text{h.c.},
\end{align}
where $\mathcal{S}$ denotes a set of stabilizers that generate $\mathcal{G}$. The unique ground state of $H_\stab$ is $|\psi_\stab \rangle$, since it is a $+1$ eigenstate of each $S \in \mathcal{S}$ and is uniquely fixed by $\mathcal{G}$. We note that $H_\stab$ might not be local.

\subsection{Definition of symmetry-protected magic} \label{sec: definition of symmetry protected magic}

Stabilizer operations, reviewed in the previous section, can be simulated efficiently on a classical computer, but the full power of quantum computation can be recovered by supplementing the stabilizer operations with ancillary non-stabilizer states. In fact, any non-stabilizer (pure) state can be used as ancillary states to promote the stabilizer operations to a universal set of operations. In this context, the non-stabilizer states are referred to as magic states.
In a precise sense, the magic of a state (or the ``non-stabilizerness'') can be treated as a resource, similar to viewing entanglement as a resource. Consequently, resource-theoretical tools have been developed to quantify the amount of magic in a state (see e.g., Refs.~\cite{VMGE14,HC17,BBCCGH19,LBT19,WWS20}), however few analytical statements have been made about magic in many-body systems. To make progress in this direction, we define the following coarse measure of the magic to help understand the large-scale structure of magic in a many-body state:

\begin{definition}[Long-range magic] \label{def: topologically protected magic}
A state $|\psi\rangle$ has long-range magic, if, for any finite-depth quantum circuit $\mathcal{U}$, the state $\mathcal{U}|\psi\rangle$ is a magic state. 
\end{definition} 

In other words, a state with long-range magic has magic that cannot be removed by any FDQC. In this sense, the state can serve as a robust source of magic.  We would like to point out that concepts similar to long-range magic have been recently introduced in  Ref.~\cite{WCS20} in the context of conformal field theories. We hope to comment on long-range magic in future work, but in the present text we focus on a restricted notion of long-range magic. 

In particular, we consider magic that cannot be removed by any FDQC composed of \textit{symmetric} gates. We say such a state has symmetry-protected magic. This is defined more precisely as:

\begin{definition}[Symmetry-protected magic] \label{def: symmetry-protected magic}
A state $|\psi\rangle$ has symmetry-protected magic, if, for any finite-depth quantum circuit $\mathcal{U}_\sym$ composed of symmetric gates, the state $\mathcal{U}_\sym|\psi\rangle$ is a magic state.
\end{definition}
As a proof of concept, we show that certain SPT states have symmetry-protected magic. 

\subsection{Symmetry-protected magic in SPT states} \label{sec: symmetry protected magic in SPT states}

Our main objective in this section is to show that SPT states, in particular those belonging to nontrivial group cohomology phases in dimensions $D\geq 2$, have symmetry-protected magic. This includes, for example, the $\ZZ_2$ SPT model introduced in Ref.~\cite{LG12}. We divide our results into two propositions. Proposition~\ref{prop: symmetry-protected magic SPT} applies to SPT states belonging to group cohomology phases protected by a symmetry of the form $\ZZ_q^m$ (a product of $m$ copies of $\ZZ_q$) and assumes that the state is defined on a system of $q$-dimensional qudits. {This is a natural assumption, as for example, the group cohomology models with a $\ZZ_q^m$ symmetry in Ref.~\cite{CGLW13} are defined on a lattice with $m$ $q$-dimensional qudits per site and a symmetry represented by tensor products of $X$ operators.} Proposition~\ref{prop: symmetry-protected magic SPT type iii} only applies to a subset of group cohomology phases, but makes no assumption on the dimension of the qudits.
In both cases, the proof relies on the anomalous boundary symmetry action characteristic of nontrivial group cohomology SPT phases (see Section~\ref{sec: anomalous boundary symmetry action}). After proving the two propositions, we comment on SPT states that fall outside of our argument -- these correspond to SPT phases that can indeed be described efficiently by the stabilizer formalism.

\begin{proposition} \label{prop: symmetry-protected magic SPT}
Any SPT state belonging to a nontrivial group cohomology phase in $D\geq 2$ dimensions protected by a $G = \ZZ_q^m$ symmetry has symmetry protected magic, if it is defined on $q$-dimensional qudits and the symmetry is represented by tensor products of Pauli operators. 
\end{proposition}

\noindent \emph{Proof of Proposition~\ref{prop: symmetry-protected magic SPT}.} 
We defined SPT phases as collections of SRE states that are equivalent under FDQCs composed of symmetric gates. Therefore, if an SPT state has symmetry-protected magic, it implies that every state in the SPT phase must be a magic state. To prove the proposition, it is then sufficient to show that there are no stabilizer states belonging to nontrivial group cohomology phases in $D \geq 2$ dimensions with a $G$ symmetry represented by Pauli strings.  

With this, we proceed by deriving a contradiction. We assume that there is a stabilizer state $|\psi_\stab \rangle$ belonging to a nontrivial group cohomology phase in $D \geq 2$ dimensions protected by a $G$ symmetry represented by a Pauli string $P(g)$ for every $g \in G$. We argue that this is in conflict with the anomalous boundary symmetry action expected in the nontrivial SPT phase. 

The first step is to find a local symmetric stabilizer Hamiltonian whose unique ground state is $|\psi_\stab \rangle$. Since $|\psi_\stab \rangle$ is an SPT state, it has a local parent Hamiltonian (albeit possibly non-stabilizer), and it is invariant under the $G$ symmetry, i.e., $P(g) |\psi_\stab \rangle = |\psi_\stab \rangle$, for all $g \in G$. In Appendix~\ref{app: Existence of a local stabilizer parent Hamiltonian}, we show that this, in fact, implies that there exists a local symmetric stabilizer Hamiltonian $H_\stab$ whose unique ground state is $|\psi_\stab \rangle$ and which commutes with the $G$ symmetry (see Lemma~\ref{lem: local stabilizer hamiltonian}). 

We can now determine the SPT phase by using $H_\stab$ to compute the anomalous symmetry action at the boundary (analogous to the calculation of the anomalous boundary symmetry action for the cluster state in Section~\ref{sec: example: cluster state}). For this purpose, we introduce a boundary by truncating the Hamiltonian $H_\stab$ to a region $M$ with boundary. We define the truncated Hamiltonian $H_\stab^M$ by removing any term whose support is not entirely contained within $M$.\footnote{To avoid pathologies, we require that $M$ is large compared to the size of the supports of the terms in $H_\stab$. More precisely, we require $d_M \gg d_S$, where $d_M$ is the diameter of the largest ball inscribing $M$, and $d_S$ is the minimum diameter such that the support of each stabilizer term fits within a ball of diameter $d_S$.} The global symmetry action $P(g)$ can also be restricted to $M$ using that $P(g)$ is a tensor product of linear representations $P_i(g)$ of $G$:
\begin{align} \label{eq: restricted sym def}
    P(g) \equiv \prod_{i} P_i(g) \rightarrow P_M(g) \equiv \prod_{i \in M} P_i(g).
\end{align}
The truncated Hamiltonian has a $G$ symmetry represented by the operators $P_M(g)$, given above. 

Similar to $H_\stab$, the truncated Hamiltonian is a sum of symmetric commuting stabilizer terms, but unlike $H_\stab$, it will generically have a ground state degeneracy, and the degenerate ground state subspace forms the boundary Hilbert space. The anomalous behavior of the symmetry, characteristic of the SPT phase, is revealed by the effective symmetry action on the boundary Hilbert space.
We recall from Section~\ref{sec: anomalous boundary symmetry action} that the effective boundary symmetry action is any operator localized near the boundary of $M$, whose action is equivalent to the symmetry action of $P_M(g)$, within the boundary Hilbert space.

\begin{figure}[t]
\centering
\includegraphics[width=.5\textwidth,trim={9cm 1.2cm 4cm 1cm},clip]{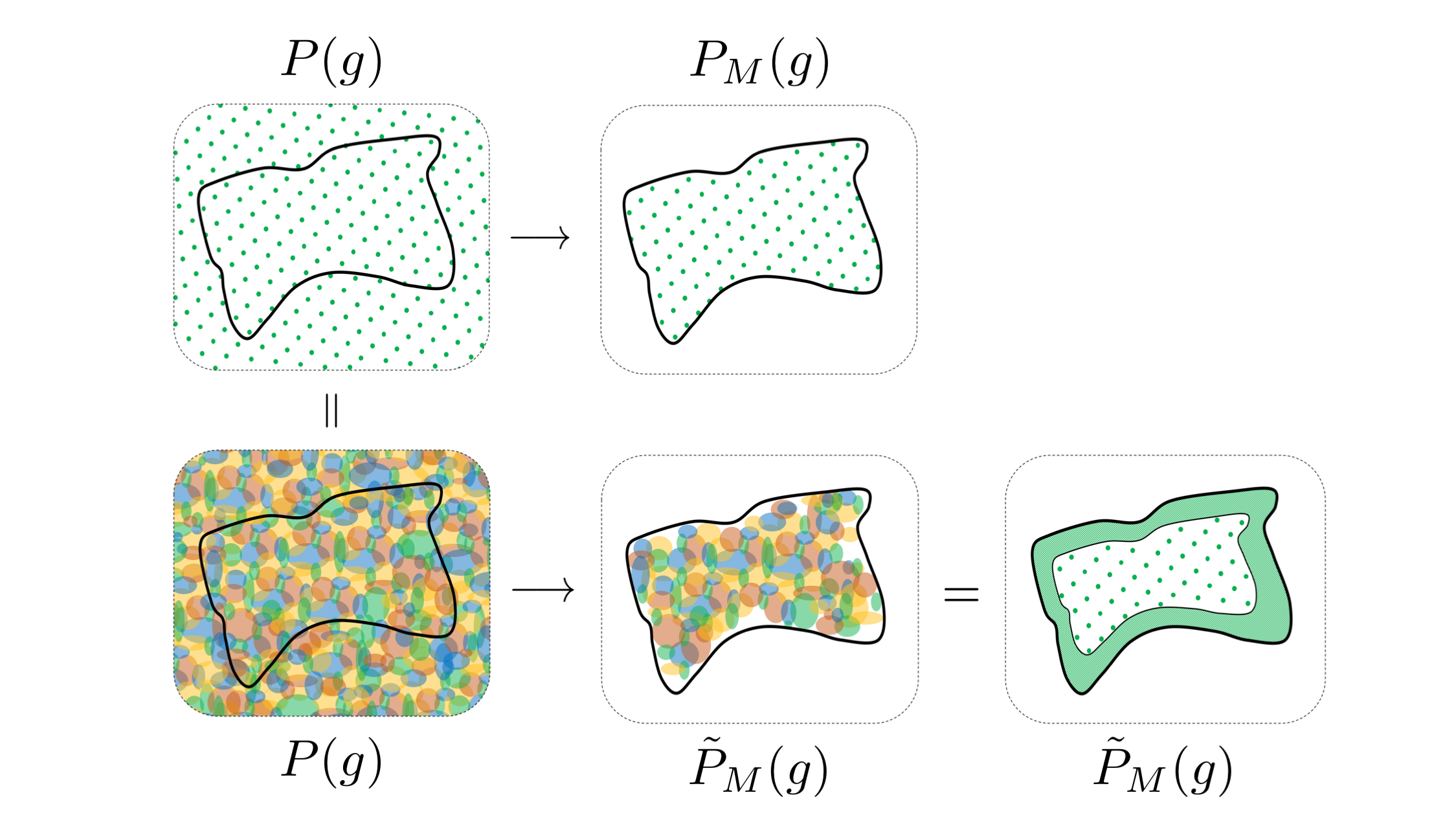}
\caption{We determine the effective boundary symmetry action from $H_\stab$ by first observing that the global symmetry $P(g)$, for any $g \in G$, can be expressed as a product of terms in $H_\stab$ (the supports of the stabilizer terms are depicted with colored ovals). The global symmetry action can be restricted to a submanifold $M$ (outlined in black) in two ways. Restricting to $M$ by using the tensor product structure of $P(g)$ results in $P_M(g)$, while restricting to $M$ using the product of stabilizer terms gives $\tilde{P}_M(g)$. $\tilde{P}_M(g)$ acts like the onsite symmetry away from the boundary of $M$.}
\label{fig: symtruncation}
\end{figure}

The effective symmetry action at the boundary can be computed by first observing that the global symmetry action $P(g)$ can be written as a product of the stabilizer terms of the un-truncated Hamiltonian $H_\stab$. This is because, $H_\stab$ commutes with $P(g)$ and has a unique ground state. Therefore, $P(g)$ is contained in the stabilizer group generated by the terms of $H_\stab$ (see Lemma~\ref{lem: Pauli string commute and stabilizer invariance} in Appendix~\ref{app: Existence of a local stabilizer parent Hamiltonian}). Consequently, $P(g)$ can be written as:
\begin{align} \label{eq: global sym as stab}
    P(g) = \prod_{S_j \in \mathcal{S}_{P(g)}}S_j,
\end{align}
where $\mathcal{S}_{P(g)}$ is defined as the set of terms in $H_\stab$ whose product is $P(g)$. By using the expression for $P(g)$ in Eq.~\eqref{eq: global sym as stab}, $P(g)$ can instead be truncated to $M$ by retaining only the stabilizers $S_j$ whose support $\text{supp}(S_j)$ is entirely contained in $M$ (see Fig.~\ref{fig: symtruncation}):
\begin{align} \label{eq: tilde P def}
    \tilde{P}_M(g) \equiv \prod_{\substack{S_j \in \mathcal{S}_{P(g)} \\ \text{supp}(S_j) \subset M}}S_j.
\end{align}
We note that $\tilde{P}_M(g)$ is a product of terms in $H^M_\stab$ and as such, acts as the identity on the boundary Hilbert space. This implies, in particular, that $P_M(g)$ is equivalent to $P_M(g)\tilde{P}_M^\dagger(g)$ in the boundary Hilbert space.
Further, by comparing Eqs.~\eqref{eq: global sym as stab} and \eqref{eq: tilde P def}, we see that the action of $\tilde{P}_M(g)$ is equivalent to that of $P(g)$ on sites in $M$ greater than a fixed distance from the boundary of $M$.
Therefore, the support of $P_M(g)\tilde{P}_M^\dagger(g)$ is contained in $M$ and localized near the boundary of $M$. As a result, we can take the effective boundary symmetry action to be:
\begin{align} \label{eq: effective boundary symmetry action}
    \mathcal{P}(g) \equiv P_M(g)\tilde{P}_M^\dagger(g).
\end{align}

The effective boundary symmetry action in Eq.~\eqref{eq: effective boundary symmetry action} forms a linear representation of $G = \ZZ_q^m$ in the boundary Hilbert space. This means that it obeys the group laws of $G$ up to products of stabilizer terms in $H^M_\stab$. In Appendix~\ref{app: modified effective boundary symmetry action}, we show that, assuming the system is defined on $q$-dimensional qudits, the effective boundary symmetry action can be modified by stabilizers in $H^M_\stab$ to guarantee that the group relations are satisfied exactly. We denote the modified effective boundary symmetry action by $\mathcal{P}'(g)$.

The characteristic group cohomology class of the SPT order can be deduced from the modified effective boundary symmetry action using the methods of Ref.~\cite{EN14}, where the group cohomology class manifests as an obstruction to realizing the effective boundary symmetry action onsite (as a tensor product of {linear} representations at each site). 
From the definition of $\mathcal{P}'(g)$ in Appendix~\ref{app: modified effective boundary symmetry action}, it can be seen that
 $\mathcal{P}'(g)$ is a tensor product of Pauli operators:
\begin{align} \label{eq: effective boundary symmetry action 2}
    \mathcal{P}'(g)=\prod_{k}\mathcal{P}'_k(g),
\end{align}
with the product over sites $k$ in $M$ close to the boundary of $M$.
% within a distance $d_S$ of the boundary of $M$. 
While $\mathcal{P}'(g)$ forms a linear representation of $G$ in the boundary Hilbert space, the operators $\mathcal{P}'_k(g)$ might only satisfy the group laws projectively. In dimensions $D \geq 2$ this does not pose an obstruction to an onsite representation of the effective boundary symmetry action. The algorithm in Ref.~\cite{EN14} shows that the effective boundary symmetry action in Eq.~\eqref{eq: effective boundary symmetry action 2} corresponds to the trivial element of $\mathcal{H}^{D+1}[G,U(1)]$ (if $D \geq 2$).

To motivate this conclusion, we argue that any projective representations formed by $\mathcal{P}'_k(g)$ can be resolved by modifying $H_\stab$ with decoupled $1$D SPT Hamiltonians acting on ancillary qudits. Importantly, the $1$D SPT Hamiltonians do not change the $D\geq 2$ SPT phase described by $H_\stab$. Moreover, the $1$D SPT Hamiltonians can always be chosen so that their projective effective boundary symmetry actions (see Section~\ref{sec: example: cluster state}, for example) compensate for the projective representations formed by the $\mathcal{P}'_k(g)$. Then by locally redefining the sites, the projective representation from $\mathcal{P}'_k(g)$ and the effective boundary symmetry of the $1$D SPT phases form a linear representation on the composite site. 

Therefore, the effective boundary symmetry action in Eq.~\eqref{eq: effective boundary symmetry action 2} is non-anomalous, and by the universality of the anomalous boundary symmetry action (see Appendix~\ref{app: universality of the anomalous symmetry action}), $|\psi_\stab \rangle$ cannot be a member of a nontrivial group cohomology SPT phase in $D \geq 2$. This contradicts the initial assumption and implies that there are no stabilizer states in nontrivial group cohomology SPT phases in $D\geq 2$ dimensions protected by a symmetry represented by Pauli strings. Thus, the nontrivial SPT states described in the proposition have symmetry-protected magic.~$\square$ \\

Note that Proposition~\ref{prop: symmetry-protected magic SPT} assumes that the SPT state is defined on a system of $q$-dimensional qudits for a symmetry $\ZZ_q^m$. For some SPT phases, we expect that the restriction on the dimension of the qudits is necessary. {This restriction is important for ensuring that the effective boundary symmetry action forms a linear representation of the symmetry group outside of the boundary Hilbert space. In future work, we will describe on a stabilizer model for a nontrivial $\ZZ_2$ SPT phase in $D=2$ dimensions defined on a system of $4$-dimensional qudits \cite{CDESTW21}. In this case, the effective boundary symmetry action is represented by a tensor product of Pauli operators, but the Pauli operators do not form a linear representation outside of the boundary Hilbert space.}

At the same time, there are group cohomology phases for which the restriction on the dimension of the qudits in Proposition~\ref{prop: symmetry-protected magic SPT} is unnecessary. A stronger statement can be made for SPT phases that can be described by a decorated domain wall model, i.e., the SPT phase is protected by a symmetry of the form $\Gone \times \Gtwo$ and characterized by an element of $\mathcal{H}^1[\Gone,\mathcal{H}^D[\Gtwo,U(1)]]$. We formulate this as Proposition~\ref{prop: symmetry-protected magic SPT type iii}.

\begin{proposition} \label{prop: symmetry-protected magic SPT type iii}
Any SPT state belonging to an $\Gone \times \Gtwo$ SPT phase in $D\geq 2$ dimensions characterized by a nontrivial element of $\mathcal{H}^1[\Gone,\mathcal{H}^D[\Gtwo,U(1)]]$ has symmetry protected magic, if the symmetry is represented by tensor products of Pauli operators. 
\end{proposition}

\noindent \emph{Proof of Proposition~\ref{prop: symmetry-protected magic SPT type iii}.} We let $|\psi_\stab \rangle$ be a stabilizer state within an $\Gone \times \Gtwo$ SPT phase in $D\geq 2$ dimensions. Following the proof of Proposition~\ref{prop: symmetry-protected magic SPT}, we identify a local stabilizer parent Hamiltonian $H_\stab$ for $|\psi_\stab \rangle$ and define an effective boundary symmetry action as in Eq.~\eqref{eq: effective boundary symmetry action}.

We argue that the effective boundary symmetry action of the stabilizer model is unable to reproduce the anomalous symmetry action of an SPT phase characterized by a nontrivial element of $\mathcal{H}^1[\Gone,\mathcal{H}^D[\Gtwo,U(1)]]$. In Appendix~\ref{app: ddw models} we prove that, the effective boundary symmetry action for some element of $\Gone \times \Gtwo$ must be a FDQC that prepares a nontrivial $(D-1)$-dimensional $\Gtwo$ SPT state from a trivial SPT state. The effective boundary symmetry action of the stabilizer model, however, is a Pauli string for every element of $\Gone \times \Gtwo$. Pauli strings are insufficient for constructing a nontrivial $(D-1)$-dimensional $\Gtwo$ SPT state from a trivial SPT state if $D \geq 2$. Therefore, $|\psi_\stab \rangle$ cannot belong to an SPT phase characterized by a nontrivial element of $\mathcal{H}^1[\Gone,\mathcal{H}^D[\Gtwo,U(1)]]$. $\square$ \\

The simplest example of an SPT phase that satisfies the conditions of Proposition~\ref{prop: symmetry-protected magic SPT type iii} is a $\ZZ_2 \times \ZZ_2 \times \ZZ_2$ SPT phase in $D=2$ dimensions. In this case, $\Gone$ and $\Gtwo$ are equal to $\ZZ_2$ and $\ZZ_2 \times \ZZ_2$, respectively. As argued in Appendix~\ref{app: ddw models}, the effective boundary symmetry action corresponding to the generator of $\Gone$ prepares a $\ZZ_2 \times \ZZ_2$ SPT state from a product state. Proposition~\ref{prop: symmetry-protected magic SPT type iii} tells us that every state belonging to this SPT phase has magic, assuming the symmetry is represented by a Pauli string.

In Propositions~\ref{prop: symmetry-protected magic SPT} and \ref{prop: symmetry-protected magic SPT type iii}, we have shown that a large class of SPT states have symmetry-protected magic. However, there are notable examples of SPT states without symmetry-protected magic. For example, the cluster state, described in Section~\ref{sec: example: cluster state}, is a stabilizer state. In this case, the anomalous boundary symmetry action corresponds to projective representations, and there is no obstruction to forming projective representations with Pauli strings. As another example, the ground state of the $2$D CZX-model in Ref.~\cite{CLW11} is a stabilizer state, but the onsite symmetry is not represented by a product of Pauli operators.

There are also well-known examples of stabilizer states in nontrivial SPT phases protected by subsystem symmetries (e.g. the $2$D cluster state) \cite{DWY18,YDBS18} or protected by higher-form symmetries (e.g. the $3$D cluster state) \cite{RBH01,Y16,RBH05,CET20,RYKB17,RB20}. In our argument, we assumed that the protecting symmetry is a $0$-form symmetry, i.e., it is supported on a codimension-$0$ manifold. The assessment of the anomalous nature of the effective boundary symmetry action was specific to $0$-form SPT phases. We expect that the proposition can be generalized by accounting for the anomalies associated to subsystem SPT phases and higher-form SPT phases, as described in Refs.~\cite{DWY18} and \cite{TW20}. Evidently, in some cases, the anomalous boundary symmetry action of these SPT phases can be described by Pauli operators. 

We have qualified that Propositions~\ref{prop: symmetry-protected magic SPT} and \ref{prop: symmetry-protected magic SPT type iii} apply to SPT phases classified by group cohomology, but our results may be more general. There are, in fact, known SPT phases in dimensions $D \geq 3$ that are outside of the group cohomology classification -- aptly named the beyond cohomology phases \cite{K14,WGW15}. In dimension $D=3$, there is a beyond cohomology phase protected by time-reversal symmetry that admits a stabilizer representation \cite{BCFV14}. However, this SPT phase falls outside of the purview of our argument, since the symmetry is anti-unitary and is not represented by Pauli strings. On the other hand, in $D=4$, there is a beyond cohomology SPT phase protected by a unitary $\ZZ_2$ symmetry represented by a Pauli string \cite{FHH20}, and we expect the proof of Proposition~\ref{prop: symmetry-protected magic SPT} applies in this case.  Indeed, it was recently argued that the effective boundary symmetry action of the SPT phase corresponds to a nontrivial $3$D quantum cellular automaton \cite{HFH18,FHH20}. The operator in Eq.~\eqref{eq: effective boundary symmetry action 2} is certainly a trivial quantum cellular automaton. \\

\section{Symmetry-protected sign problem} \label{sec: symmetry protected sign problem}

The sign problem is a notorious obstacle in efficiently simulating many-body quantum systems using Monte Carlo methods. Often, the sign problem refers to a difficulty in writing the partition function of a quantum system as a classical partition function with non-negative Boltzmann weights. Here, however, our focus is on a sign problem that manifests in the sign structure of a quantum state \cite{H16}, i.e., in the complex amplitudes of a wave function. 
To get started, we define the sign problem in terms of quantum states. We then describe a symmetry-constrained variant of the sign problem, which we call the symmetry-protected sign problem. We illustrate this concept by showing that a subset of SPT states exhibit a symmetry-protected sign problem.

\subsection{Definition of symmetry-protected sign problem} \label{sec: definition of symmetry protected sign problem}

Complex probability amplitudes are a key feature of quantum states and are essential for describing non-classical phenomena such as quantum interference. For this reason, non-negative wave functions can be regarded as more classical. Indeed, the amplitudes of a non-negative wave function correspond to (the square root of) a classical probability distribution. Whether a state has non-negative amplitudes, however, is basis dependent, i.e., it may be possible to remove a complex sign structure by making a local basis change. This motivates defining the following sign problem at the level of probability amplitudes:

\begin{definition}[Sign problem] \label{def: wave function sign problem}
A state $|\psi\rangle$ has a sign problem relative to a basis $\{|\alpha \rangle\}$, if, for any finite-depth quantum circuit $\mathcal{U}$, at least one amplitude of the state $\mathcal{U}|\psi \rangle$ in the basis $\{|\alpha \rangle\}$ is outside of the set $\mathbb{R}_{\geq 0}$.
\end{definition}

\noindent It is natural to take the basis $\{|\alpha \rangle\}$ to be the computational basis and to interpret the FDQC  $\mathcal{U}$ as a local basis change -- then, a state has a sign problem if there is no local basis in which the amplitudes of the state are all non-negative. We make the basis $\{|\alpha \rangle\}$ explicit here to more readily generalize to a symmetry-protected sign problem below.

It remains an open question as to whether any many-body system exhibits a sign problem in the sense above. We note that, in Refs.~\cite{RK17,GSR20,SGR20}, it is shown that certain topological phases of matter have an obstruction to finding a parent Hamiltonian that is stoquastic - i.e., where the off-diagonal matrix elements of the Hamiltonian are all non-positive \cite{BDOT06,GH20}. While a stoquastic parent Hamiltonian is sufficient to guarantee that the ground state is non-negative, it is not necessary.
Nonetheless, it is natural to conjecture that these same phases of matter exhibit a sign problem related to the sign structure of a ground state wave function.

Notably, SPT states do not have a sign problem. This is because SPT states can be disentangled into a product state by applying a FDQC (see also \cite{S15}).
However, we consider a symmetry-protected variant of the sign problem, and show that certain SPT states indeed exhibit a \textit{symmetry-protected} sign problem, defined as:

\begin{samepage}
\begin{definition}[Symmetry-protected sign problem] \label{def: symmetry-protected sign problem}
A state $|\psi\rangle$ has a symmetry-protected sign problem relative to a basis $\{|\alpha \rangle\}$, if, for any finite-depth quantum circuit $\mathcal{U}_\sym$ composed of symmetric gates, at least one amplitude of the state $\mathcal{U}_\sym|\psi \rangle$ in the basis $\{|\alpha \rangle\}$ is outside of the set $\mathbb{R}_{\geq 0}$.
\end{definition}
\end{samepage}

\noindent In other words, relative to the reference basis $\{|\alpha \rangle\}$, there are no symmetry-preserving local basis changes that make the wave function non-negative. With this simplification of the sign problem to symmetry-preserving basis changes, we are able to show that particular SPT phases have a symmetry-protected sign problem.

\subsection{Symmetry-protected sign problem for SPT states} \label{sec: symmetry protected sign problem for SPT states}

{In this section, we argue that SPT states in dimension $D=1$ have a symmetry-protected sign problem relative to the symmetry-charge basis, i.e., the basis of product states in which the symmetry is diagonal \cite{RS15,SR16}. We provide two proofs. The first proof relies on strange correlators, defined in Section~\ref{sec: strange correlators}, while the second makes use of the quantum wire property of $1$D SPT states. Before stating the first proposition, we would like to emphasize:}\\

{\noindent \textbf{Remark:} The SPT states considered in this work (defined in Section~\ref{sec: definition of SPT phases}) are assumed to have zero correlation length. This means that, for every SPT state $|\psi_\spt \rangle$ there is a distance $\lambda$, independent of system size, such that the correlator $\langle \psi_\spt | \mathcal{O}_1 \mathcal{O}_2 | \psi_\spt \rangle$ vanishes for every pair of operators $\mathcal{O}_1$, $\mathcal{O}_2$ whose supports are separated by a distance greater than $\lambda$. In some contexts, SPT states may include states with nonzero correlation length. The proof below holds for the restricted class of SPT states considered in this text.} \\

\begin{proposition} \label{prop: 1d sym sign problem}
{Any state $|\psi_\spt\rangle$ belonging to a nontrivial 1D SPT phase protected by a finite Abelian symmetry has a symmetry-protected sign problem relative to the product state bases in which the symmetry is represented by products of diagonal Pauli operators.}
\end{proposition}

\noindent {\emph{Proof of Proposition~\ref{prop: 1d sym sign problem}.}}
Without loss of generality, we assume the symmetry is represented by $X$-type Pauli strings. In this case, the symmetry-charge basis is the X-basis.  
In what follows, we write the basis states as $\{|x \rangle\}$. Note that, in this basis, Pauli Z operators are permutations with positive matrix elements.

We derive a contradiction by assuming that $|\psi_\spt \rangle$ \textit{does not} have a symmetry-protected sign problem relative to the symmetry-charge basis. This implies that there is a FDQC $\mathcal{U}_\sym$ composed of symmetric gates with the property that $|\varphi_\spt \rangle \equiv {\mathcal{U}_\sym |\psi_\spt \rangle}$ has only non-negative amplitudes in the X-basis. 
$|\varphi_\spt \rangle$ can then be written as:
\begin{align}
    |\varphi_\spt \rangle = \sum_{x} \sqrt{p(x)} | x \rangle,
\end{align}
for some probability distribution $p(x)$. 

From the arguments in Appendix~\ref{app: strange order parameters for 1D SPT phases}, $|\varphi_\spt \rangle$ must admit a strange order parameter {$\{\mathcal{O}_i,\mathcal{O}_j\}$} satisfying:
\begin{align} \label{eq: strange correlator magnitude}
    \left| \frac{\langle x | \mathcal{O}_i\mathcal{O}_j|\varphi_\spt \rangle}{\langle x | \varphi_\spt \rangle} \right| = 1,  \quad \forall x,i,j.
\end{align}
As shown below, the existence of the strange order parameter {$\{\mathcal{O}_i,\mathcal{O}_j\}$} leads to pairs of operators with long-range correlations in the non-negative state $|\varphi_\spt \rangle$. This contradicts the assumption that $|\varphi_\spt \rangle$ has zero correlation length.

To start, we simplify the expression in Eq.~\eqref{eq: strange correlator magnitude} by acting with the Pauli X operators in $\mathcal{O}_i$ and $\mathcal{O}_j$ on the state $\langle x |$. This leaves us with:
\begin{align} \label{eq: strange correlator X generated}
     \frac{\langle x | \mathcal{O}_i \mathcal{O}_j |\varphi_\spt\rangle}{\langle x |\varphi_\spt\rangle} = \frac{\langle x | \mathcal{Z}_i \mathcal{Z}_j |\varphi_\spt \rangle}{\langle x |\varphi_\spt \rangle},
\end{align}
where $\mathcal{Z}_i$ and $\mathcal{Z}_j$ are charged operators generated by sums of Z-type Pauli strings. To be explicit, $\mathcal{Z}_i$ and $\mathcal{Z}_j$ are of the form:
\begin{align}
    \mathcal{Z}_i = \sum_{P^Z}C^i_{P^Z}P^Z, \quad \mathcal{Z}_j = \sum_{P^Z}C^j_{P^Z}P^Z,
\end{align}
where the sums are over all Z-type Pauli strings $P^Z$, and $C^i_{P^Z}$ and $C^j_{P^Z}$ are some complex valued coefficients. Due to the locality of the $\mathcal{O}_i$ and $\mathcal{O}_j$, $\mathcal{Z}_i$ and $\mathcal{Z}_j$ are localized near $i$ and $j$, respectively. Moreover, there are only finitely many nonzero coefficients $C^i_{P^Z}$ and $C^j_{P^Z}$. We use $t_i$ ($t_j$) to denote the number of nonzero coefficients in the expansion of $\mathcal{Z}_i$ ($\mathcal{Z}_j$), and we define $t$ to be the product of $t_i$ and $t_j$:
\begin{align}
    t \equiv t_i t_j.
\end{align}
Furthermore, note that the coefficients $C^i_{P^Z}$ and $C^j_{P^Z}$ have bounded norm. This follows from the fact $\mathcal{O}_i$ and $\mathcal{O}_j$ are local and have bounded norm. We also note that neither $\mathcal{Z}_i$ nor $\mathcal{Z}_j$ is the identity, since $\mathcal{O}_i$ and $\mathcal{O}_j$ are charged.

Next, we expand the right-hand side of Eq.~\eqref{eq: strange correlator X generated} using the expressions for $\mathcal{Z}_i$ and $\mathcal{Z}_j$:
\begin{eqs} \label{eq: expansion of strange correlator}
      \sum_{P_i^Z}\sum_{P_j^Z} C^i_{P_i^Z}C^j_{P_j^Z}\frac{\langle x | P_i^Z P_j^Z |\varphi_\spt \rangle}{\langle x|\varphi_\spt \rangle},
\end{eqs}
There are at most $t$ nonzero coefficients in Eq.~\eqref{eq: expansion of connected correlator}, all of which are bounded from above. Since the magnitude of the expression in Eq.~\eqref{eq: expansion of strange correlator} is equal to $1$ [by Eq.~\eqref{eq: strange correlator magnitude}], we see that, for every $i,j$, there must be a choice of $P_i^Z$ and $P_j^Z$ such that:
\begin{align}\label{eq: PP correlator constant}
    \frac{\langle x | P_i^Z P_j^Z |\varphi_\spt \rangle}{\langle x |\varphi_\spt \rangle} > c,
\end{align}
for some constant $c > 0$ independent of $i,j$. Note that the expression on the left hand side is always non-negative, since $|\varphi_\spt \rangle$ is a non-negative state and, in the X basis, each $P_i^Z$ is a permutation. 
Let us denote a choice of $P_i^Z$ and $P_j^Z$ that satisfies Eq.~\eqref{eq: PP correlator constant} by $Q_i^Z(m)$ and $Q_j^Z(m)$, respectively. Here, $m$ indexes the $t$ pairs of $P_i^Z$ and $P_j^Z$ with nonzero coefficients in Eq.~\eqref{eq: expansion of strange correlator}. 
Explicitly, for every pair of sites $i,j$ and basis state $|x\rangle$, there exists an $m$ such that:
\begin{align}\label{eq: simplified strange correlator}
    \frac{\langle x | Q_i^Z(m) Q_j^Z(m) |\varphi_\spt \rangle}{\langle x |\varphi_\spt \rangle} > c.
\end{align}
Given a pair of sites $i,j$, let us label each basis state $|x\rangle$ by an choice $m$ for which Eq.~\eqref{eq: simplified strange correlator} is satisfied. This divides the basis states into $t$ sets, labeled by $m$. We denote the $m^\text{th}$ set of basis states by $\mathfrak{X}_m$.

%%%%%%%%%%%%%%%%%%%%%%%%%%%%%%%%%%%%%%%%%%%%%%%%%%%%%%%%%%%%%%%%%%%%%%%%%%%%%%%%%%%%%%%%%%%%%%%%%%%%%%%%%%%%%%%%%%%%%%%%%%%%%%%%%%%%%%%%%%%%%%%%%%%%%%%%%

We are now prepared to derive a contradiction by considering the correlations of $Q_i^Z(m')$ and $Q_j^Z(m')$ (for an arbitrary $m'$) in the state $|\varphi_\spt \rangle$:
\begin{align} \label{eq: connected correlator}
    {\langle \varphi_\spt | Q_i^Z(m') Q_j^Z(m') |\varphi_\spt  \rangle}.
\end{align}
We evaluate the correlator
by expanding $\langle \varphi_\spt |$ in the X-basis. This gives us:
\begin{multline} \label{eq: expansion of connected correlator}
   {\langle \varphi_\spt | Q_i^Z(m') Q_j^Z(m') |\varphi_\spt  \rangle} \\ = \sum_{x} \sqrt{p(x)} \langle x | Q_i^Z(m') Q_j^Z(m') |\varphi_\spt \rangle,
\end{multline}
which can be written in terms of strange correlators as:
\begin{multline}
   {\langle \varphi_\spt | Q_i^Z(m') Q_j^Z(m') |\varphi_\spt  \rangle} \\ = \sum_x {p(x)} \frac{\langle x | Q_i^Z(m') Q_j^Z(m') |\varphi_\spt \rangle}{\langle x |\varphi_\spt \rangle}.
\end{multline}
Next, we partition the basis states into the $t$ sets $\mathfrak{X}_m$:
\begin{multline} \label{eq: expansion of connected correlator2}
   {\langle \varphi_\spt | Q_i^Z(m') Q_j^Z(m') |\varphi_\spt  \rangle} \\
= \sum_{m=1}^t \sum_{x \in \mathfrak{X}_m} {p(x)} \frac{\langle x | Q_i^Z(m') Q_j^Z(m') |\varphi_\spt \rangle}{\langle x |\varphi_\spt \rangle},
\end{multline}
and substitute Eq.~\eqref{eq: simplified strange correlator} for every $x \in \mathfrak{X}_{m'}$:
\begin{multline} \label{eq: expansion of connected correlator3}
   {\langle \varphi_\spt | Q_i^Z(m') Q_j^Z(m') |\varphi_\spt  \rangle}
> \sum_{x \in \mathfrak{X}_{m'}} {p(x)} c \\
+ \sum_{m\neq m'} \sum_{x \in \mathfrak{X}_m} {p(x)} \frac{\langle x | Q_i^Z(m') Q_j^Z(m') |\varphi_\spt \rangle}{\langle x |\varphi_\spt \rangle}.
\end{multline}
Notice that each term in the second line of Eq.~\eqref{eq: expansion of connected correlator3} is non-negative. This is because $p(x) \geq 0$ for all $x$, and the Z-type operators $Q_i^Z(m'),Q_j^Z(m')$ preserve non-negativity (in the X-basis). Consequently, the correlator can be lower bounded by:
\begin{align} \label{eq: correlator lower bound}
    {\langle \varphi_\spt | Q_i^Z(m') Q_j^Z(m') |\varphi_\spt  \rangle}
> \sum_{x \in \mathfrak{X}_{m'}} {p(x)} c.
\end{align}

For some choice of $m'$, the right-hand side of Eq.~\eqref{eq: correlator lower bound} is nonzero. Indeed, since $p(x)$ is a probability distribution:
\begin{align}
    \sum_x p(x) = \sum_{m=1}^t \sum_{x \in \mathfrak{X}_m} p(x) = 1,
\end{align}
there must be a $\tilde{m}$ such that:
\begin{align}
    \sum_{x \in \mathfrak{X}_{\tilde{m}}} p(x) \geq \frac{1}{t}.
\end{align}
By taking $m'$ to be $\tilde{m}$ in Eq.~\eqref{eq: correlator lower bound}, we obtain:
\begin{align}
        {\langle \varphi_\spt | Q_i^Z(\tilde{m}) Q_j^Z(\tilde{m}) |\varphi_\spt  \rangle} > c/t.
\end{align}
This tells us that $|\varphi_\spt  \rangle$ has long-range correlations, since, for every $i,j$, we can find local operators $Q_i^Z(\tilde{m})$ and $Q_j^Z(\tilde{m})$ satisfying the inequality above. 
Long-range correlations contradict the assumption that $|\varphi_\spt \rangle$ can be prepared from a product state by a FDQC. Therefore, $|\psi_\spt \rangle$ must have a symmetry-protected sign problem relative to the X-basis (i.e., the symmetry charge basis).~$\square$ \\

{\noindent \textbf{Remark:} It is tempting to extend Proposition~\ref{prop: 1d sym sign problem} to SPT states belonging to two-dimensional SPT phases. Indeed, it is conjectured that every nontrivial SPT state in $2$D has a strange order parameter with strange correlations that are either constant or decay as a power law (see the argument in Section~\ref{sec: strange correlators}). With this, it can be argued that at least one term in the expansion of the correlator [Eq.~\eqref{eq: expansion of connected correlator}] decays slowly with $|i-j|$. This is sufficient for showing that any $2$D SPT state on a \textit{finite-sized system} has a symmetry-protected sign problem relative to the symmetry-charge basis. However, it is unclear whether this holds in the limit of infinite system size, where the amplitude $\sqrt{p(x)}$ of $|\varphi_\spt \rangle$ may tend to $0$.} \\

For a concrete application of Proposition~\ref{prop: 1d sym sign problem}, we can consider the cluster state, discussed in Section~\ref{sec: example: cluster state}. The $\ZZ_2 \times \ZZ_2$ symmetry of the cluster state is represented by Pauli X operators, and therefore, the cluster state has a symmetry-protected sign problem relative to the X-basis. This is to say, in the symmetry-charge basis, at least one amplitude of the cluster state is outside of $\RR_{\geq 0}$, and moreover, there are no symmetry-preserving local basis changes from the X-basis that make all of the amplitudes of the cluster state non-negative. 

We note that, although the cluster state has a symmetry-protected sign problem relative to the X-basis, there still exists product state bases in which the amplitudes of the cluster state are non-negative. In particular, the amplitudes of the cluster state are non-negative if the Z-basis is used on even sites and the X-basis is used on odd sites.\footnote{In fact, it can be checked that the Hamiltonian is stoquastic in this basis.} Proposition~\ref{prop: 1d sym sign problem} does not apply in this case, because, in this basis, the symmetry is not diagonal, and it cannot be mapped to the X-basis by symmetry-preserving local transformations.\footnote{Specifically, the bases are related by applying Hadamard gates on the even sites. Hadamard gates do not commute with the symmetry formed by products of Pauli X operators.}

To highlight the quantum nature of states with a symmetry-protected sign problem, we would like to describe an alternative proof of Proposition~\ref{prop: 1d sym sign problem} that applies to $1$D SPT states. The idea is to make use of the quantum wire property, where, for certain $1$D SPT states, measurements in the symmetry-charge basis can be used to generate long-range entanglement \cite{ESBD12prl,RWPWS17,M17}. We argue that, if a state can serve as a quantum wire and it is non-negative in the symmetry-charge basis, then it contradicts the results of Ref.~\cite{H16}, in which a bound is set on the entanglement created by making measurements of a non-negative state. More formally, we show the following:

%%%%%%%%%%%%%%%%%%%%%%%%%%%%%%%%%%%%%%%%%%%%%%%%%%%%%%%%%%%

\begin{customthm}{3$\:\!{'}$}\label{prop: 1d sym sign problem2} Let $|\psi_\spt\rangle$ be a state belonging to a 1D SPT phase protected by an Abelian symmetry and corresponding to a maximally non-commutative cohomology class.\footnote{A cohomology class $[\omega]\in \mathcal{H}^2[G,U(1)]$ is maximally non-commutative if for every element $g\in G$ other than the identity, there exists an $h\in G$ such that $\omega(g,h)\neq \omega(h,g)$.} 
Then $|\psi_\spt \rangle$ has a symmetry-protected sign problem relative to product state bases in which the symmetry is represented by products of diagonal unitaries.
\end{customthm}

\noindent \textbf{Remark:} Here, we make the technical assumption that $|\psi_\spt \rangle$ belongs to an SPT phase labeled by a maximally non-commutative cohomology class to guarantee that $|\psi_\spt \rangle$ exhibits the quantum wire property \cite{ESBD12prl}. {In fact, the proof below is sufficient if a subgroup $H$ of the symmetry $G$ is represented by a product of diagonal unitaries and the cohomology class is maximally non-commutative when restricted to $H$.} We note that we are also working under the assumption that $|\psi_\spt \rangle$ is defined on a lattice with periodic boundary conditions and can be prepared exactly from a product state by a FDQC, as established in Section~\ref{sec: definition of SPT phases}. \\

\noindent \emph{Proof of Proposition~\ref{prop: 1d sym sign problem2}.} First, we use the results of Ref.~\cite{ESBD12prl} to show that long-range entanglement can be generated from measurements of $|\psi_\spt \rangle$. To see this, we consider a matrix product state (MPS) representation of $|\psi_\spt \rangle$:
\begin{align}
        \vcenter{\hbox{\includegraphics[scale=.1,trim={8cm 13cm 7cm 16cm},clip]{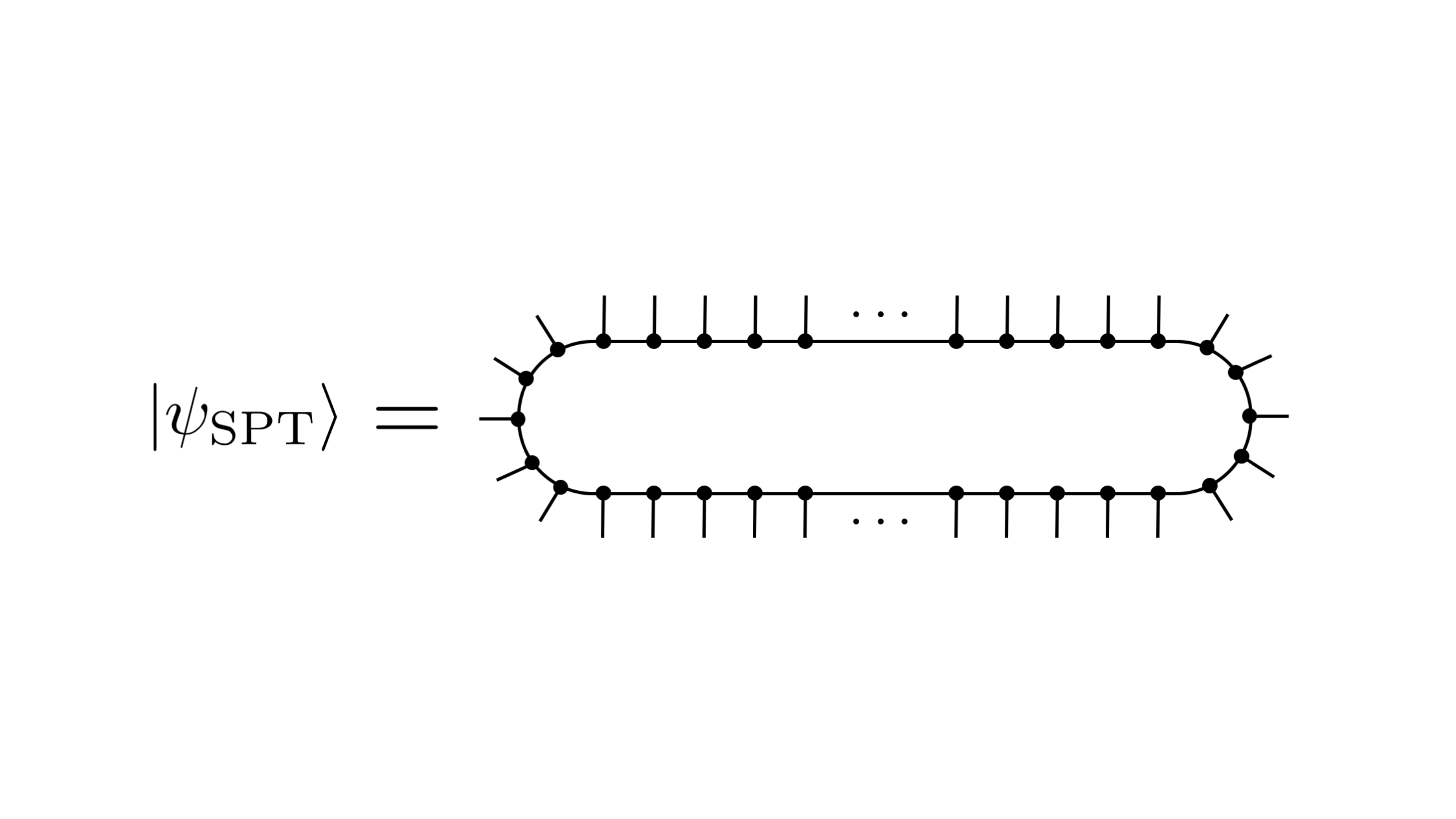}}}\raisebox{-3pt}{.}
\end{align}
We then coarse grain the lattice by combining a constant number of neighboring sites into super-sites,
% , depending on the Lieb-Robinson length of a FDQC that prepares $|\psi_\spt \rangle$ from a product state. 
such that, for each local tensor of the coarse grained MPS, there exists an isometry $W$ that graphically satisfies: 
\begin{align} \label{eq: isometry local tensor}
    \vcenter{\hbox{\includegraphics[scale=.1,trim={15cm 17cm 16cm 13cm},clip]{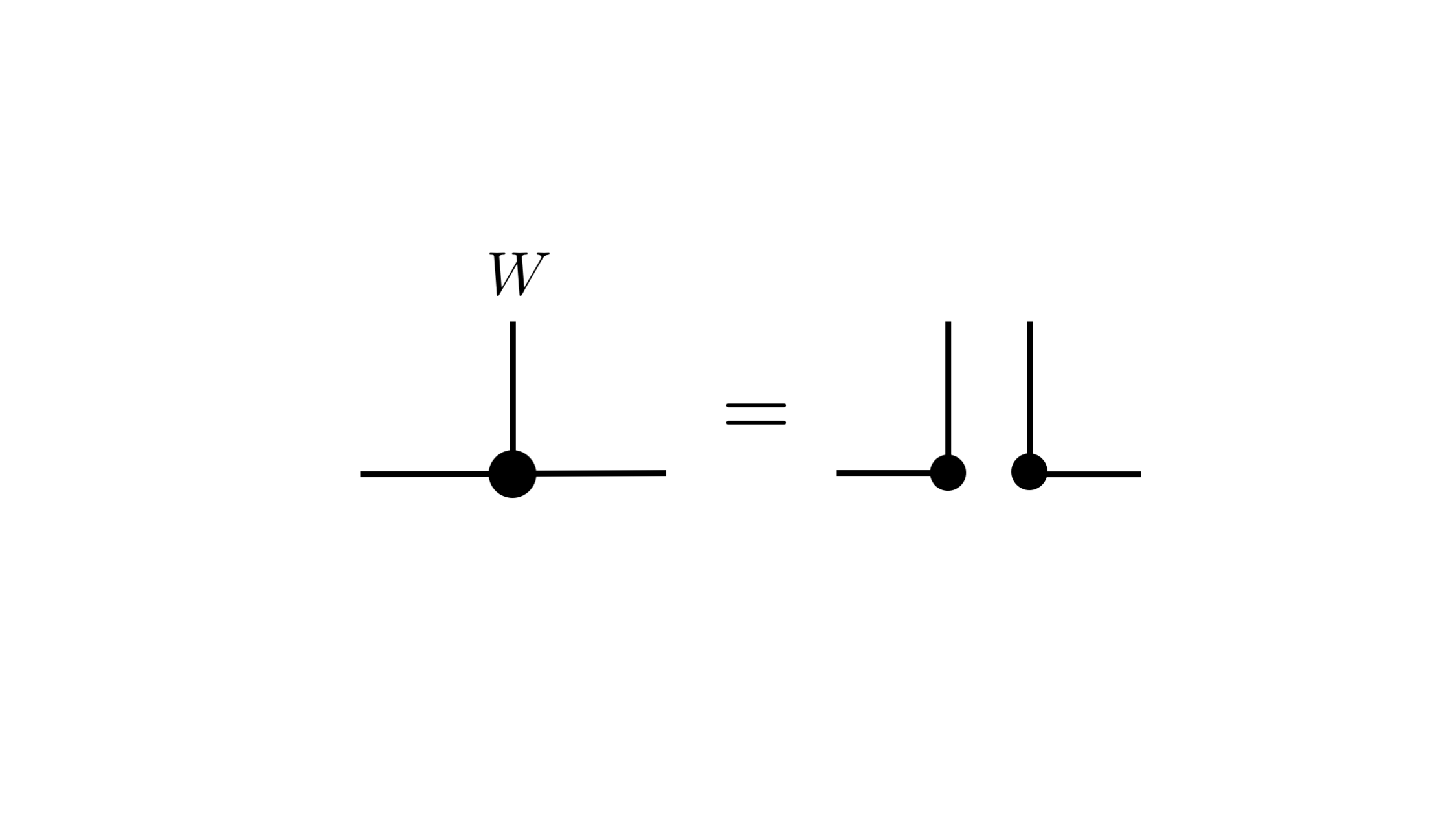}}}\raisebox{-10pt}{.}
\end{align}
Here, $W$ is an isometry that maps from the $d$-dimensional physical Hilbert space to a pair of Hilbert spaces of dimension $\chi_L$ and $\chi_R$, where $\chi_L$ and $\chi_R$ are the dimensions of the left and right virtual Hilbert spaces, respectively. Importantly, $W$ disentangles the states in the left virtual Hilbert space from the states in the right virtual Hilbert space. Heuristically, $W_A$ at the super-site $A$ can be interpreted as first acting with a unitary operator supported on $A$ that locally disentangles $|\psi_\spt \rangle$ and then subsequently removing the unentangled degrees of freedom. 

\begin{figure}[t]
\centering
\includegraphics[width=.40\textwidth,trim={20cm 2cm 19cm 3cm},clip]{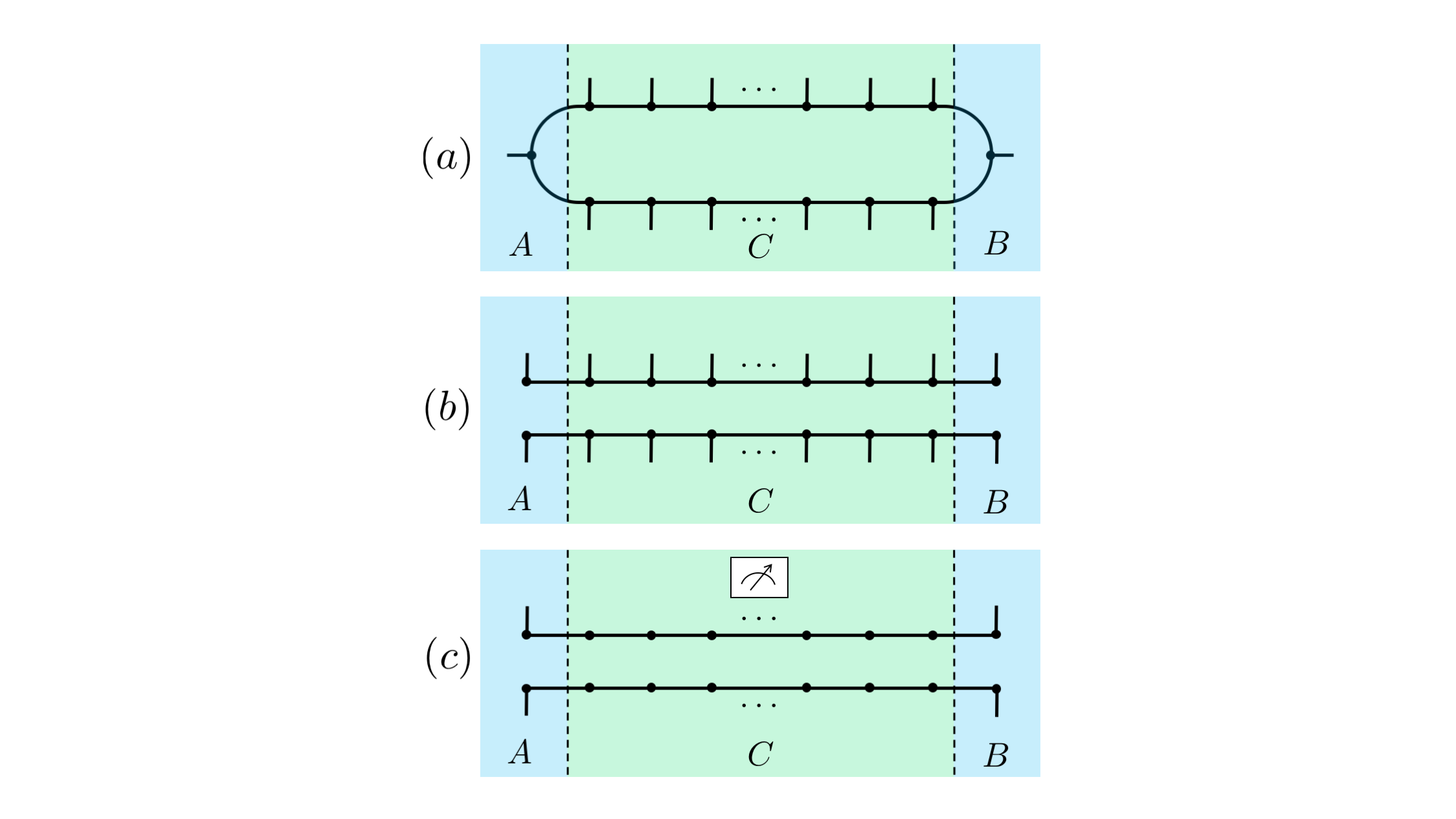}
\caption{(a) We partition the coarse grained MPS into regions $A$, $B$, and $C$ by choosing super-sites $A$ and $B$. (b) Applying the isometry $W_A \otimes W_B$ to $|\psi_\spt \rangle$ splits it into two unentangled MPS: $|\psi^1_\spt \rangle$ and $|\psi^2_\spt \rangle$. (c) The measurement on the super-sites in $C$ fixes the physical indices in the region $C$ according to the measurement outcome $|x\rangle$ and leaves us with the state $|\psi_{AB}^x\rangle$ on $A \cup B$. $|\psi_{AB}^x\rangle$ is entangled between $A$ and $B$ through the virtual bonds, as described in Ref.~\cite{ESBD12prl}.}
\label{fig: quantumwire}
\end{figure}

The next step is to choose well-separated super-sites $A$ and $B$ and make measurements in the symmetry-charge basis on the complement of $A\cup B$, denoted by $C$ (Fig.~\ref{fig: quantumwire}). We claim that measurements in the symmetry-charge basis on $C$ generate entanglement between $A$ and $B$ lower bounded by a value that is independent of the separation of $A$ and $B$. To show this, we apply the isometry $W_A \otimes W_B$ to $|\psi_\spt \rangle$, with $W_A$ and $W_B$ defined as in Eq.~\eqref{eq: isometry local tensor}. The isometry splits $|\psi_\spt \rangle$ into two independent MPS, as shown in  Fig.~\ref{fig: quantumwire}:
\begin{align}
    W_A \otimes W_B |\psi_\spt \rangle =|\psi_\spt^1\rangle \otimes |\psi_\spt^2\rangle.
\end{align}
Note that, after applying the isometry $W_A \otimes W_B$, the degrees of freedom at $A$ and $B$ correspond to the virtual bonds of the MPS, up to local positive diagonal operators (see Cororally 3.12 of Ref.~\cite{CP-GSV17}). This is important given that the results of Ref.~\cite{ESBD12prl} show that measurements of an SPT state create entanglement at the level of the virtual bonds.

We now measure the sites in $C$ in the symmetry-charge basis and with probability $p_x$ obtain the product state $|x\rangle$ on $C$. We define $|\psi^x_{AB} \rangle$ to be the state on $A \cup B$ given by fixing the degrees of freedom of $|\psi_\spt^1\rangle \otimes |\psi_\spt^2\rangle$ on $C$ according to the product state $|x\rangle$. By the assumption that $|\psi_\spt \rangle$ belongs to an SPT phase corresponding to a maximally non-commutative cohomology class, Theorem $1$ of Ref.~\cite{ESBD12prl} tells us that $|\psi^x_{AB} \rangle$ can be written in the form:
\begin{eqs} \label{eq: entanglement from quantum wire}
    |\psi^{x}_{AB}\rangle=&\left(|\varphi^1_{\rm junk}\rangle\otimes U^{x,1}_B|\Phi^1_{\max}\rangle\right) \\ \otimes &\left(|\varphi^2_{\rm junk}\rangle\otimes U^{x,2}_B|\Phi^2_{\max}\rangle\right).
\end{eqs}
Here, $|\varphi^1_{\rm junk}\rangle$,$|\varphi^2_{\rm junk}\rangle$ are unimportant states that depend on the details of $|\psi_\spt \rangle$, $|\Phi^1_{\max}\rangle$,$|\Phi^2_{\max}\rangle$ are $\sqrt{|G|}$-dimensional maximally entangled states between $A$ and $B$, and $U^{x,1}_B$,$U^{x,2}_B$ are some unitary operators supported only on $B$. Note that the first and second lines of Eq.~\eqref{eq: entanglement from quantum wire} correspond to independent contributions from $|\psi^1_\spt \rangle$ and $|\psi^2_\spt \rangle$. For any measurement outcome $|x\rangle$, the entanglement entropy of $|\psi_{AB}^x\rangle$ between $A$ and $B$ is therefore bounded below as:
\begin{align}\label{eq:constEEmeas}
    S\left(\rho^{x}_A\right)\geq 2\log_2 \sqrt{|G|},
\end{align}
with $\rho^x_A$ denoting the reduced density matrix of $|\psi^x_{AB}\rangle$ on $A$.
Since $W_A \otimes W_B$ has no affect on the entanglement generated between $A$ and $B$, we see that making measurements of $|\psi_\spt\rangle$ on $C$ induces entanglement between $A$ and $B$ with a constant lower bound, as claimed.

On the other hand, Proposition 4.1 of Ref.~\cite{H16} implies that, if $|\psi_\spt\rangle$ is non-negative in the symmetry-charge basis, then the average entanglement entropy after the measurements is bounded from above as:
\begin{equation} \label{eq: average entanglement entropy bound}
\sum_x p_xS\left(\rho^{x}_A\right) \leq f(L),
\end{equation}
where $L$ is the distance between the super-sites $A$ and $B$, and $f(L)$ is a function that decays rapidly to zero (faster than any polynomial). For a sufficiently large $L$, the bound in Eq.~\eqref{eq: average entanglement entropy bound} conflicts with Eq.~\eqref{eq:constEEmeas}.

Therefore, $|\psi_\spt \rangle$ cannot be non-negative in the symmetry-charge basis. Furthermore, the argument applies to any state constructed from $|\psi_\spt \rangle$ by a FDQC composed of symmetric gates, since the quantum wire property is shared by states in the same phase. We can thus conclude that $|\psi_\spt \rangle$ has a symmetry-protected sign problem relative to the symmetry-charge basis.~$\square$

\section{Discussion} \label{sec: discussion}

We have introduced the concepts of symmetry-protected magic and a symmetry-protected sign problem to facilitate the study of many-body magic and the sign structure of wave functions. We have applied these concepts to SPT states to assess their quantum complexity. Using the universal properties of certain nontrivial group cohomology phases in $D \geq 2$ dimensions, we showed that the corresponding SPT states have symmetry-protected magic, assuming the symmetry is represented by products of Pauli operators. This implies that there is no stabilizer state representative in these SPT phases {(with the restrictions on the dimensions of the qudits stated in the proposition)}. We also argued that SPT states in 1D have a symmetry-protected sign problem in bases where the symmetry is diagonal. Consequently, in this basis, there is an obstruction to a description of the SPT phase by a non-negative wave function. 

By imposing symmetry constraints, we were able to make analytic statements about the complexity of quantum states, including the first verification of a sign problem at the level of probability amplitudes. We note that a restriction to symmetric systems has also been beneficial for studying the No Low-energy Trivial States conjecture in Ref.~\cite{BKKT19}. We anticipate that, moving forwards, symmetry constraints will be a valuable tool for addressing outstanding quantum information problems.

We would like to emphasize that our assessment of the symmetry-protected magic in SPT phases applies to systems with even-dimensional qudits.
This is noteworthy given that magic is more easily quantified and better understood in systems of odd-dimensional qudits (especially odd prime dimensions), thanks to the discrete Wigner formalism \cite{Wootters87,G06}. %\zw{Mark} 
The associated discrete Wigner function maps states to quasi-probabilities, and for systems of odd-dimensional qudits, the negative quasi-probabilities can be used to define a measure of the amount of magic in the state \cite{VMGE14}.

In the case of odd-dimensional qudits, symmetry-protected magic can be interpreted as a sign problem, manifesting through the quasi-probability distribution of the discrete Wigner function (known as the discrete Hudson's theorem \cite{G06}). 
Symmetry-protected magic says that the signs in the quasi-probability distribution cannot be removed by making symmetry-preserving unitary local changes to the state. We point out that this sign problem has appeared in the simulation of random quantum circuits as in Ref.~\cite{PWB15}. Our work therefore deals with two different notions of a sign problem -- one relates to the quasi-probability distribution of a discrete Wigner function, while the other corresponds to the complex probability amplitudes of a state (Section~\ref{sec: definition of symmetry protected sign problem}).

{The precise relation between these two notions of a sign problem is unclear. The cluster state in Section~\ref{sec: example: cluster state}, shows that a symmetry-protected sign problem in the symmetry-charge basis does not imply symmetry-protected magic. {As for the converse, our results are inconclusive -- it is not clear whether every SPT state with symmetry-protected magic also has a symmetry-protected sign problem relative to the symmetry-charge basis.} However, more generally, symmetry-protected magic does not necessitate a symmetry-protected sign problem relative to arbitrary product state bases. For example, the $\ZZ_2 \times \ZZ_2 \times \ZZ_2$ SPT state defined in Ref.~\cite{Y15} has symmetry-protected magic by Proposition~\ref{prop: symmetry-protected magic SPT type iii}, but it does not have a symmetry-protected sign problem relative to the product state basis in which the X-basis is used on the $A$ sublattice and the Z-basis is used on the $B$ and $C$ sublattices.} 

{It should be noted that the sign problems in this text are distinct from the usual notion of a sign problem related to the ``stoquasticity'' of a Hamiltonian, discussed in the context quantum Monte Carlo methods. That being said, a sign problem at the level of the wave function implies that the system suffers from a sign problem in the stoquastic sense (assuming the Hamiltonian is gapped, see Ref.~\cite{H16} and Appendix A of Ref.~\cite{SGR20}). 
Thus, a sign problem, as defined in Section~\ref{sec: definition of symmetry protected sign problem}, poses an obstruction to Monte Carlo simulation. A symmetry-protected sign problem, in contrast, does not imply any fundamental obstruction to Monte Carlo simulation. In practice, one is free to use arbitrary local basis changes to find a basis in which the system is amenable to Monte Carlo methods -- there is no reason to restrict to symmetry-preserving local basis changes. Nonetheless, a symmetry-protected sign problem informs us about bases that are inefficient for Monte Carlo simulation. Hence, the operational significance of a symmetry-protected sign problem is as a no-go for Monte Carlo simulation in certain bases.}

To conclude, we would like to further comment on related work, make a few conjectures, and discuss some promising directions for future work.

\vspace{1.5mm}
\noindent \begin{center}\emph{Symmetry-protected magic:}\end{center}
\vspace{1.5mm}

Propositions~\ref{prop: symmetry-protected magic SPT} and \ref{prop: symmetry-protected magic SPT type iii} show that stabilizer operations are insufficient for simulating certain SPT states. It is important to note that this also implies that those SPT states can be used as a source of magic for quantum computing.
% Moreover, our results show that magic is present in every state belonging to group cohomology SPT phases in dimensions $D \geq 2$ (protected by a symmetry represented by Pauli strings). 
It would be interesting if a quantized universal property of the corresponding SPT phases - say, their responses to probing with symmetry defects - could be exploited to reliably produce a standard magic state (e.g. a $\text{CCZ}$ state), independent of the microscopic details of the systems. 
We also speculate that there is a series of adaptive measurements that produces a standard magic state on a large length scale, similar to how a series of local measurements of $1$D SPT states can create entanglement between distant sites \cite{M17}.

It is also interesting to consider the implications of our work for the use of group cohomology SPT states as resources for measurement-based quantum computing (MBQC). Remarkable progress has been made in identifying computationally universal phases of matter protected by subsystem symmetries \cite{ROWSN19,SNBER19}, but much remains to be understood about the MBQC utility of SPT phases with global ($0$-form) symmetries. In Refs.~\cite{MM16} and \cite{MM18}, it was recognized that a particular ``fixed point'' wave function in a $2D$ $\ZZ_2 \times \ZZ_2 \times \ZZ_2$ group cohomology SPT phase harbors magic. Furthermore, it was shown that the state can be used as a resource for universal MBQC using only Pauli measurements. It is natural to wonder whether the entire phase can be used for universal MBQC with Pauli measurements. Our results are consistent with this conjecture and suggest that other group cohomology SPT states may be able to serve as universal resources as well. For a related discussion on the quantification of magic, see Ref.~\cite{LW20}.

According to Propositions~\ref{prop: symmetry-protected magic SPT} and \ref{prop: symmetry-protected magic SPT type iii}, certain SPT states must be non-stabilizer and, as such, cannot be prepared from $|0 \ldots 0 \rangle$ by a Clifford unitary. Inspired by Ref.~\cite{Y17}, we speculate that the higher levels of the Clifford hierarchy may also be useful for understanding the complexity of SPT states. The Clifford unitaries form the second level of the hierarchy $\mathcal{C}_2$, and the higher levels of the hierarchy are obtained recursively as:
\begin{align}
    \mathcal{C}_{D+1}\equiv \left\{ U : UPU^\dag \in \mathcal{C}_{D}, \forall P \in \mathcal{C}_1 \right\},
\end{align}
where $\mathcal{C}_1$ denotes the set of Pauli strings. The results of Ref.~\cite{Y17} imply that particular finely tuned SPT states in $D$-dimensions cannot be prepared from $|0 \ldots 0 \rangle$ by any FDQC belonging to the $D^\text{th}$ level of the hierarchy. It may be interesting to study the extent to which this applies to other states in the SPT phase.  

In this text, we focused on the magic in quantum phases characterized by SRE states, but an important avenue for future work is to study magic in systems with long-range entanglement, such as in conformal field theories (CFTs) and intrinsic topological orders. Refs.~\cite{WCS20} and \cite{SMB20} have made the first steps in numerically studying the emergence of magic at a critical point, and Ref.~\cite{WCS20} conjectured that CFTs generically have magic at large length scales, detectable by correlations.

Regarding the magic inherent in topologically ordered phases, Ref.~\cite{H18} provided a classification of systems in $2$D that can be described by a local stabilizer Hamiltonian, assuming the stabilizer Hamiltonian is translationally invariant and defined on a Hilbert space built of prime dimensional qudits. These results place important restrictions on the phases of matter that admit a representation by a stabilizer state. However, more work is needed to lift the assumptions and better understand long-range magic (Definition~\ref{def: topologically protected magic}) in phases with intrinsic topological order. We conjecture that states defined on qubits in the double semion phase, for example, have long-range magic, and we look forward to commenting further on this conjecture in upcoming work.    

\vspace{1.5mm}
\noindent \begin{center}\emph{Symmetry-protected sign problem:}\end{center}
\vspace{1.5mm}

We argued that nontrivial SPT states in dimension $D =1 $ have a symmetry-protected sign problem relative to the symmetry-charge basis, where the symmetry is diagonal. It is unclear whether these SPT states have a symmetry-protected sign problem relative to other bases, as our techniques are specialized for the symmetry-charge basis. For instance, does the cluster state have a symmetry-protected sign problem relative to the Z-basis? A complete characterization of the symmetry-protected sign problems might lead to new tools useful for tackling the sign problem in the absence of symmetry constraints. New techniques are also needed to study the symmetry-protected sign problem in SPT states in dimensions $D \geq 3$, since the strange correlations may no longer be a reliable way to diagnose the SPT order.

In Section~\ref{sec: symmetry protected magic in SPT states}, we also argued that the quantum wire property of nontrivial $1$D SPT states is incompatible with a non-negative wave function in the symmetry-charge basis. This suggests a potential operational consequence of a symmetry-protected sign problem. In particular, for nontrivial $1$D SPT states, entanglement can be generated between any two regions by making measurements on the complement in the symmetry-charge basis. We speculate that, more generally, a symmetry-protected sign problem relative to a basis $|\{\alpha\}\rangle$ implies that measurements in the $|\{\alpha\}\rangle$ basis can be used to create entanglement between distant regions. In any event, further work is needed to build off of the results of Ref.~\cite{H16} and to fully understand the connection between the sign structure of a quantum state and its localizable entanglement \cite{PVMC05}. 

\vspace{0.1in}
\noindent{\it Acknowledgements -- } We would like to acknowledge Tomotaka Kuwahara for helpful conversations related to Proposition~\ref{prop: 1d sym sign problem}, and we thank Sergey Bravyi for insights that led us to Lemma~\ref{lem: local stabilizer hamiltonian}. TDE thanks Yu-An Chen, Kyle Kawagoe, Alex Kubica, and Beni Yoshida for clarifying discussions, and he is especially grateful to Lukasz Fidkowski and Nathanan Tantivasadakarn for providing feedback on the manuscript. TDE is also appreciative of the hospitality of Perimeter Institute, where this work was initiated. ZWL is supported by Perimeter Institute for Theoretical Physics. Perimeter Institute is supported in part by the Government of Canada through the Department of Innovation, Science and Economic Development Canada and by the Province of Ontario through the Ministry of Economic Development, Job Creation and Trade. KK is supported by MEXT Quantum Leap Flagship Program (MEXT Q-LEAP) Grant Number JPMXS0120319794.  TH acknowledges support from the Natural Sciences and Engineering Research Council of Canada (NSERC) through a
Discovery Grant. 

\appendix

\section{Universality of the anomalous symmetry action} \label{app: universality of the anomalous symmetry action}

In Section~\ref{sec: anomalous boundary symmetry action}, we stated that group cohomology SPT phases can be characterized by anomalies -- i.e., obstructions to finding an effective boundary symmetry action that is onsite.\footnote{More specifically, the obstructions are to finding an effective boundary symmetry action that is onsite up to taking tensor products with effective actions of lower-dimensional SPT phases and by conjugating the effective symmetry action with a FDQC.} The calculation of the anomaly, as described in Section~\ref{sec: anomalous boundary symmetry action}, (seemingly) depends on the following choices: (i) a representative SPT state, (ii) a parent SPT Hamiltonian for the SPT state, and (iii) an effective boundary symmetry action derived from the parent Hamiltonian. Following Appendix C of Ref.~\cite{EN14}, we sketch an argument below that the computation of the anomaly ultimately does not depend on these choices. In other words, we argue that the anomaly is well-defined as an invariant of the SPT phase. The strategy is to study the anomaly at the interface between two different possible choices of (i), (ii), and (iii). 
To simplify the discussion, we assume that the parent Hamiltonians are un-frustrated. (This is sufficient for our purposes in the main text.) We expect that the argument can be generalized to show that the anomaly is well-defined for any choice of parent SPT Hamiltonian.

%%%%%%%%%%%%%%%%%%%%%%%

We consider two states $|\psi_1\rangle$ and $|\psi_2 \rangle$ belonging to the same $D$-dimensional SPT phase along with a choice of corresponding SPT Hamiltonians $H_1$ and $H_2$. 
For concreteness, we take $|\psi_1\rangle$ and $|\psi_2\rangle$ to be defined on a $D$-sphere $S^D$. Since $|\psi_1 \rangle$ and $|\psi_2 \rangle$ are in the same phase, there exists a FDQC $\mathcal{U}_\sym$ composed of symmetric gates such that:
\begin{align}
    \mathcal{U}_\sym |\psi_2 \rangle = |\psi_1 \rangle. 
\end{align}
We can then construct the SPT Hamiltonian $\tilde{H}_2$, defined as:
\begin{align}
    \tilde{H}_2 \equiv \mathcal{U}_\sym H_2 \, \mathcal{U}^\dagger_\sym,
\end{align}
which has the unique ground state $|\psi_1 \rangle$. 

\begin{figure}[t]
\centering
\includegraphics[width=.45\textwidth,trim={10cm 10.7cm 10cm 8.5cm},clip]{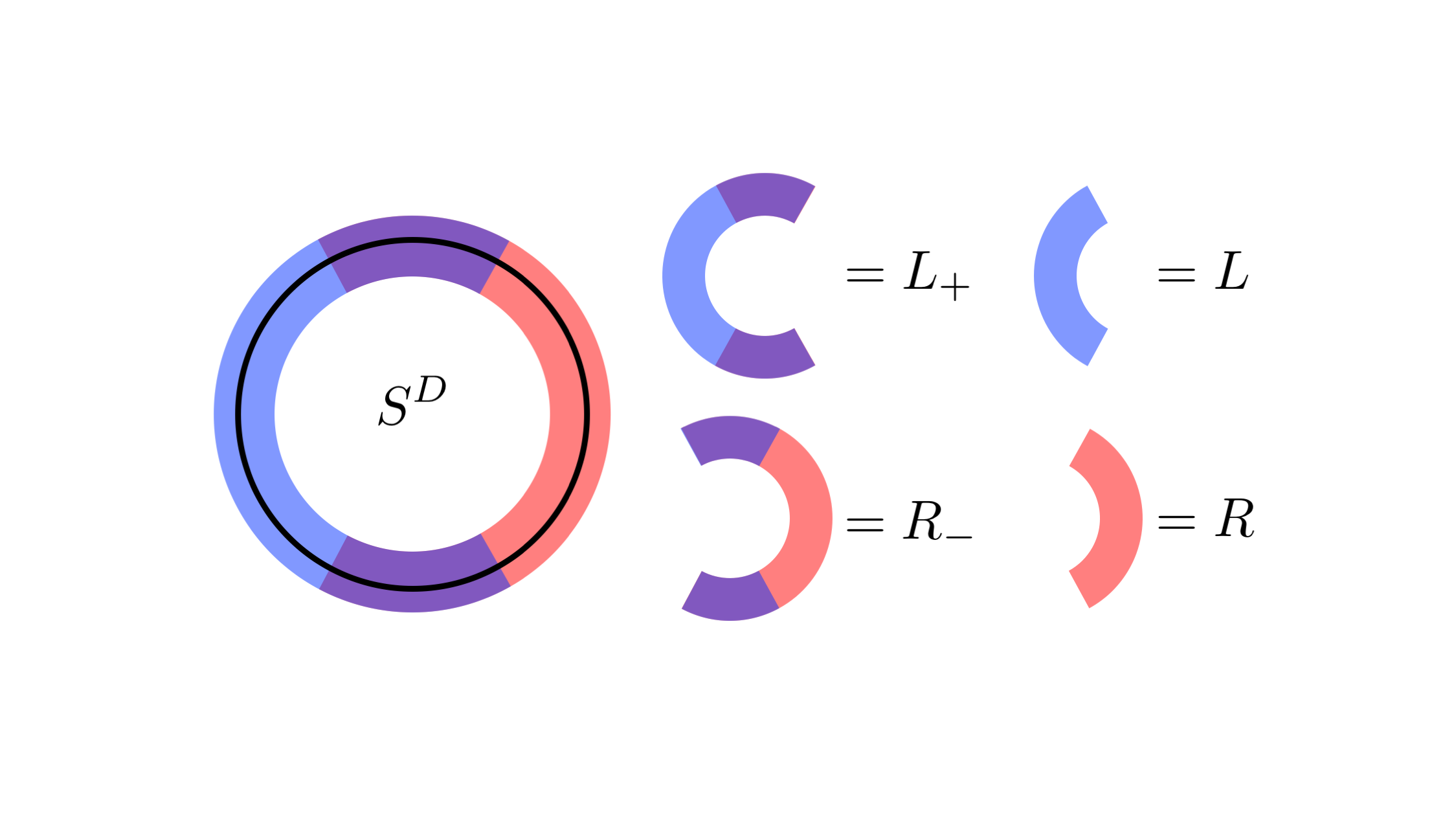}
\caption{$|\psi_1\rangle$ and $|\psi_2 \rangle$ are defined on a $D$-sphere $S^D$. We partition $S^D$ into overlapping regions $L_+$ (union of blue and purple) and $R_-$ (union of red and purple). $L_+$ contains the subregion $L$ (blue), and $R_-$ contains the subregion $R$ (red).  }
\label{fig: anomaliesoverlap}
\end{figure}

Now, we combine terms from $H_1$ and $\tilde{H}_2$ to form a new SPT Hamiltonian $H$ on $S^D$ whose ground state is also $|\psi_1 \rangle$. Roughly speaking, $H$ is comprised of the terms in $H_1$ on the left half of the sphere and the terms in $\tilde{H}_2$ on the right half.
More specifically, we divide the $D$-sphere into two overlapping regions $L_+$ and $R_-$, as shown in Fig.~\ref{fig: anomaliesoverlap}. $L_+$ and $R_-$ cover the $D$-sphere, and their intersection is a thickened $(D-1)$-sphere. We take the ``width'' of the intersection to be large compared with the Lieb-Robinson length of some (arbitrary) FDQC that prepares $|\psi_1\rangle$. To construct $H$, we truncate $H_1$ to $L_+$, to define $H_1^{L_+}$, and we truncate $\tilde{H}_2$ to $R_-$, to define $\tilde{H}_2^{R_-}$. The SPT Hamiltonian $H$ is then:
\begin{align} \label{eq: H definition}
    H \equiv H_1^{L_+} + \tilde{H}_2^{R_-},
\end{align}
with the unique ground state $|\psi_1 \rangle$. The fact that $|\psi_1 \rangle$ is the ground state follows from the assumption that $H_1$ and $H_2$ are un-frustrated.

Next, we study the possible anomaly at the interface between $H_1$ and $\tilde{H}_2$. In particular, we introduce a boundary by truncating $H$ to the region $L \cup R$, where $L$ and $R$ are defined as:
\begin{align}
    L \equiv L_+ - L_+\cap R_-, \quad R \equiv R_- - L_+ \cap R_-,
\end{align}
as depicted in Fig.~\ref{fig: anomaliesoverlap}.
In other words, we remove any term in $H$ that is supported (in part) on the overlap between $L_+$ and $R_-$. We are left with the truncated Hamiltonian $H^{L\cup R}$:
\begin{align}
    H^{L\cup R} \equiv H_1^L + \tilde{H}_2^R,
\end{align}
where $H_1^L$ is the truncation of $H_1$ to $L$ and $\tilde{H}_2^R$ is the truncation of $\tilde{H}_2$ to $R$. Importantly, we consider $H^{L \cup R}$ as a Hamiltonian on the full Hilbert space of $S^D$.

The boundary Hilbert space of $H^{L\cup R}$ can be decomposed into a tensor product of the following three Hilbert spaces:
(i) the low-energy Hilbert space of $H_1^L$ on $L$, (ii) the full Hilbert space on the intersection $L_+ \cap R_-$, and (iii) the low-energy Hilbert space of $\tilde{H}_2^R$ on $R$.
Accordingly, we can construct an effective symmetry action near the boundary of $L\cup R$ by multiplying an effective action on $L$, an onsite symmetry on $L_+ \cap R_-$, and an effective action on $R$.
More explicitly, we can choose the effective boundary symmetry action representing $g \in G$ to be $v^L(g)$ on $L$ and $v^R(g)$ on $R$, so that an effective boundary symmetry action $v(g)$ on $L\cup R$ is: 
\begin{align} \label{eq: effective boundary symmetry on sphere}
    v(g) \equiv v^L(g)\left(\prod_{i \in L_+ \cap R_-}u_{i}(g)\right)v^R(g).
\end{align}
We note that the effective boundary symmetry action in Eq.~\eqref{eq: effective boundary symmetry on sphere} is localized near $L_+ \cap R_-$.

The algorithm defined in Ref.~\cite{EN14} can now be applied to $v(g)$ to identify potential obstructions to making $v(g)$ onsite. The obstruction corresponds to an element $[\omega] \in \mathcal{H}^{D+1}[G,U(1)]$, and one can show that it can be divided into a contribution $[\omega^L]\in \mathcal{H}^{D+1}[G,U(1)]$ from $v^L(g)$ and a contribution $[\omega^R]\in \mathcal{H}^{D+1}[G,U(1)]$ from $v^R(g)$, so that:
\begin{align}
    [\omega] = [\omega^L] \cdot [\omega^R].
\end{align}

The last step is to argue that $[\omega]$ calculated from $v(g)$ using the procedure in Ref.~\cite{EN14} must correspond to the trivial class in $\mathcal{H}^{D+1}[G,U(1)]$. Therefore, regardless of the choices made in determining $v^L(g)$ and $v^R(g)$, we have $[\omega^L]=[\omega^R]^{-1}$. This constraint implies that the anomaly is well-defined, since $v^L(g)$ and $v^R(g)$ can be chosen independently. For simplicity, we show that $[\omega]$ is the trivial class for only the $1$D case. We note that the $2$D case can be found in Appendix~C of Ref.~\cite{EN14}, and we fully expect that the argument can be generalized straightforwardly to higher dimensions.

To show that $[\omega]$ must be trivial in the $1$D case, we consider the state $|\psi_1\rangle$. Since $|\psi_1 \rangle$ belongs to the boundary Hilbert space of $H^{L \cup R}$, the symmetry action $u(g)$, for any $g \in G$, can be replaced by $v(g)$ when acting on $|\psi_1 \rangle$. Therefore, we have the equality:
\begin{align} \label{eq: anomalous symmetry on gapped SRE}
     v(g) |\psi_1 \rangle = |\psi_1 \rangle, \quad \forall g \in G.
\end{align}
In $1$D, the support of $v(g)$ can be partitioned into two connected components, which we label as $A$ and $B$. Consequently, $v(g)$ can be split\footnote{The operator can be split unambiguously up to a $g$ dependent phase.} into an operator $v_A(g)$ supported on $A$ and $v_B(g)$ supported on $B$. From Eq.~\eqref{eq: anomalous symmetry on gapped SRE}, we have:
\begin{align}
    v_A(g)v_B(g)|\psi_1 \rangle = |\psi_1 \rangle, \quad \forall g \in G.
\end{align}
Furthermore, we can always define $v_A(g)$ and $v_B(g)$ so that:
\begin{align}
    v_A(g)|\psi_1\rangle = |\psi_1\rangle, \quad \forall g \in G.
\end{align}
It follows that $v_A(g)$ forms a linear representation of $G$ on $|\psi_1 \rangle$:
\begin{align}
    v_A(g)v_A(h)|\psi_1 \rangle = |\psi_1 \rangle = v_A(gh) |\psi_1 \rangle, \quad \forall g,h \in G.
\end{align}
Since $v_A(g)$ forms a trivial projective representation (i.e., a linear representation), the associated element of $\mathcal{H}^2[G,U(1)]$ must be the trivial class.

\vspace{1.5mm}
\noindent \begin{center}\emph{Cluster state example:}\end{center}
\vspace{1.5mm}

Using the ideas above, we argue that the cluster state belongs to a nontrivial SPT phase. In particular, the projective representation satisfied by the effective boundary symmetry action poses an obstruction to finding a FDQC with symmetric gates that can disentangle the cluster state.  We show this by deriving a contradiction. 

Suppose that $|\psi_\cs \rangle$ can be disentangled by a FDQC $\mathcal{U}_\sym$ composed of symmetric gates:
\begin{align}
    \mathcal{U}_\sym |\psi_\cs \rangle = |\! + \ldots + \rangle.
\end{align}
Then the Hamiltonian $\tilde{H}_\cs \equiv \mathcal{U}_\sym H_\cs\, \mathcal{U}^\dagger_\sym$ has the unique product state ground state $|\! + \ldots + \rangle$. Further, we can identify an effective boundary symmetry action for $\tilde{H}_\cs$ by conjugating the effective action computed using $H_\cs$ [copied from Eq.~\eqref{eq: define CS effective boundary symmetry action}]:
\begin{eqs} \label{eq: define CS effective boundary symmetry action 2}
    &v\boldsymbol{(}(g,1)\boldsymbol{)} = Z_1 (Z_{2M-1}X_{2M}), \\ &v\boldsymbol{(}(1,g)\boldsymbol{)} = (X_1 Z_2)  Z_{2M},
\end{eqs}
by the FDQC $\mathcal{U}_\sym$:
\begin{eqs}
   &\tilde{v}\boldsymbol{(}(g,1)\boldsymbol{)} \equiv \mathcal{U}_\sym v\boldsymbol{(}(g,1)\boldsymbol{)} \mathcal{U}^\dagger_\sym , \\ &\tilde{v}\boldsymbol{(}(1,g)\boldsymbol{)} \equiv \mathcal{U}_\sym {v}\boldsymbol{(}(1,g)\boldsymbol{)} \mathcal{U}^\dagger_\sym.
\end{eqs}
Similar to the effective action in Eq.~\eqref{eq: define CS effective boundary symmetry action 2}, when $\tilde{v}\boldsymbol{(}(g,1)\boldsymbol{)}$ and $\tilde{v}\boldsymbol{(}(1,g)\boldsymbol{)}$ are restricted to a region near either the endpoint $1$ or $2M$, they form a projective representation of $\ZZ_2 \times \ZZ_2$, corresponding to the nontrivial element of ${\mathcal{H}^2[\ZZ_2 \times \ZZ_2,U(1)]}$.

We compare $\tilde{H}_\cs$ to the paramagnet SPT Hamiltonian $H_0$:
\begin{align}
    H_0 \equiv -\sum_i X_i,
\end{align}
which also has $|\! + \ldots + \rangle$ as its unique ground state.
An effective boundary symmetry action computed with respect to $H_0$ is given by:
\begin{eqs}
&v^0\boldsymbol{(}(g,1)\boldsymbol{)} \equiv X_2 X_{2M} \\
&v^0\boldsymbol{(}(1,g)\boldsymbol{)} \equiv X_1 X_{2M-1}.
\end{eqs}
The restrictions of $v^0\boldsymbol{(}(g,1)\boldsymbol{)}$ and $v^0\boldsymbol{(}(1,g)\boldsymbol{)}$ to an endpoint forms a linear representation of $\ZZ_2 \times \ZZ_2$, corresponding to the trivial element of ${\mathcal{H}^2[\ZZ_2 \times \ZZ_2,U(1)]}$.

We see that the quantized invariant for the SPT phase containing $|\! + \ldots +\rangle$ computed using $\tilde{H}_\cs$ differs from the quantized invariant computed using $H_0$. This contradicts the fact the anomaly is well-defined. Therefore, the cluster state $|\psi_\cs \rangle$ cannot be disentangled using a FDQC composed of symmetric gates.

\section{Decorated domain wall models} \label{app: ddw models}

In this section, we define decorated domain wall models corresponding to {$D$-dimensional} $\Gone \times \Gtwo$ SPT phases characterized by an element of $\mathcal{H}^1[\Gone,\mathcal{H}^D[\Gtwo,U(1)]]$.
To illustrate the construction, we consider an example of a $1$D SPT phase protected by a $\ZZ_2 \times \ZZ_2$ symmetry.
We then demonstrate that the decorated domain wall models have an effective boundary symmetry action that prepares a $(D-1)$-dimensional $\Gtwo$ SPT state from a trivial SPT state. This property of the decorated domain wall models allows us to complete the proof of Proposition~\ref{prop: symmetry-protected magic SPT type iii}. 

\vspace{1.5mm}
\noindent \begin{center}\emph{Review of $\mathcal{H}^1[\Gone,\mathcal{H}^D[\Gtwo,U(1)]]$:}\end{center}
\vspace{1.5mm}

Our decorated domain wall models are constructed directly from the data of the group cohomology group $\mathcal{H}^1[\Gone,\mathcal{H}^D[\Gtwo,U(1)]]$. Therefore, before describing the decorated domain wall models, we review the essential details of $\mathcal{H}^1[\Gone,\mathcal{H}^D[\Gtwo,U(1)]]$. We refer to Ref.~\cite{CGLW13} for more information on group cohomology.

We start by reviewing the relevant data of the group cohomology $\mathcal{H}^D[\Gtwo,U(1)]$. The elements of $\mathcal{H}^D[\Gtwo,U(1)]$ {are equivalence classes (see Ref.~\cite{CGLW13}) of functions} from $\Gtwo^{D+1}$ to $U(1)$ that satisfy certain constraints. More specifically, an element $[\nu]$ in $\mathcal{H}^D[\Gtwo,U(1)]$ is labeled by a function $\nu:\Gtwo^{D+1} \to U(1)$ that is closed and homogeneous. By closed, we mean that $\nu$ satisfies the condition:
\begin{align} \label{eq: nu is closed}
    \prod_{i=0}^{D} \nu(k_0, \ldots, \widehat{k_i}, \ldots, k_{D+1})^{(-1)^i} = 1,
\end{align}
where $\widehat{k_i}$ denotes that ${k_i}$ is omitted. By homogeneous, we mean that $\nu$ satisfies:
\begin{align}
    \nu(k_0,\ldots,k_D) = \nu(kk_0, \ldots, kk_D),
\end{align}
for all $k\in K$.

The elements of $\mathcal{H}^D[\Gtwo,U(1)]$ form a finite Abelian group under multiplication with the product of $[\nu]$ and $[\nu']$ in $\mathcal{H}^D[\Gtwo,U(1)]$ given by:
\begin{align}
   [\nu] \cdot [\nu'] = [\nu\nu'].
\end{align}
Accordingly, the group $\mathcal{H}^D[\Gtwo,U(1)]$ takes the general form:
\begin{align}
    \mathcal{H}^D[\Gtwo,U(1)] = \prod_{j=1}^p \ZZ_{n_j},
\end{align}
where $p$ gives the number of generators of $\mathcal{H}^D[\Gtwo,U(1)]$. We label the $j^\text{th}$ generator of the cohomology by the function $\nu_{j}$.

With this, we can describe the data of the group cohomology $\mathcal{H}^1[\Gone,\mathcal{H}^D[\Gtwo,U(1)]]$. The elements of the group $\mathcal{H}^1[\Gone,\mathcal{H}^D[\Gtwo,U(1)]]$ can be labeled by functions of the form:
\begin{align} \label{eq: mixed anomaly cocycle}
    \eta \equiv \prod_{j=1}^p \nu_{j}^{\phi_{j}}: K^{D+1} \times H^2 \to U(1).
\end{align}
Here, $\phi_j$ is a function from $H^2$ to $\ZZ_{n_j}$ that is closed and homogeneous. In this case, closed means that $\phi_j$ satisfies:
\begin{align} \label{eq: phi closed}
    \phi_j(h_1,h_2) - \phi_j(h_0,h_2) + \phi_j(h_0,h_1) = 0, 
\end{align}
and homogeneous means that $\phi_j$ satisfies:
\begin{align} \label{eq: phi homogeneous}
    \phi_j(h_0,h_1) = \phi_j(hh_0,hh_1), \quad \forall h \in H.
\end{align}
Explicitly, $\eta$ evaluated on the group elements $k_0,\ldots,k_D$ in $K$ and $h_0,h_1$ in $H$ is:
\begin{eqs}
    \eta(k_0,\ldots,k_D;h_0,h_1) = \prod_{j=1}^p \nu_j(k_0,\ldots,k_D)^{\phi_j(h_0,h_1)}.
\end{eqs}
In the next section, we show that the function $\eta$ above can be used to construct a model belonging to the SPT phase characterized by $[\eta] \in \mathcal{H}^1[\Gone,\mathcal{H}^D[\Gtwo,U(1)]]$.

\vspace{1.5mm}
\noindent \begin{center}\emph{Construction of the decorated domain wall models:}\end{center}
\vspace{1.5mm}

\begin{figure}[t]
\centering
\includegraphics[width=.5\textwidth,trim={14cm 4cm 12cm 3.5cm},clip]{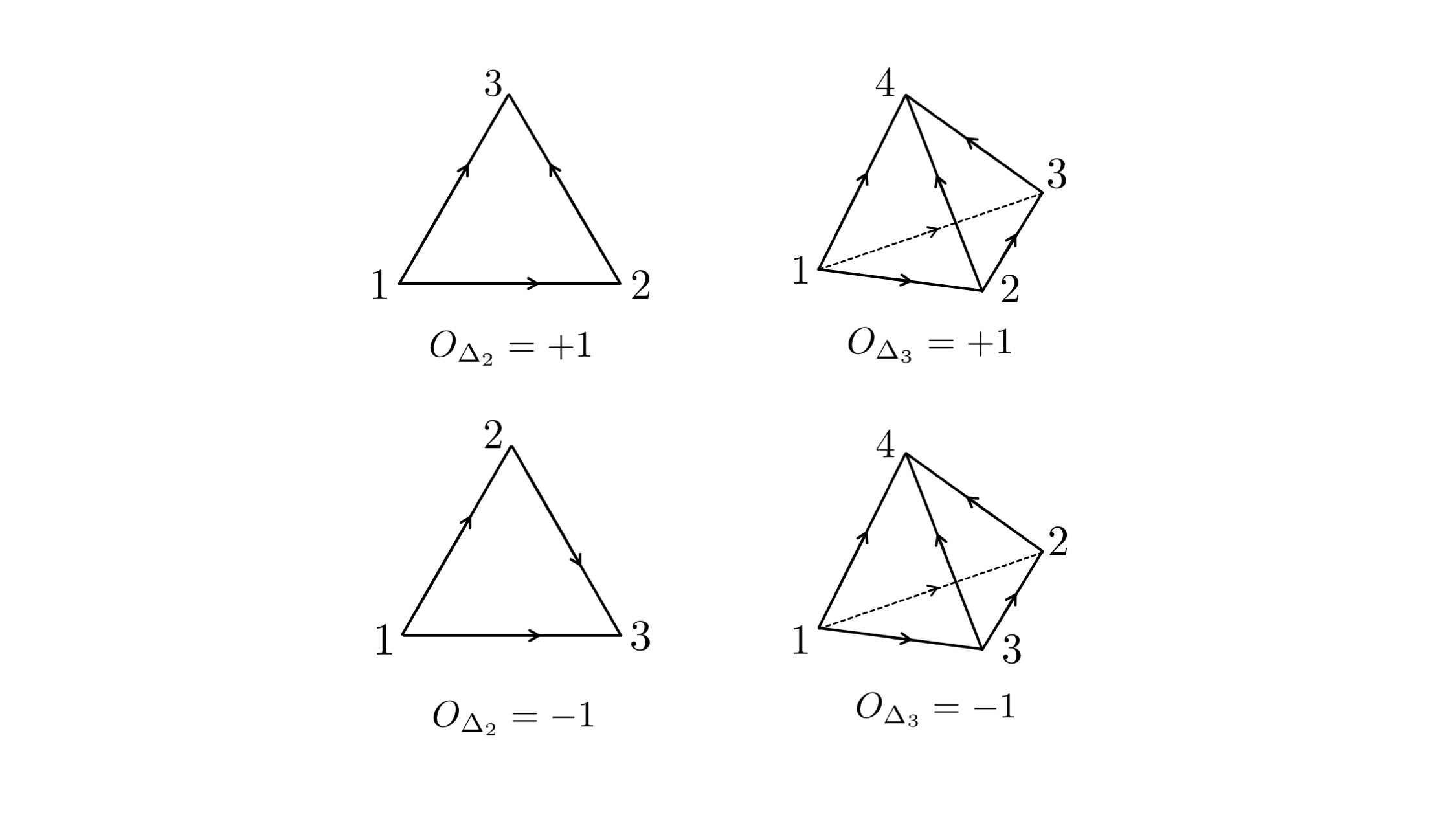}
\caption{The branching structure induces an ordering of the vertices of a simplex according to the number of edges pointing towards a vertex. It also defines an orientation of $d$-simplices in a $d$-manifold relative to the orientation of the manifold. We denote the $\{\pm 1\}$-valued orientation of a $d$-simplex $\Delta_d$ by $O_{\Delta_d}$.
 }
\label{fig:branchingstructure}
\end{figure}

We define our decorated domain wall models on an arbitrary triangulation of an orientable closed $D$-dimensional manifold $N$. The triangulation specifies a decomposition of $N$ into simplices (e.g., vertices, edges, faces, etc.), and we denote $d$-dimensional simplices ($d$-simplices) by $\simp_d$. We also require that the triangulation is equipped with a branching structure, i.e., an assignment of an orientation to each edge so that there are no cycles around any face. The branching structure defines a local ordering of the vertices and an orientation of the $d$-simplices (for $d\geq 1$), as shown in Fig.~\ref{fig:branchingstructure}. We occasionally write a $d$-simplex in terms of its vertices, such as: $\simp_d = \langle 1, \ldots , d +1 \rangle$, where the vertices $1,\ldots,d+1$ are ordered according to the branching structure.

\begin{figure}[t]
\centering
\includegraphics[width=.4\textwidth,trim={14cm 6cm 14cm 4cm},clip]{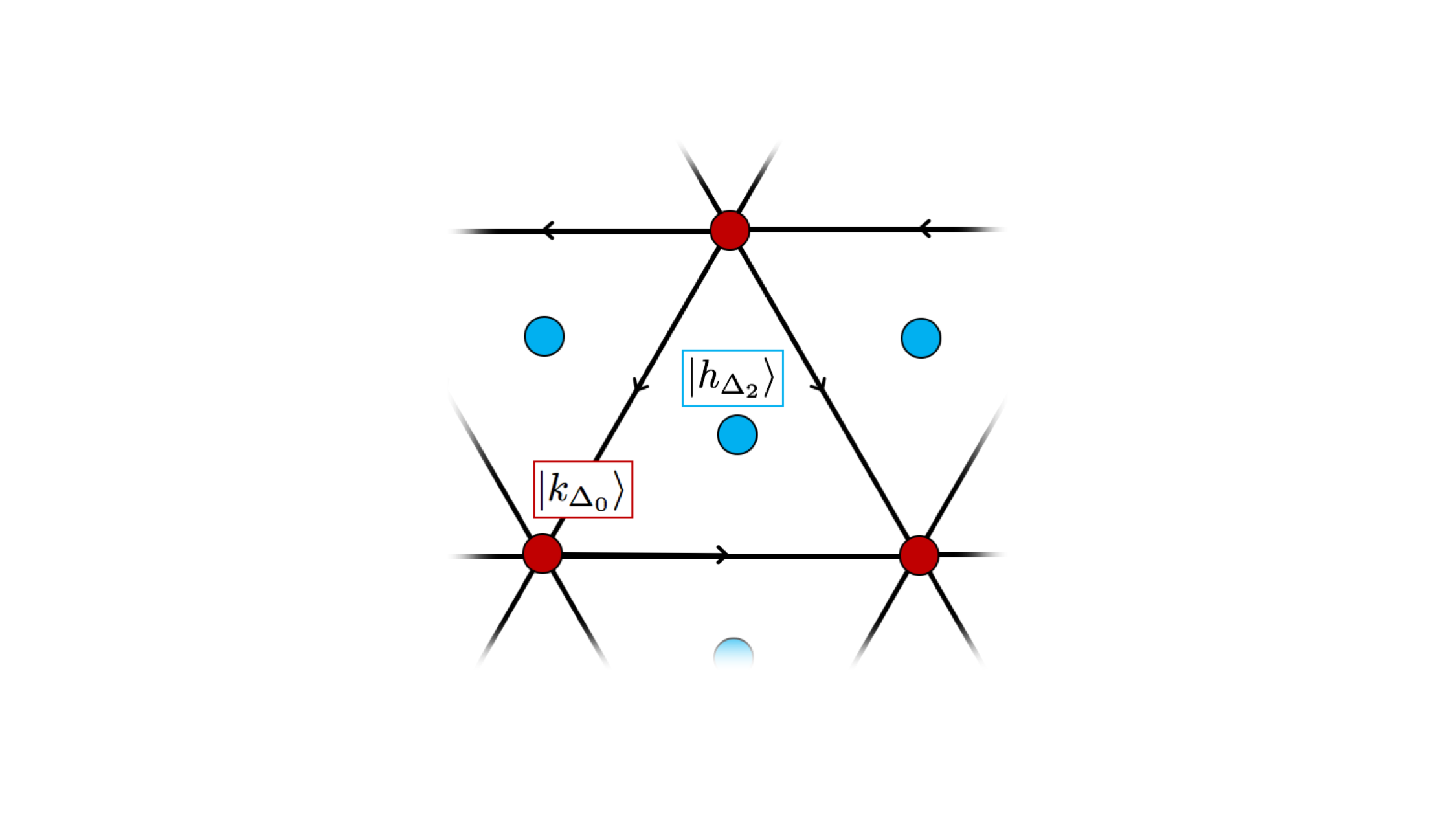}
\caption{In $2$D, the decorated domain wall model for an $H \times K$ SPT phase is defined on a triangulation of a $2$D lattice with an $|H|$-dimensional Hilbert space (blue) at each $2$-simplex and a $|K|$-dimensional Hilbert space (red) at each vertex. A product state basis for the total Hilbert space is given by states labeled by $(H \times K)$-configurations $\{h_{\simp_{{\scriptscriptstyle 2}}}\},\{k_{\simp_{{\scriptscriptstyle 0}}}\}$. For the basis state $|\{h_{\simp_{{\scriptscriptstyle 2}}}\},\{k_{\simp_{{\scriptscriptstyle 0}}}\}\rangle$, the state at the $2$-simplex $\Delta_2$ is $|h_{\Delta_2}\rangle$ and the state at the vertex $\Delta_0$ is $|k_{\Delta_0}\rangle$. }
\label{fig:ddwdof}
\end{figure}

To construct a model for an SPT phase protected by $\Gone \times \Gtwo$ symmetry, we place an $|\Gone|$-dimensional qudit at each $D$-simplex and a $|\Gtwo|$-dimensional qudit at each vertex. We label the basis states of a qudit at a $D$-simplex by elements of $\Gone$ and the basis states of a qudit at a vertex by elements of $\Gtwo$. A basis for the full Hilbert space is then given by product states of the form $|\{h_{\simp_{{\scriptscriptstyle D}}}\},\{k_{\simp_{{\scriptscriptstyle 0}}}\}\rangle$, where the state at $\simp_{{\scriptscriptstyle D}}$ is $|h_{\simp_{{\scriptscriptstyle D}}}\rangle$ and the state at $\simp_{{\scriptscriptstyle 0}}$ is $|k_{\simp_{{\scriptscriptstyle 0}}}\rangle$ (Fig.~\ref{fig:ddwdof}). Furthermore, we define the $H \times K$ symmetry to act as left multiplication, so for any element $(h,k) \in H \times K$, the onsite symmetry action $u\boldsymbol{(}(h,k)\boldsymbol{)}$ is defined by:
\begin{align}
    u\boldsymbol{(}(h,k)\boldsymbol{)}|\{h_{\simp_{{\scriptscriptstyle D}}}\},\{k_{\simp_{{\scriptscriptstyle 0}}}\}\rangle=|\{hh_{\simp_{{\scriptscriptstyle D}}}\},\{kk_{\simp_{{\scriptscriptstyle 0}}}\}\rangle.
\end{align}

Next, we introduce an $H \times K$ paramagnet Hamiltonian with a ground state in the trivial SPT phase. We construct the decorated domain wall models starting with the paramagnet Hamiltonian. The paramagnet Hamiltonian is:
\begin{align}\label{eq: ddw paramagnet}
    H_0 \equiv -\sum_{\simp_{{\scriptscriptstyle D}}} \Pi_{\simp_{{\scriptscriptstyle D}}} - \sum_{\simp_{{\scriptscriptstyle 0}}} \Pi_{\simp_{{\scriptscriptstyle 0}}},
\end{align}
where $\Pi_{\simp_{{\scriptscriptstyle D}}}$ and $\Pi_{\simp_{{\scriptscriptstyle 0}}}$ are projectors onto symmetric states at $\simp_{{\scriptscriptstyle D}}$ and $\simp_{{\scriptscriptstyle 0}}$ (tensored with the identity on all other sites):
\begin{align}
    \Pi_{\simp_{{\scriptscriptstyle D}}} &\equiv \frac{1}{|H|}\bigg(\sum_{h_{\simp_{{\scriptscriptstyle D}}}} |h_{\simp_{{\scriptscriptstyle D}}}\rangle\bigg) \bigg( \sum_{h_{\simp_{{\scriptscriptstyle D}}}} \langle h_{\simp_{{\scriptscriptstyle D}}}|\bigg), \\
    \Pi_{\simp_{{\scriptscriptstyle 0}}} &\equiv \frac{1}{|K|}\bigg(\sum_{k_{\simp_{{\scriptscriptstyle 0}}}} |k_{\simp_{{\scriptscriptstyle 0}}}\rangle\bigg) \bigg(\sum_{k_{\simp_{{\scriptscriptstyle 0}}}}\langle k_{\simp_{{\scriptscriptstyle 0}}}|\bigg).
\end{align}
The ground state of $H_0$ is the symmetric product state:
\begin{align}
    |\psi_0 \rangle \equiv \sum_{\{h_{\simp_{{\scriptscriptstyle D}}}\},\{k_{\simp_{{\scriptscriptstyle 0}}}\}}|\{h_{\simp_{{\scriptscriptstyle D}}}\},\{k_{\simp_{{\scriptscriptstyle 0}}}\}\rangle,
\end{align}
with the sum over $\{h_{\simp_{{\scriptscriptstyle D}}}\}$ and $\{k_{\simp_{{\scriptscriptstyle 0}}}\}$ configurations and with the normalization suppressed. To simplify the notation, we use $\mathcal{C}$ to denote a configuration of $\{h_{\simp_{{\scriptscriptstyle D}}}\}$ and $\{k_{\simp_{{\scriptscriptstyle 0}}}\}$. $|\psi_0\rangle$ can then be written as:
\begin{align}
    |\psi_0 \rangle = \sum_\mathcal{C} |\mathcal{C}\rangle.
\end{align}

\begin{figure}[t]
\centering
\includegraphics[width=.5\textwidth,trim={9cm 11cm 5cm 10cm},clip]{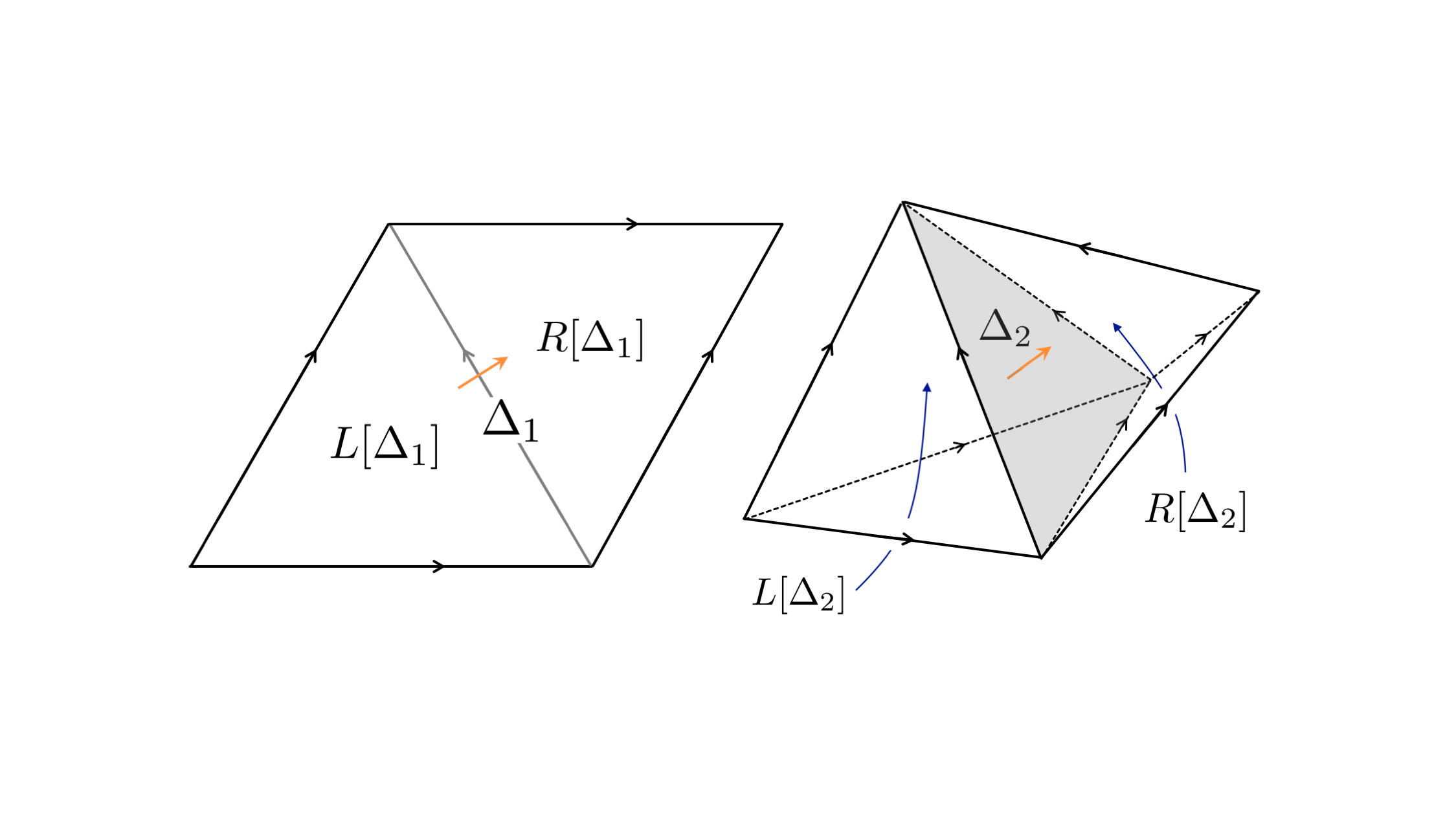}
\caption{The $D$-simplices $L[\Delta_{\scriptscriptstyle D-1}]$ and $R[\Delta_{\scriptscriptstyle D-1}]$ are the two $D$-simplices that neighbor the $(D-1)$-simplex $\Delta_{\scriptscriptstyle D-1}$. The normal vector of $\Delta_{\scriptscriptstyle D-1}$ (orange) is determined by the orientation of $\Delta_{\scriptscriptstyle D-1}$ (Fig.~\ref{fig:branchingstructure}) and points from $L[\Delta_{\scriptscriptstyle D-1}]$ to $R[\Delta_{\scriptscriptstyle D-1}]$.}
\label{fig:RLsimplices}
\end{figure}

The decorated domain wall models are constructed from the $H \times K$ paramagnet Hamiltonian by conjugating with a FDQC $\mathcal{U}_\eta$. The FDQC is built from a choice of $\eta$ in Eq.~\eqref{eq: mixed anomaly cocycle} and takes the form:
\begin{align} \label{eq: Ueta def}
    \mathcal{U}_\eta \equiv \prod_{\simp_{{\scriptscriptstyle D-1}}} U_{\simp_{{\scriptscriptstyle D-1}}},
\end{align}
where $U_{\simp_{{\scriptscriptstyle D-1}}}$ are the local gates: 
\begin{align} \label{eq: Ueta gate}
      U_{\simp_{{\scriptscriptstyle D-1}}} \equiv \sum_{\mathcal{C}} \bar{\eta}_{\mathcal{C}}(\simp_{{\scriptscriptstyle D-1}})  |\mathcal{C}\rangle\langle\mathcal{C}|.
\end{align}
Here, $\bar{\eta}_\mathcal{C}(\simp_{{\scriptscriptstyle D-1}})$ is a $U(1)$-valued phase that depends on the configuration $\mathcal{C}$ in the vicinity of  $\simp_{{\scriptscriptstyle D-1}}$. For a $(D-1)$-simplex $\simp_{{\scriptscriptstyle D-1}}= \langle 1, \ldots, D \rangle$, $\bar{\eta}_\mathcal{C}(\simp_{{\scriptscriptstyle D-1}})$ is explicitly:
\begin{eqs} \label{eq: ddw phase}
    \bar{\eta}_\mathcal{C}(\simp_{{\scriptscriptstyle D-1}}) &\equiv \eta(1,k_1,\ldots,k_{D};h_{L[{\simp_{{\scriptscriptstyle D-1}}}]},h_{R[{\simp_{{\scriptscriptstyle D-1}}}]}) \\
    &=  \prod_{j=1}^p \nu_j(1,k_1,\ldots,k_{D})^{\phi_j\left(h_{L[{\simp_{{\scriptscriptstyle D-1}}}]},h_{R[{\simp_{{\scriptscriptstyle D-1}}}]}\right)},
\end{eqs}
where $k_1,\ldots,k_{D-1}$ are the $\Gtwo$ labels at the vertices of $\simp_{{\scriptscriptstyle D-1}}$ in the configuration $\mathcal{C}$, and $L[\simp_{{\scriptscriptstyle D-1}}]$ and $R[\simp_{{\scriptscriptstyle D-1}}]$ are the $D$-simplices on either side of $\simp_{{\scriptscriptstyle D-1}}$ as depicted in Fig.~\ref{fig:RLsimplices}.

We are now able to define the decorated domain wall Hamiltonian $H_\eta$ corresponding to the element $[\eta]$ in the group cohomology $\mathcal{H}^1[\Gone,\mathcal{H}^D[\Gtwo,U(1)]]$. We define $H_\eta$ to be the Hamiltonian:
\begin{align}
    H_\eta \equiv \mathcal{U}_\eta^\dagger H_0 \mathcal{U}_\eta.
\end{align}
$H_\eta$ is local due to the finite Lieb-Robinson length of $\mathcal{U}_\eta$, and it is symmetric due to the fact that $\mathcal{U}_\eta$ commutes with the global symmetry $u\boldsymbol{(}(h,k)\boldsymbol{)}$, for every $(h,k)$ in $H \times K$. The symmetry of $\mathcal{U}_\eta$ can be checked by using that $\nu_j$ and $\phi_j$ are closed and homogeneous for every $j$. Note that the local gates $U_{\simp_{{\scriptscriptstyle D-1}}}$ are not symmetric for nontrivial $\eta$, however. The ground state of $H_\eta$ can be constructed by applying $\mathcal{U}_\eta$ to the ground state of $H_0$; this defines:
\begin{eqs} \label{eq: ddw ground state}
    |\psi_\eta \rangle &\equiv \mathcal{U}_\eta |\psi_0\rangle = \sum_{\mathcal{C}}\prod_{\simp_{{\scriptscriptstyle D-1}}}\bar{\eta}_{\mathcal{C}}(\simp_{{\scriptscriptstyle D-1}})|\mathcal{C}\rangle.
\end{eqs}

\begin{figure}[t]
\centering
\includegraphics[width=.5\textwidth,trim={18cm 15cm 20cm 12cm},clip]{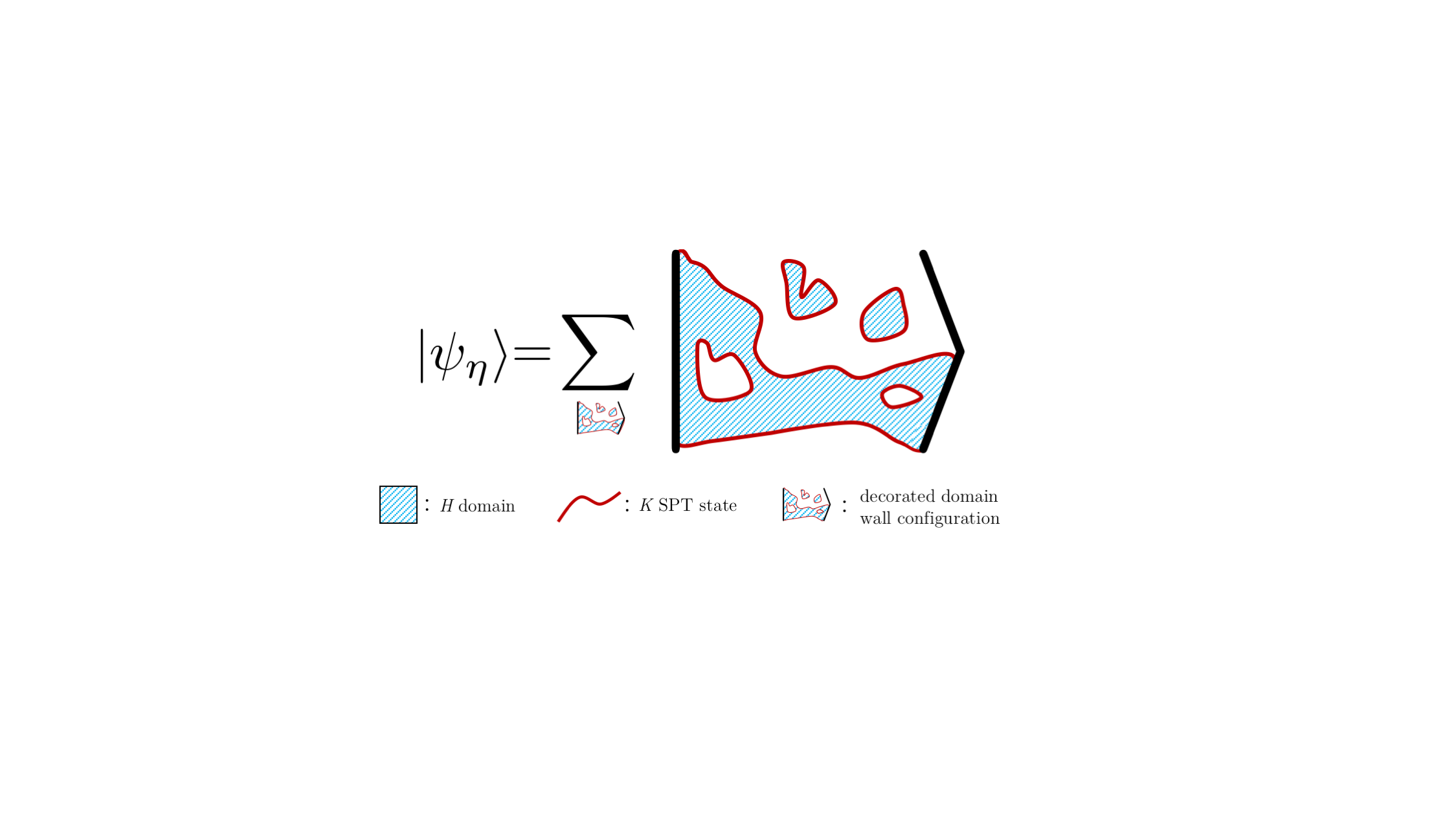}
\caption{The ground state $|\psi_\eta\rangle$ is a superposition of decorated domain wall configurations, where $H$ domains (blue) are decorated with $K$ SPT states (red) along the domain walls.}
\label{fig:ddwstate}
\end{figure}

The ground state $|\psi_\eta \rangle$ can be understood in terms of $H$ domain walls decorated by $(D-1)$-dimensional $K$ SPT states. To see this, we expand the amplitude in Eq.~\eqref{eq: ddw ground state} using the expression for $\bar{\eta}_{\mathcal{C}}(\simp_{{\scriptscriptstyle D-1}})$ in Eq.~\eqref{eq: ddw phase}. The amplitude corresponding to the configuration $\mathcal{C}$ is:
\begin{align}
   \prod_{\simp_{{\scriptscriptstyle D-1}}} \prod_{j=1}^p \nu_j(1,k_1,\ldots,k_{D})^{\phi_j\left(h_{L[{\simp_{{\scriptscriptstyle D-1}}}]},h_{R[{\simp_{{\scriptscriptstyle D-1}}}]}\right)}.
\end{align}
The functions $\phi_j$ count the $H$ domain walls between the $D$-simplices $L[{\simp_{{\scriptscriptstyle D-1}}}]$ and $R[{\simp_{{\scriptscriptstyle D-1}}}]$ bordering $\simp_{{\scriptscriptstyle D-1}}$. Factors of $\nu_j$ are assigned to the $(D-1)$-simplices according to the number of $H$ domain walls. The product of $\nu_j$ along an $H$ domain wall gives a $(D-1)$-dimensional $\Gtwo$ SPT state defined in Ref.~\cite{CGLW13}. The ground state $|\psi_\eta \rangle$ is depicted in Fig.~\ref{fig:ddwstate}.

\vspace{1.5mm}
\noindent \begin{center}\emph{Cluster state example:}\end{center}
\vspace{1.5mm}

To make the construction of the decorated domain wall models less abstract, we consider a $1$D example with a $\ZZ_2 \times \ZZ_2$ symmetry. We show that the decorated domain wall model reproduces the cluster state model in Section~\ref{sec: example: cluster state}. 

For $D=1$ with a $\ZZ_2 \times \ZZ_2$ symmetry, the decorated domain wall model is characterized by an element of the group cohomology $\mathcal{H}^1[\ZZ_2,\mathcal{H}^1[\ZZ_2,U(1)]]$. In this case, $\mathcal{H}^1[\ZZ_2,U(1)]$ forms a $\ZZ_2$ group, and the generator can be represented by the function $\nu:\ZZ_2^2 \to U(1)$ defined by:
\begin{align}
    \nu(k_0,k_1)=(-1)^{k_0+k_1}.
\end{align}
The nontrivial element of the group cohomology group $\mathcal{H}^1[\ZZ_2,\mathcal{H}^1[\ZZ_2,U(1)]]=\ZZ_2$ can be labeled by the function:
\begin{align}
    \eta = \nu^\phi : \ZZ_2^2 \times \ZZ_2^2 \to U(1),
\end{align}
where $\phi:\ZZ_2^2 \to \ZZ_2$ is:
\begin{align}
    \phi(h_0,h_1) = h_0 + h_1.
\end{align}
More explicitly, $\eta$ is defined by:
\begin{align}
    \eta(k_0,k_1;h_0,h_1) = \nu(k_0,k_1)^{\phi(h_0,h_1)} = (-1)^{(k_0+k_1)(h_0 + h_1)}.
\end{align}

We can now build the decorated domain wall model on a periodic $1$D lattice with a qubit at each edge $\simp_1$ and at each vertex $\simp_{{\scriptscriptstyle 0}}$. To make contact with the cluster state model in Section~\ref{sec: example: cluster state}, we label edges by even integers ($\simp_1 = 2j$) and vertices by odd integers ($\simp_{{\scriptscriptstyle 0}}=2j+1$). The symmetry action corresponding to $(h,k)$ in $\ZZ_2 \times \ZZ_2$ is:
\begin{align}
    u\boldsymbol{(}(h,k)\boldsymbol{)} |\{h_{2j}\},\{k_{2j+1}\}\rangle = |\{h+h_{2j}\},\{k+k_{2j+1}\}\rangle,
\end{align}
where the addition is modulo $2$.
This can be represented by the product of Pauli X operators:
\begin{align}
    u\boldsymbol{(}(h,k)\boldsymbol{)} = \Big(\prod_{j} X_{2j}\Big)^h \Big(\prod_{j} X_{2j+1}\Big)^k,
\end{align}
in agreement with the symmetry action in Section~\ref{sec: example: cluster state}.

We construct the Hamiltonian for the decorated domain wall model starting with the $\ZZ_2 \times \ZZ_2$ paramagnet Hamiltonian $H_0$:
\begin{align}
    H_0 = -\sum_{i} X_{i}.
\end{align}
Note that the paramagnet Hamiltonian above only differs from the Hamiltonian described in Eq.~\eqref{eq: ddw paramagnet} by a constant shift of energy. The ground state of the paramagnet Hamiltonian is the product state:
\begin{align}
    |\psi_0 \rangle = \sum_\mathcal{C} |\mathcal{C}\rangle = |\!+ \cdots + \rangle.
\end{align}

The last step in the construction is to conjugate $H_0$ by the FDQC $\mathcal{U}_\eta$. Here, $\mathcal{U}_\eta$ is given by: 
\begin{align}
    \mathcal{U}_\eta = \prod_{j} \sum_{\mathcal{C}} (-1)^{k_{2j+1}(h_{2j}+h_{2j+2})}|\mathcal{C}\rangle \langle \mathcal{C} |,
\end{align}
which can be written using Pauli Z operators as:
\begin{align} \label{eq: Ueta Pauli Z}
    \mathcal{U}_\eta = \prod_{j} Z_{2j+1}^{\frac{1}{2}(1-Z_{2j}Z_{2j+2})}.
\end{align}
The exponent in Eq.~\eqref{eq: Ueta Pauli Z} detects domain walls between the sites $2j$ and $2j+2$. The operator $Z_{2j+1}$ then creates a charge at $2j+1$ if there is a domain wall between the neighboring even sites. The charge can be interpreted as a nontrivial $0$-dimensional SPT state.

$U_\eta$ can equivalently be expressed as a product of control Z operators:
\begin{eqs}
    \mathcal{U}_\eta &= \prod_j CZ_{2j(2j+1)}CZ_{(2j+1)(2j+2)} \\
    &= \prod_i CZ_{i(i+1)}.
\end{eqs}
In this form, one can see that $\mathcal{U}_\eta$ is precisely the FDQC $\mathcal{U}_\cs$ from Section~\ref{sec: example: cluster state}. Furthermore, the Hamiltonian $H_\eta = \mathcal{U}_\eta H_0 \mathcal{U}^\dagger_\eta$ is the same as the cluster state Hamiltonian $H_\cs$ in Eq.~\eqref{eq: cluster state Hamiltonian}.

\vspace{1.5mm}
\noindent \begin{center}\emph{Effective boundary symmetry action of the decorated domain wall models:}\end{center}
\vspace{1.5mm}

Our decorated domain wall models are designed so that the effective boundary symmetry action corresponding to some element $(h,1)$ in $H \times K$ is a FDQC that prepares a nontrivial $K$ SPT state from a trivial SPT state. In fact, the effective boundary symmetry action prepares the representative group cohomology $K$ SPT states constructed in Ref.~\cite{CGLW13}. For simplicity, we assume that $H$ is Abelian. The computation can be generalized to non-Abelian symmetries straightforwardly.

As described in Section~\ref{sec: anomalous boundary symmetry action}, the first step in computing the effective boundary symmetry action is to define the boundary Hilbert space. To do so, we truncate the Hamiltonian $H_\eta$ to a manifold $M$ with boundary by removing any terms in $H_\eta$ that are supported on sites outside of $M$. The boundary Hilbert space $\mathcal{H}_\eta^\text{low}$ is then the low-energy Hilbert space of the truncated Hamiltonian $H^M_\eta$. 

We also define a truncation of the FDQC $\mathcal{U}_\eta$ to $M$. $\mathcal{U}_\eta$ is truncated by removing all of the gates supported on sites outside of $M$. This gives the FDQC:
\begin{align}
    \mathcal{U}^M_\eta \equiv \prod_{\Delta_{\scriptscriptstyle D-1} \in M \setminus \partial M} U_{\Delta_{\scriptscriptstyle D-1}},
\end{align}
where the product is over $(D-1)$-simplices in $M$ that do not belong to the boundary of $M$. Importantly, for any state $|\psi_\text{low}\rangle$ in the low-energy Hilbert space of $H^M_\eta$, the FDQC $(\mathcal{U}^M_\eta)^\dagger$ disentangles $|\psi_\text{low}\rangle$ away from the boundary of $M$. More specifically, the action of $(\mathcal{U}^M_\eta)^\dagger$ on $|\psi_\text{low}\rangle$ is:
\begin{align} \label{eq: disentangle bulk with truncated circuit}
    (\mathcal{U}^M_\eta)^\dagger|\psi_\text{low}\rangle = |\psi \rangle_{\scriptscriptstyle M  \setminus M_\circ} \otimes |\psi_\text{prod} \rangle_{\scriptscriptstyle M_\circ}.
\end{align}
Here, $M_\circ$ is a submanifold of $M$ that contains the sites greater than two Lieb-Robinson lengths from the boundary of $M$ but does not contain any sites within one Lieb-Robinson length of the boundary. Furthermore, in Eq.~\eqref{eq: disentangle bulk with truncated circuit} $|\psi\rangle_{M\setminus M_\circ}$ is some state defined on the sites in $M$ outside of $M_\circ$, and $|\psi_{\text{prod}}\rangle$ is a symmetric product state on the sites in $M_\circ$. Eq.~\eqref{eq: disentangle bulk with truncated circuit} agrees with the intuition that states in the boundary Hilbert space resemble the ground state of $H_\eta$ far away from the boundary of $M$.

We now define the effective boundary symmetry action corresponding to $(h,k)$ in $H \times K$ to be:
\begin{align} \label{eq: ddw effective boundary sym}
    v\boldsymbol{(}(h,k)\boldsymbol{)} \equiv u_{\scriptscriptstyle M}\boldsymbol{(}(h,k)\boldsymbol{)} \mathcal{U}^M_\eta u^\dagger_{\scriptscriptstyle M_\circ}\boldsymbol{(}(h,k)\boldsymbol{)} (\mathcal{U}_\eta^M)^\dagger,
\end{align}
where $u_{\scriptscriptstyle M}\boldsymbol{(}(h,k)\boldsymbol{)}$ and $u^\dagger_{\scriptscriptstyle M_\circ}\boldsymbol{(}(h,k)\boldsymbol{)}$ are the onsite symmetry actions restricted to $M$ and $M_\circ$, respectively. To show that this is an appropriate choice for the effective boundary symmetry action, we consider acting with $v\boldsymbol{(}(h,k)\boldsymbol{)}$ on an arbitrary state $|\psi_\text{low}\rangle$ in the boundary Hilbert space:
\begin{eqs}
   v\boldsymbol{(}(h,k)\boldsymbol{)} |\psi_\text{low}\rangle
    = u_{\scriptscriptstyle M}\boldsymbol{(}(h,k)\boldsymbol{)} \mathcal{U}^M_\eta u^\dagger_{\scriptscriptstyle M_\circ}\boldsymbol{(}(h,k)\boldsymbol{)} (\mathcal{U}_\eta^M)^\dagger |\psi_\text{low}\rangle.
\end{eqs}
Using Eq.~\eqref{eq: disentangle bulk with truncated circuit} and that $|\psi_\text{prod}\rangle_{\scriptscriptstyle M_\circ}$ is symmetric, we find:
\begin{eqs}
   v\boldsymbol{(}(h,k)\boldsymbol{)} |\psi_\text{low}\rangle 
    &= u_{\scriptscriptstyle M}\boldsymbol{(}(h,k)\boldsymbol{)} \mathcal{U}^M_\eta \Big(|\psi \rangle_{\scriptscriptstyle M  \setminus M_\circ} \otimes |\psi_\text{prod} \rangle_{\scriptscriptstyle M_\circ}\Big) \\
    &= u_{\scriptscriptstyle M}\boldsymbol{(}(h,k)\boldsymbol{)}|\psi_\text{low}\rangle.
\end{eqs}
% \begin{eqs}
%   v\boldsymbol{(}(h&,k)\boldsymbol{)} |\psi_\text{low}\rangle \\
%     &= u_{\scriptscriptstyle M}\boldsymbol{(}(h,k)\boldsymbol{)} \mathcal{U}^M_\eta u^\dagger_{\scriptscriptstyle M_\circ}\boldsymbol{(}(h,k)\boldsymbol{)} \Big(|\psi \rangle_{\scriptscriptstyle M  \setminus M_\circ} \otimes |\psi_\text{prod} \rangle_{\scriptscriptstyle M_\circ}\Big) \\
%     &= u_{\scriptscriptstyle M}\boldsymbol{(}(h,k)\boldsymbol{)} \mathcal{U}^M_\eta \Big(|\psi \rangle_{\scriptscriptstyle M  \setminus M_\circ} \otimes |\psi_\text{prod} \rangle_{\scriptscriptstyle M_\circ}\Big) \\
%     &= u_{\scriptscriptstyle M}\boldsymbol{(}(h,k)\boldsymbol{)}|\psi_\text{low}\rangle.
% \end{eqs}
Therefore, the action of $v\boldsymbol{(}(h,k)\boldsymbol{)}$ on states in the boundary Hilbert space is the same as the action of $ u_{\scriptscriptstyle M}\boldsymbol{(}(h,k)\boldsymbol{)}$. This implies that $v\boldsymbol{(}(h,k)\boldsymbol{)}$ forms a linear representation of the symmetry on the boundary Hilbert space. The support of $v\boldsymbol{(}(h,k)\boldsymbol{)}$ is also localized to the boundary of $M$. This is because $\mathcal{U}_\eta$ is symmetric, and $\mathcal{U}^M_\eta$ commutes with $u_{\scriptscriptstyle M_\circ}^\dagger\boldsymbol{(}(h,k)\boldsymbol{)}$ up to an operator supported within three Lieb-Robinson lengths of the boundary of $M$. 

Finally, we compute the effective boundary symmetry action for an element $(h,1)$ in $H \times K$ using Eq.~\eqref{eq: ddw effective boundary sym}. We start by conjugating $\mathcal{U}^M_\eta$ by the symmetry action restricted to $M_\circ$:
\begin{align}
    u_{\scriptscriptstyle M_\circ}\boldsymbol{(}(h,1)\boldsymbol{)}\mathcal{U}_\eta^M u_{\scriptscriptstyle M_\circ}^\dagger\boldsymbol{(}(h,1)\boldsymbol{)}.
\end{align}
The FDQC $\mathcal{U}_\eta^M$ only depends on the $|H|$-dimensional qudits through the functions $\phi_j$ in Eq.~\eqref{eq: ddw phase}. Given that each $\phi_j$ is homogeneous [Eq.~\eqref{eq: phi homogeneous}], $\mathcal{U}_\eta^M$ is only affected by the symmetry action near the boundary of $M_\circ$. In particular, one can use that $\phi_j$ is closed [Eq.~\eqref{eq: phi closed}] to show:
\begin{eqs} \label{eq: ddw conjugate truncated circuit}
    u_{\scriptscriptstyle M_\circ}\boldsymbol{(}(h,1)\boldsymbol{)}\mathcal{U}_\eta^M u_{\scriptscriptstyle M_\circ}^\dagger\boldsymbol{(}(h,1)\boldsymbol{)} = \mathcal{V}\boldsymbol{(}(h,1)\boldsymbol{)} \mathcal{U}_\eta^M,
\end{eqs}
where $\mathcal{V}\boldsymbol{(}(h,1)\boldsymbol{)}$ is the operator:
\begin{eqs}
      \mathcal{V}\boldsymbol{(}(h,1)\boldsymbol{)} = \prod_{\Delta_{\scriptscriptstyle D-1} \in \partial M_\circ} \sum_{\mathcal{C}}  \bar{\eta}^{(h)}_\mathcal{C}(\Delta_{\scriptscriptstyle D-1})^{O_{\Delta_{\scriptscriptstyle D-1}}}|\mathcal{C}\rangle \langle \mathcal{C} |.
\end{eqs}
In the expression above, the product is over $(D-1)$-simplices in the boundary of $M_\circ$, $O_{\Delta_{\scriptscriptstyle D-1}}$ is the $\{\pm 1\}$-valued orientation of the simplex $\Delta_{\scriptscriptstyle D-1}$ relative to the boundary of $M_\circ$,\footnote{If $D=1$, then $O_{\Delta_{\scriptscriptstyle 0}}$ is $+1$ if ${\Delta_{\scriptscriptstyle 0}}$ is a left endpoint of $M_\circ$ and $-1$ if ${\Delta_{\scriptscriptstyle 0}}$ is a right endpoint of $M_\circ$.} and for $\Delta_{\scriptscriptstyle D-1} = \langle 1, \ldots, D-1 \rangle$, $\bar{\eta}^{(h)}_\mathcal{C}(\Delta_{\scriptscriptstyle D-1})$ is the $U(1)$ phase:
\begin{align} \label{eq: ddw ebsa Abelian}
      \bar{\eta}^{(h)}_\mathcal{C}(\Delta_{\scriptscriptstyle D-1}) \equiv
      \prod_{j=1}^p\nu_j(1,k_1, \ldots , k_{D})^{\phi_j(1,h)}.
\end{align}
If $\phi_j(1,h)$ is nontrivial, then $\mathcal{V}\boldsymbol{(}(h,1)\boldsymbol{)}$ is precisely the FDQC described in Ref.~\cite{CGLW13} that prepares a $K$ SPT state characterized by the element of $H^D[K,U(1)]$ labeled by $\nu_j$.

To finish the calculation of the effective boundary symmetry action, we substitute Eq.~\eqref{eq: ddw conjugate truncated circuit} into Eq.~\eqref{eq: ddw effective boundary sym}. This gives us:
\begin{align}
    v\boldsymbol{(}(h,1)\boldsymbol{)} = u_{\scriptscriptstyle M \setminus M_\circ}\boldsymbol{(} (h,1) \boldsymbol{)} \mathcal{V}\boldsymbol{(}(h,1)\boldsymbol{)},
\end{align}
where $u_{\scriptscriptstyle M \setminus M_\circ}\boldsymbol{(} (h,1) \boldsymbol{)}$ is the restriction of the onsite symmetry to the sites in $M$ that are outside of $M_\circ$. We now see that, for any $(h,1)$ in $H \times K$ for which $\phi_j(1,h)$ is nontrivial, the effective boundary symmetry action prepares a $(D-1)$-dimensional $K$ SPT state corresponding to $\nu_j$ along the boundary of $M_\circ$. If the decorated domain wall model corresponds to a nontrivial element of the group cohomology $\mathcal{H}^1[\Gone,\mathcal{H}^D[\Gtwo,U(1)]]$ then $\phi_j(1,h)$ is nontrivial for some group element $(h,1)$ and $\nu_j$ characterizes a nontrivial $K$ SPT phase.

\vspace{1.5mm}
\noindent \begin{center}\emph{Completing the proof of Proposition~\ref{prop: symmetry-protected magic SPT type iii}:}\end{center}
\vspace{1.5mm}

Thus far, we have shown that the decorated domain wall models characterized by a nontrivial element of the group cohomology $\mathcal{H}^1[\Gone,\mathcal{H}^D[\Gtwo,U(1)]]$ have an effective boundary symmetry action that prepares a nontrivial SPT state in $(D-1)$-dimensions. We now use the decorated domain wall models to complete the proof of Proposition~\ref{prop: symmetry-protected magic SPT type iii}. In particular, we argue that a stabilizer state $|\psi_\stab \rangle$ in dimension $D\geq 2$ cannot belong to the same SPT phase as a nontrivial decorated domain wall model (assuming the symmetry is represented by a product of Pauli operators). The key observation is that the effective boundary symmetry action of the stabilizer model cannot prepare a nontrivial $(D-1)$-dimensional $K$ SPT state for $D\geq 2$. 

% since it shows that the effective boundary symmetry action of these SPT phases cannot be represented by a Pauli string. A tensor product of Pauli operators is unable to prepare a nontrivial SPT state from a product state in dimensions $D\geq 1$. 

% We begin by describing how the effective boundary symmetry action can be used to construct a symmetry defect. Here, a symmetry defect corresponds to a $(D-1)$-dimensional submanifold where the symmetry action has \textit{effectively} been applied to one side of the submanifold \cite{BBCW19}. A closed symmetry defect around the boundary of a region $M$, as pictured in Fig. [FIGURE], can be created by applying the onsite symmetry action to the region $M$. Alternatively, this can implemented by applying the effective boundary symmetry action around the boundary of $M$. More generally, a closed symmetry defect may wrap around a nontrivial cycle, as in Fig. [FIGURE]. In this case, the symmetry defect cannot be created by restricting the symmetry action, but it can be created by applying the effective boundary symmetry action along the nontrivial cycle. 

To derive a contradiction, suppose $|\psi_\stab \rangle$ belongs to the same SPT phase as the ground state $|\psi_\eta \rangle$ of a nontrivial decorated domain wall model. Then there exists a FDQC $\mathcal{U}_\sym$ composed of symmetric gates with the property:
\begin{align}
    \mathcal{U}_\sym |\psi_\stab \rangle = |\psi_\eta \rangle.
\end{align}
With this, we can define a Hamiltonian $\widetilde{H}_\eta$ as:
\begin{align}
    \widetilde{H}_\eta \equiv \mathcal{U}_\sym H_\stab \mathcal{U}_\sym^\dagger,
\end{align}
where $H_\stab$ is the local stabilizer parent Hamiltonian for $|\psi_\stab \rangle$.
Notably, $\widetilde{H}_\eta$ has the same ground state as the decorated domain wall Hamiltonian $H_\eta$. In what follows, we use $\widetilde{H}_\eta$ to compute an alternative effective boundary symmetry action $\widetilde{\mathcal{V}}\boldsymbol{(} (h,k) \boldsymbol{)}$ for the decorated domain wall model. We then show that $\widetilde{\mathcal{V}}\boldsymbol{(} (h,k) \boldsymbol{)}$ is inconsistent with the effective boundary symmetry action $\mathcal{V}\boldsymbol{(} (h,k) \boldsymbol{)}$ derived using $H_\eta$.

The first step in deriving an effective boundary symmetry action for the decorated domain wall model using $\widetilde{H}_\eta$ is to truncate $\widetilde{H}_\eta$ and define the corresponding boundary Hilbert space. Given a manifold with boundary $M$, we truncate $\widetilde{H}_\eta$ to the submanifold $M_\circ$, with $M_\circ$ defined below Eq.~\eqref{eq: disentangle bulk with truncated circuit}. 
By truncating to $M_\circ$, we ensure that the low-energy Hilbert space $\widetilde{\mathcal{H}}_\eta^{\text{low}}$ of the truncated Hamiltonian $\widetilde{H}^{M_\circ}_\eta$ contains the low-energy Hilbert space $\mathcal{H}^{\text{low}}_\eta$ of the Hamiltonian $H^M_\eta$, i.e., $\mathcal{H}^{\text{low}}_\eta \subset \widetilde{\mathcal{H}}^{\text{low}}_\eta$.
% \begin{align}
%     \mathcal{H}^{\text{low}}_\eta \subset \widetilde{\mathcal{H}}^{\text{low}}_\eta.
% \end{align}
Therefore, any effective boundary symmetry action defined on $\widetilde{\mathcal{H}}^{\text{low}}_\eta$ gives an effective boundary symmetry action on $\mathcal{H}^{\text{low}}_\eta$.

Recall that, in the proof of Proposition~\ref{prop: symmetry-protected magic SPT}, we argued that an effective boundary symmetry action for $|\psi_\stab\rangle$ with a local stabilizer parent Hamiltonian $H_\stab$ is given by a tensor product of Pauli operators denoted as $\mathcal{P}\boldsymbol{(} (h,k) \boldsymbol{)}$. Using $ \mathcal{P}\boldsymbol{(} (h,k) \boldsymbol{)}$ constructed on the submanifold $M_\circ$, we define $\widetilde{\mathcal{V}}\boldsymbol{(} (h,k) \boldsymbol{)}$ as:
\begin{align} \label{eq: tilde EBSA}
    \widetilde{\mathcal{V}}\boldsymbol{(} (h,k) \boldsymbol{)} \equiv \mathcal{U}_\sym \mathcal{P}\boldsymbol{(} (h,k) \boldsymbol{)} \mathcal{U}_\sym^\dagger.
\end{align}
$\widetilde{\mathcal{V}}\boldsymbol{(} (h,k) \boldsymbol{)}$ gives an effective boundary symmetry action on $\widetilde{\mathcal{H}}^{\text{low}}_\eta$, and thus, it gives an effective boundary symmetry action on the boundary Hilbert space $\mathcal{H}^{\text{low}}_\eta$. This means that, for any $(h,k) \in H \times K$ and any state $|\psi_\text{low}\rangle \in \mathcal{H}^{\text{low}}_\eta$, we have:
\begin{align} \label{eq: both are EBSA}
     {\mathcal{V}}\boldsymbol{(} (h,k) \boldsymbol{)} |\psi_\text{low}\rangle = \widetilde{\mathcal{V}}\boldsymbol{(} (h,k) \boldsymbol{)} |\psi_\text{low}\rangle.
\end{align}

We argue below that Eq.~\eqref{eq: both are EBSA} leads to a contradiction. The main idea is that ${\mathcal{V}}\boldsymbol{(} (h,k) \boldsymbol{)}$ prepares a nontrivial {$(D-1)$-dimensional} $K$ SPT state, while $\widetilde{\mathcal{V}}\boldsymbol{(} (h,k) \boldsymbol{)}$ is unable to prepare a nontrivial $(D-1)$-dimensional $K$ SPT state from a trivial SPT state. To make this explicit, we use dimensional reduction, as described in Ref.~\cite{T17}.

\begin{figure}[t]
\centering
\includegraphics[width=.5\textwidth,trim={5cm 8cm 3cm 5cm},clip]{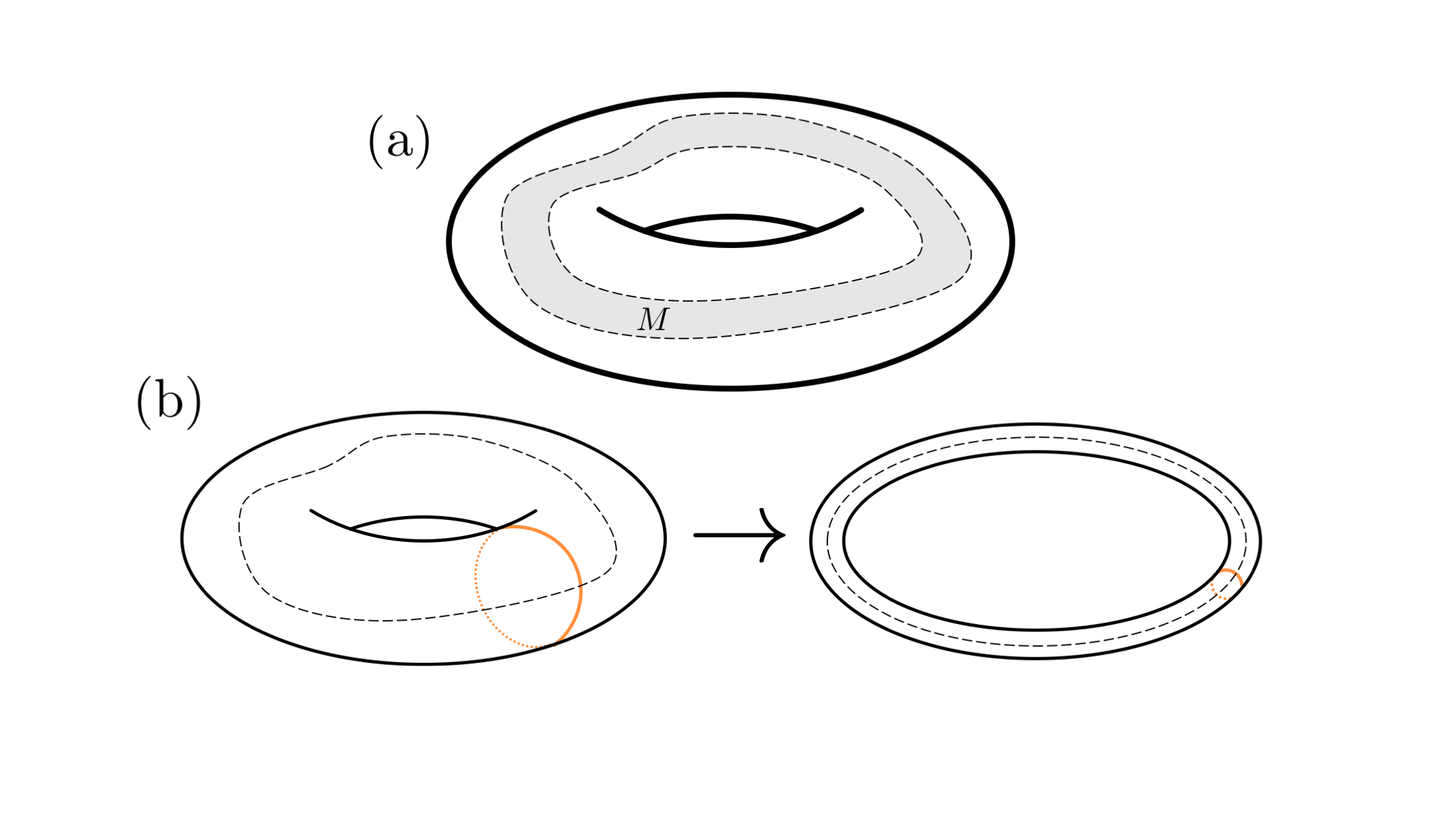}
\caption{(a) The submanifold $M$ of the torus is topologically equivalent to a thickened $1$-dimensional torus, and the boundary of $M$ has two components (dashed lines). (b) We compactify the $2$-dimensional torus to a (quasi) $1$-dimensional torus by making the meridian of the torus (orange) finite. Note that the component of the boundary of $M$ (dashed line) forms a non-contractible submanifold of the compactified torus. }
\label{fig:TD-1}
\end{figure}

We consider the decorated domain wall model on a $D$-dimensional torus $T^D$. We then construct a submanifold $M$ with boundary, by thickening a non-contractible $(D-1)$-dimensional torus embedded in $T^D$ (see Fig.~\ref{fig:TD-1} for an example). The boundary of $M$ has two components, each of which is topologically equivalent to a $(D-1)$-dimensional torus. We truncate to the submanifold $M$ in the usual way and define the effective boundary symmetry actions ${\mathcal{V}}\boldsymbol{(} (h,k) \boldsymbol{)}$ and $\widetilde{\mathcal{V}}\boldsymbol{(} (h,k) \boldsymbol{)}$. Since the boundary of $M$ has two disconnected components, the effective boundary symmetry actions split as:
\begin{eqs}
      {\mathcal{V}}\boldsymbol{(} (h,k) \boldsymbol{)} &= {\mathcal{V}}_a\boldsymbol{(} (h,k) \boldsymbol{)} {\mathcal{V}}_b\boldsymbol{(} (h,k) \boldsymbol{)} \\
      \widetilde{\mathcal{V}}\boldsymbol{(} (h,k) \boldsymbol{)} &= \widetilde{\mathcal{V}}_a\boldsymbol{(} (h,k) \boldsymbol{)} \widetilde{\mathcal{V}}_b\boldsymbol{(} (h,k) \boldsymbol{)},
\end{eqs}
where $a$ and $b$ denote the two components of the boundary of $M$.

We focus on the operators ${\mathcal{V}}_a\boldsymbol{(} (h,k) \boldsymbol{)}$ and $\widetilde{\mathcal{V}}_a\boldsymbol{(} (h,k) \boldsymbol{)}$. These create a symmetry branch cut (or symmetry defect \cite{BBCW19}) along $a$, and can intepreted as inserting symmetry flux through the $D$-torus. It is important to note that, for some group element $(h,1)$, ${\mathcal{V}}_a\boldsymbol{(} (h,1) \boldsymbol{)}$ prepares a nontrivial $(D-1)$-dimensional $K$ SPT state on the non-contractible submanifold $a$. Also note that, for every $(h,k)$ in $H \times K$, $\widetilde{\mathcal{V}}_a\boldsymbol{(} (h,k) \boldsymbol{)}$ takes the form of a tensor product of Pauli operators $\mathcal{P}_a\boldsymbol{(} (h,k) \boldsymbol{)}$ conjugated by $\mathcal{U}_\sym$, as in Eq.~\eqref{eq: tilde EBSA}: 
\begin{align}
        \widetilde{\mathcal{V}}_a\boldsymbol{(} (h,k) \boldsymbol{)} = \mathcal{U}_\sym \mathcal{P}_a\boldsymbol{(} (h,k) \boldsymbol{)} \mathcal{U}_\sym^\dagger.
\end{align}
Assuming the components $a$ and $b$ are well separated in terms of the Lieb-Robinson length of $\mathcal{U}_\eta$, the action of ${\mathcal{V}}_a\boldsymbol{(} (h,k) \boldsymbol{)}$ and $\widetilde{\mathcal{V}}_a\boldsymbol{(} (h,k) \boldsymbol{)}$ agree on any state belonging to $\mathcal{H}^\text{low}_\eta$:
\begin{align}
         {\mathcal{V}}_a\boldsymbol{(} (h,k) \boldsymbol{)} |\psi_\text{low}\rangle = \widetilde{\mathcal{V}}_a\boldsymbol{(} (h,k) \boldsymbol{)} |\psi_\text{low}\rangle.
\end{align}
In particular, the ground state $|\psi_\eta \rangle$ on $T^D$ is in the low-energy Hilbert space $\mathcal{H}^\text{low}_\eta$, so the operators satisfy:
\begin{align} \label{eq: symmetry defects the same}
         {\mathcal{V}}_a\boldsymbol{(} (h,k) \boldsymbol{)} |\psi_\eta\rangle = \widetilde{\mathcal{V}}_a\boldsymbol{(} (h,k) \boldsymbol{)} |\psi_\eta\rangle.
\end{align}

Finally, we consider a compactification of $T^D$ to a (quasi) $({D-1})$-dimensional torus $T^{D-1}$ by making one of the directions in $T^D$ finite, as shown in Fig.~\ref{fig:TD-1}. In particular, we compactify $T^D$ to $T^{D-1}$ in such a way that $a$ forms a non-contractible submanifold of $T^{D-1}$. The equality in Eq.~\eqref{eq: symmetry defects the same} also holds on the $(D-1)$-dimensional torus.

On the $(D-1)$-dimensional torus, we can argue that Eq.~\eqref{eq: symmetry defects the same} gives a contradiction. It can be checked using the methods of Ref.~\cite{EN14} that ${\mathcal{V}}_a\boldsymbol{(} (h,1) \boldsymbol{)}$ prepares a nontrivial $(D-1)$-dimensional $K$ SPT state for some choice of $h$, while $\widetilde{\mathcal{V}}_a\boldsymbol{(} (h,1) \boldsymbol{)}$ cannot change the $(D-1)$-dimensional SPT phase of the compactified state $|\psi_\eta\rangle$ for any $h$. The latter is made explicit by writing $\widetilde{\mathcal{V}}_a\boldsymbol{(} (h,1) \boldsymbol{)}$ as: 
\begin{align}
        \widetilde{\mathcal{V}}_a\boldsymbol{(} (h,1) \boldsymbol{)} = \mathcal{U}_\sym \mathcal{P}_a\boldsymbol{(} (h,1) \boldsymbol{)} \mathcal{U}_\sym^\dagger.
\end{align}
Since $\mathcal{U}_\sym$ and $\mathcal{U}_\sym^\dagger$ are FDQCs composed of symmetric gates, they do not change the SPT phase. Furthermore, the tensor product of Pauli operators $\mathcal{P}_a\boldsymbol{(} (h,1) \boldsymbol{)}$ is able to create charges, but it is unable change the SPT phase of a $(D-1)$-dimensional SPT state, for $D\geq 2$. Therefore, the states ${\mathcal{V}}_a\boldsymbol{(} (h,1) \boldsymbol{)}|\psi_\eta\rangle$ and $\widetilde{\mathcal{V}}_a\boldsymbol{(} (h,1) \boldsymbol{)} |\psi_\eta\rangle$
belong to distinct $K$ SPT phases for some $h \in H$. 

We have shown that Eq.~\eqref{eq: symmetry defects the same} is inconsistent. This implies that the stabilizer state $|\psi_\stab \rangle$ cannot belong to the same phase as the ground state $|\psi_\eta \rangle$ of a nontrivial decorated domain wall model, assuming the symmetry is represented by a Pauli string. In other words, the stabilizer state must belong to an SPT phase characterized by the trivial element of $\mathcal{H}^1[\Gone,\mathcal{H}^D[\Gtwo,U(1)]]$. This completes the proof of Proposition~\ref{prop: symmetry-protected magic SPT type iii}.

\section{Existence of a local stabilizer parent Hamiltonian} \label{app: Existence of a local stabilizer parent Hamiltonian}

In the proof of Proposition~\ref{prop: symmetry-protected magic SPT}, we claimed that a stabilizer SPT state admits a local symmetric stabilizer Hamiltonian. Here, we justify this claim. We start by proving a lemma regarding stabilizer groups that stabilize a unique state. 

{\begin{lemma} \label{lem: Pauli string commute and stabilizer invariance}
Let $\mathcal{G}$ be a stabilizer group that uniquely fixes the stabilizer state $|\psi_\stab \rangle$. If a Pauli string $P$ satisfies one of the following, then $P$ is contained in $\mathcal{G}$:
\begin{enumerate}[label=(\roman*)]
    \item $P$ commutes with every element of $\mathcal{G}$
    \item $P$ stabilizes $|\psi_\stab \rangle$, i.e., $P|\psi_\stab \rangle = |\psi_\stab \rangle$.
\end{enumerate}
\end{lemma}}

{\noindent \emph{Proof of Lemma~\ref{lem: Pauli string commute and stabilizer invariance}:} We first prove that condition (i) implies condition (ii). Assuming (i), then $P$ commutes with every element of $\mathcal{G}$, and we have:
\begin{align}
    SP|\psi_\stab \rangle = PS |\psi_\stab \rangle = P |\psi_\stab \rangle, \quad \forall S \in \mathcal{G}.
\end{align}
This implies that $P|\psi_\stab \rangle$ is stabilized by $\mathcal{G}$. Since $\mathcal{G}$ stabilizes a unique state, it must be that $P|\psi_\stab \rangle=|\psi_\stab \rangle$.}

{With this, it suffices to prove condition (ii) of Lemma~\ref{lem: Pauli string commute and stabilizer invariance}. We do so by considering a Pauli string $E$ with the property that $E$ commutes with every $S \in \mathcal{G}$, but fails to commute with $P$ by a root of unity (not equal to $+1$). The existence of such a Pauli string follows from a straightforward generalization of Proposition 10.4 in Ref.~\cite{NC10}. Consequently, the state $E|\psi_\stab \rangle$ satisfies:
\begin{align} \label{eq: SE identity}
      SE |\psi_\stab \rangle &= E |\psi_\stab \rangle, \quad \forall S \in \mathcal{G} \\ \label{eq: PE identity}
      PE|\psi_\stab \rangle &\neq E|\psi_\stab \rangle.
\end{align}
Since $\mathcal{G}$ stabilizes a unique state, Eq.~\eqref{eq: SE identity} implies that $E|\psi_\stab \rangle=|\psi_\stab \rangle$. Substituting $E|\psi_\stab \rangle$ for $|\psi_\stab \rangle$ in condition (ii), we find that $PE|\psi_\stab\rangle = E|\psi_\stab \rangle$. However, this contradicts Eq.~\eqref{eq: PE identity}. Hence, $P$ must belong to $\mathcal{G}$. $\square$} \\

Now, we can prove the following statement about the existence of a local symmetric stabilizer Hamiltonian. 

\begin{lemma} \label{lem: local stabilizer hamiltonian}
Let $|\psi_\stab \rangle$ be a stabilizer state which is a unique ground state of a geometrically local Hamiltonian $H_\loc$.\footnote{We recall that a geometrically local Hamiltonian is a sum of terms such that the support of each term can be contained within a ball of fixed finite diameter.} Then, there exists a local stabilizer Hamiltonian $H_\stab$ such that $|\psi_\stab\rangle$ is the unique ground state of $H_\stab$.  Furthermore, if $|\psi_\stab\rangle$ is invariant under a Pauli string $P$, i.e., $P |\psi_\stab \rangle = |\psi_\stab \rangle$, then $H_\stab$ commutes with $P$. 
\end{lemma}

\noindent \emph{Proof of Lemma~\ref{lem: local stabilizer hamiltonian}.} Since $|\psi_\stab\rangle$ is a stabilizer state, there is a stabilizer group $\mathcal{G}$ that uniquely fixes $|\psi_\stab \rangle$. We claim that the generators of $\mathcal{G}$ can always be chosen to be geometrically local. To see this, we imagine minimizing the largest support of the generators over all possible choices for generators of $\mathcal{G}$. We let $d_S$ denote the minimum length such that each stabilizer term can be contained in a ball of diameter $d_S$. We argue that $d_S$ is constant, i.e., independent of the system size. 

If there exists a generator $S$ that is not contained in a constant size ball, then there is a stabilizer state $|\phi_\stab \rangle$ that has a $+1$ eigenvalue for all the generators except $S$, for which the eigenvalue is some root of unity (not equal to $+1$). $|\psi_\stab \rangle$ and $|\phi_\stab \rangle$ are orthogonal, and yet they have the same reduced density matrices on any constant-size region. The latter follows from the fact that $S$ is not contained in any constant-size region and that the reduced density matrices of stabilizer states on a region $M$ depend solely on the stabilizer group elements whose support is contained in $M$ \cite{FCYBC04,LMRW13}. (Note that if there is another generator $T$ such that $ST$ is supported in a constant-sized region $M$, then we can use $ST$ as a generator instead of $S$. This contradicts the assumption that we have minimized the maximum support of the generators.)

Since $|\psi_\stab \rangle$ and $|\phi_\stab \rangle$ have the same reduced density matrices on constant-sized regions, $|\phi_\stab \rangle$ gives another ground state of $H_\loc$. This conflicts with the assumption that $H_\loc$ has a unique ground state. Thus, the support of $S$ must be contained in a constant-size ball, and $d_S$ can be chosen independent of the system size. Therefore, it is possible to find a set of generators for $\mathcal{G}$, which are geometrically local. We define $H_\stab$ to be the negative sum of the local generators (with their Hermitian conjugates). $|\psi_\stab \rangle$ is the unique ground state of $H_\stab$, since the terms of $H_\stab$ span $\mathcal{G}$.

Lastly, if $|\psi_\stab \rangle$ is invariant under a Pauli operator $P$, then $P$ is an element of $\mathcal{G}$, according to Lemma~\ref{lem: Pauli string commute and stabilizer invariance}. Since $\mathcal G$ is a commuting group, $P$ commutes with $H_\stab$.~$\square$

\section{Modified effective boundary symmetry action} \label{app: modified effective boundary symmetry action}

To prove Proposition~\ref{prop: symmetry-protected magic SPT}, we derived an effective boundary symmetry action $\mathcal{P}(g)$ for the stabilizer state $|\psi_\stab \rangle$ using a stabilizer parent Hamiltonian $H_\stab$. While the effective boundary symmetry action forms a linear representation of the $G = \ZZ_q^m$ symmetry in the boundary Hilbert space, it is only guaranteed to satisfy the group laws of $G$ up to products of stabilizer terms belonging to the truncated Hamiltonian $H^M_\stab$. Here, we complete the proof of Proposition~\ref{prop: symmetry-protected magic SPT} by showing that the effective boundary symmetry action can be modified so that is satisfies the group laws exactly. This assumes that the state $|\psi_\stab \rangle$ is defined on a Hilbert space composed of $q$-dimensional qudits.

To begin, we verify that the effective boundary symmetry action satisfies the group laws up to elements of $\mathcal{G}^M$, where $\mathcal{G}^M$ is the stabilizer group generated by the terms in $H^M_\stab$. Recall that the effective boundary symmetry action derived in the proof of Proposition~\ref{prop: symmetry-protected magic SPT} is:
\begin{align} \label{eq: eff boundary def app}
    \mathcal{P}(g) = P_M(g)\tilde{P}_M^\dagger(g),
\end{align}
where $g$ is in $G$, $P_M(g)$ is the restriction of the onsite symmetry to $M$, and $\tilde{P}_M^\dagger(g)$ belongs to $\mathcal{G}^M$. The product of $\mathcal{P}(g)$ and $\mathcal{P}(h)$ for arbitrary elements $g,h$ in $G$ is then:
\begin{eqs} \label{eq: eff boundary group laws 1}
    \mathcal{P}(g) \mathcal{P}(h) &= \left[P_M(g)\tilde{P}_M^\dagger(g)\right]\left[P_M(h)\tilde{P}_M^\dagger(h)\right].
\end{eqs}
This can be simplified by commuting $\tilde{P}_M^\dagger(g)$ past $P_M(h)$. These commute because $P_M(h)$ is the onsite symmetry action restricted to $M$, and $\tilde{P}_M^\dagger(g)$ is a product of symmetric stabilizer terms whose supports are contained in $M$. This gives us:
\begin{align} \label{eq: eff boundary group laws 2}
    \mathcal{P}(g) \mathcal{P}(h) = P_M(gh)\tilde{P}_M^\dagger(g)\tilde{P}_M^\dagger(h).
\end{align}
The product ${\tilde{P}_M^\dagger(g)\tilde{P}_M^\dagger(h)}$ agrees with $\tilde{P}_M^\dagger(gh)$ away from the boundary of $M$ [see the definition in Eq.~\eqref{eq: tilde P def}]. Moreover, it only differs from $\tilde{P}_M^\dagger(gh)$ by elements of $\mathcal{G}^M$ that are supported near the boundary of $M$. We define $S(g,h)$ to be the product of stabilizer terms in $\mathcal{G}^M$ near the boundary of $M$ such that:
\begin{align} \label{eq: Sgh def}
    \tilde{P}_M^\dagger(g)\tilde{P}_M^\dagger(h) = S(g,h) \tilde{P}_M^\dagger(gh).
\end{align}
Substituting Eq.~\eqref{eq: Sgh def} into Eq.~\eqref{eq: eff boundary group laws 2}, we arrive at:
\begin{align} \label{eq: eff boundary group laws 3}
    \mathcal{P}(g) \mathcal{P}(h) = S(g,h) \mathcal{P}(gh),
\end{align}
which implies that the effective boundary symmetry action obeys the group laws up to elements of $\mathcal{G}^M$ supported near the boundary of $M$.

We note that the operators in Eq.~\eqref{eq: Sgh def} are mutually commuting. This is a consequence of the fact that, for any $g \in G$, $\tilde{P}_M^\dagger(g)$ is a product of symmetric stabilizer terms whose supports are contained in $M$.  We also point out that the effective boundary symmetry action $\mathcal{P}(1)$, corresponding to the identity in $G$, can be taken to be the identity. This implies that $S(g,1)=S(1,g)=1$, for any $g \in G$. 

The extra stabilizer terms $S(g,h)$ in Eq.~\eqref{eq: eff boundary group laws 3} can be eliminated by modifying the effective boundary symmetry action. In particular, we are free to modify $\mathcal{P}(g)$ by any product of stabilizers in $H^M_\stab$ supported near the boundary of $M$. We define $\mathcal{P}'(g)$ to be the effective boundary symmetry action modified by stabilizer terms $T(g)$, i.e.:
\begin{align}
    \mathcal{P}'(g) = T(g) \mathcal{P}(g).
\end{align}
In what follows, we describe a particular choice of $T(g)$ for every $g \in G$ such that the modified effective boundary symmetry action $\mathcal{P}'(g)$ satisfies the group laws exactly.

To specify our choice of $T(g)$, we introduce some notation for the elements of $G=\ZZ_q^m$. We denote the $m$ generators of $\ZZ_q^m$ as $k_1,\ldots, k_m$, and the product of two generators $k_i$, $k_j$ as $k_ik_j$. A general element $g$ of $\ZZ_q^m$ can then be written as:
\begin{align}
    g = k_1^{a_1}k_2^{a_2}\cdots k_m^{a_m},
\end{align}
where $a_1,\ldots,a_m$ are integers belonging to $\{0,\ldots, q-1\}$. 
{Using this notation, we take $T(k_1^{a_1}k_2^{a_2}\cdots k_m^{a_m})$ to be the product of stabilizer terms given by:
\begin{eqs} \label{eq: general T def}
    T(k_1^{a_1}k_2^{a_2} \cdots k_m^{a_m}) = \prod_{\substack{i=1 \\ a_i \neq 0}}^m \prod_{c_i=1}^{a_i} S(k_i,k_i^{a_i-c_i} g_{i+1}^{a_{i+1}} \cdots k_m^{a_m}).
\end{eqs}}

{To motivate the expression above, we first consider $T(k_i)$ for a generator $k_i$. According to Eq.~\eqref{eq: general T def}, $T(k_i)$ is equal to $S(k_i,1)$, which simplifies to $T(k_i)=1$, since $S(g,1)$ is the identity for any $g \in \ZZ_q^m$. Thus, $\mathcal{P}'(k_i)$ is defined as:
\begin{align} \label{eq: pprime generator}
    \mathcal{P}'(k_i) = \mathcal{P}(k_i).
\end{align}
Further, we consider $T(k_ik_j)$ with $k_ik_j \neq 1$. In this case, $T(k_ik_j)$ is equal to $S(k_i,k_j)$, and $\mathcal{P}'(k_ik_j)$ is given by:
\begin{eqs} \label{eq: pprime two generators}
      \mathcal{P}'(k_ik_j) = S(k_i,k_j) \mathcal{P}(k_ik_j).
\end{eqs}
Combining Eqs.~\eqref{eq: pprime generator} and \eqref{eq: pprime two generators}, we see that the group laws are satisfied by $\mathcal{P}'(k_i)$ and $\mathcal{P}'(k_j)$:
\begin{eqs}
     \mathcal{P}'(k_i) \mathcal{P}'(k_j) &= \mathcal{P}(k_i) \mathcal{P}(k_j) \\
     &= S(k_i,k_j)\mathcal{P}(k_ik_j) \\
     &= \mathcal{P}'(k_ik_j).
\end{eqs}
More generally, the product in Eq.~\eqref{eq: general T def} is designed so that:
\begin{eqs} \label{eq: decomposition of modified eff boundary}
     \mathcal{P}'(k_1)^{a_1}\mathcal{P}'(k_2)^{a_2} \cdots \mathcal{P}'(k_m)^{a_m}  = \mathcal{P}'(k_1^{a_1}k_2^{a_2} \cdots k_m^{a_m}).
\end{eqs}}

The general expression in Eq.~\eqref{eq: decomposition of modified eff boundary} allows us to verify the group laws for the modified effective boundary symmetry actions defined using the choice of $T(g)$ in Eq.~\eqref{eq: general T def}. The product of $\mathcal{P}'(g)$ and $\mathcal{P}'(h)$ for general group elements $g=k_1^{a_1}k_2^{a_2} \cdots k_m^{a_m}$ and $h=k_1^{b_1}k_2^{b_2} \cdots k_m^{b_m}$ is:
\begin{eqs} \label{eq: group law modified eff boundary 1}
     \mathcal{P}'(g) \mathcal{P}'(h) = &\left[\mathcal{P}'(k_1)^{a_1}\mathcal{P}'(k_2)^{a_2} \cdots \mathcal{P}'(k_m)^{a_m}\right]\\
     \times &\left[\mathcal{P}'(k_1)^{b_1}\mathcal{P}'(k_2)^{b_2} \cdots \mathcal{P}'(k_m)^{b_m}\right],
\end{eqs}
where we have used the identity in Eq.~\eqref{eq: decomposition of modified eff boundary}. Since the operators on the right hand side of Eq.~\eqref{eq: group law modified eff boundary 1} are mutually commuting, we can write $\mathcal{P}'(g) \mathcal{P}'(h)$ as:
\begin{eqs} \label{eq: group law modified eff boundary 2}
     \mathcal{P}'(g) \mathcal{P}'(h) = \mathcal{P}'(k_1)^{a_1+b_1}\mathcal{P}'(k_2)^{a_2+b_2} \cdots \mathcal{P}'(k_m)^{a_m+b_m}.
\end{eqs}
We evaluate the expression in Eq.~\eqref{eq: group law modified eff boundary 2} further by writing $a_i+b_i$ as:
\begin{eqs} \label{eq: modulo q rewrite}
    a_i+b_i &= \left(a_i+b_i - [a_i+b_i]_q\right) + [a_i+b_i]_q \\
            &= n_i q + [a_i+b_i]_q,
\end{eqs}
where $[\cdot]_q$ denotes addition modulo $q$, and $n_i q$ is an integer multiple of $q$. Substituting Eq.~\eqref{eq: modulo q rewrite} into Eq.~\eqref{eq: group law modified eff boundary 2}, we find:
\begin{multline} 
     \mathcal{P}'(g) \mathcal{P}'(h) = \mathcal{P}'(k_1)^{n_1q}\mathcal{P}'(k_2)^{n_2q} \cdots \mathcal{P}'(k_m)^{n_mq} \\
                                     \times \mathcal{P}'(k_1)^{[a_1+b_1]_q}\mathcal{P}'(k_2)^{[a_2+b_2]_q} \cdots \mathcal{P}'(k_m)^{[a_m+b_m]_q}.
\end{multline}
According to Eq.~\eqref{eq: decomposition of modified eff boundary}, this is equivalent to:
\begin{eqs} \label{eq: group law modified eff boundary 3}
     \mathcal{P}'(g) \mathcal{P}'(h) = \left[\mathcal{P}'(k_1)^{n_1q}\mathcal{P}'(k_2)^{n_2q} \cdots \mathcal{P}'(k_m)^{n_mq} \right]\mathcal{P}'(gh).
\end{eqs}

Lastly, we argue that the pre-factor on the right hand side of Eq.~\eqref{eq: group law modified eff boundary 3} is the identity. To see this, we expand each factor of $\mathcal{P}'(a_i)$ using the definition of $\mathcal{P}(g)$ in Eq.~\eqref{eq: eff boundary def app} and use that $P_M(g)$ forms a linear representation of $\ZZ_q^m$:
\begin{multline}
     \mathcal{P}'(k_1)^{n_1q}\mathcal{P}'(k_2)^{n_2q} \cdots \mathcal{P}'(k_m)^{n_mq} = \\   \tilde{P}_M^\dagger(k_1)^{n_1q}  \tilde{P}_M^\dagger(k_2)^{n_2q} \cdots  \tilde{P}_M^\dagger(k_m)^{n_mq}
\end{multline}
Since the system is defined on $q$-dimensional qudits, any product of stabilizer terms raised to the power of $q$, such as $\tilde{P}^\dagger_M(a_i)^{n_iq}$, must be the identity. Therefore, we have:
\begin{align}
    \mathcal{P}'(k_1)^{n_1q}\mathcal{P}'(k_2)^{n_2q} \cdots \mathcal{P}'(k_m)^{n_mq} = 1,
\end{align}
and according to Eq.~\eqref{eq: group law modified eff boundary 3}, the modified effective boundary symmetry action satisfies the group laws exactly:
\begin{align}
    \mathcal{P}'(g) \mathcal{P}'(h) = \mathcal{P}'(gh), \quad \forall g,h \in G.
\end{align}

\section{{Strange order parameters for 1D SPT phases}} \label{app: strange order parameters for 1D SPT phases}

We show that every $1$D SPT state $|\psi_\spt \rangle$ protected by a finite Abelian symmetry admits a strange order parameter $\{\mathcal{O}_i, \mathcal{O}_j\}$ with the property:
\begin{align} \label{eq: norm of strange correlator}
    \left| \frac{\langle \Omega | \mathcal{O}_i \mathcal{O}_j | \psi_\spt \rangle}{\langle \Omega | \psi_\spt \rangle} \right| = 1, 
\end{align}
for every $i,j$ and unentangled eigenstate $| \Omega \rangle$ of the symmetry. This is accomplished by finding a representative SPT state, for each SPT phase, that satisfies Eq.~\eqref{eq: norm of strange correlator}. We then show that this property can be extended to every state in the SPT phase. 

More specifically, we show that the ground state $|\psi_\eta \rangle$ of a $1$D decorated domain wall model in Appendix~\ref{app: ddw models} satisfies:
\begin{align} \label{eq: restricted symmetry on eta}
    u_R(g)^\dagger \mathcal{O}_a \mathcal{O}_b |\psi_\eta \rangle = |\psi_\eta \rangle,
\end{align}
where $u_R(g)$ is the symmetry action restricted to an arbitrary interval $R$, and $\mathcal{O}_a$,$\mathcal{O}_b$ are charged unitary operators localized near the endpoints $a$ and $b$ of $R$. This implies that $\mathcal{O}_a$ and $\mathcal{O}_b$ can be used to form a strange order parameter. For an arbitrary tensor product eigenstate $|\Omega \rangle$ of the symmetry, Eq.~\eqref{eq: restricted symmetry on eta} leads to:
\begin{align}
    1 = \left| \frac{\langle \Omega |u_R(g)^\dagger| \psi_\eta \rangle}{\langle \Omega |\psi_\eta \rangle} \right| = \left| \frac{\langle \Omega |\mathcal{O}_a \mathcal{O}_b| \psi_\eta \rangle}{\langle \Omega |\psi_\eta \rangle} \right|.
\end{align}
Therefore, if a state satisfies Eq.~\eqref{eq: restricted symmetry on eta} for every interval $R$ and a pair of operators $\mathcal{O}_a$, $\mathcal{O}_b$, then 
it will satisfy Eq.~\eqref{eq: norm of strange correlator}.

Given a state $|\psi_\spt\rangle$ in the same SPT phase as $|\psi_\eta \rangle$, $|\psi_\spt\rangle$ must satisfy:
\begin{align}
     \mathcal{U}_\sym \left(u_R(g)^\dagger \mathcal{O}_a \mathcal{O}_b\right) \mathcal{U}_\sym^\dagger |\psi_\spt \rangle = |\psi_\spt \rangle,
\end{align}
where $\mathcal{U}_\sym$ is the FDQC composed of symmetric gates that maps $|\psi_\eta \rangle$ to $|\psi_\spt \rangle$:
\begin{align}
    \mathcal{U}_\sym |\psi_\eta \rangle = |\psi_\spt \rangle. 
\end{align}
Since $\mathcal{U}_\sym$ is composed of symmetric gates, $u_R(g)$ commutes with $\mathcal{U}_\sym$ up to operators near the endpoints $a$ and $b$. Thus, we have:
\begin{align}
    u_R(g)^\dagger \mathcal{O}'_a \mathcal{O}'_b |\psi_\spt \rangle = |\psi_\spt \rangle,
\end{align}
for some local unitary operators $\mathcal{O}_a'$, $\mathcal{O}_b'$. $|\psi_\spt \rangle$ then satisfies Eq.~\eqref{eq: norm of strange correlator} using $\mathcal{O}_a'$, $\mathcal{O}_b'$ as a strange order parameter.

Now, we only need to show that the ground state $|\psi_\eta \rangle$ of a $1$D decorated domain wall model obeys Eq.~\eqref{eq: restricted symmetry on eta}. To do so, we freely use the notation introduced in Appendix~\ref{app: ddw models}. By replacing $M$ with the full $1$D system and $M_\circ$ with the interval $R$, Eq.~\eqref{eq: ddw conjugate truncated circuit} gives us:
\begin{align}
   u_R\boldsymbol{(}(h,1)\boldsymbol{)} |\psi_\eta \rangle = \mathcal{O}_a \mathcal{O}_b |\psi_\eta \rangle,
\end{align}
with $\mathcal{O}_a$, $\mathcal{O}_b$ defined as:
\begin{align}
    \mathcal{O}_a &\equiv \sum_\mathcal{C} \prod_{j=1}^p \nu_j (1,k_a)^{\phi_j(1,h)}|\mathcal{C} \rangle \langle \mathcal{C} | \\
    \mathcal{O}_b &\equiv \sum_\mathcal{C} \prod_{j=1}^p \nu_j (1,k_b)^{-\phi_j(1,h)}|\mathcal{C} \rangle \langle \mathcal{C} |.
\end{align}
$\mathcal{O}_a$ and $\mathcal{O}_b$ are charged local unitaries. One can check that they are charged by using that $\nu_j$ is closed [Eq.~\eqref{eq: nu is closed}].

Therefore, $|\psi_\eta \rangle$ satisfies Eq.~\eqref{eq: restricted symmetry on eta}, and it follows that every SPT state belonging to the same SPT phase as $|\psi_\eta\rangle$ admits a strange order parameter obeying Eq.~\eqref{eq: norm of strange correlator}. It follows from the K\"unneth theorem that the decorated domain wall models provide a representative SPT state for all $1$D SPT phases protected by finite Abelian symmetries. Indeed, let $G$ be an arbitrary finite Abelian group of the form $G=\prod_{i=1}^p\ZZ_{n_i}$. Since $H^2[\ZZ_{n_i},U(1)]$ is trivial for all $i$, the K\"unneth theorem gives:
\begin{align}
    H^2[G,U(1)] = \prod_{i=1}^p H^1[\ZZ_{n_i},H^1[\prod_{j>i}^p\ZZ_{n_j},U(1)]].
\end{align}
Hence, all $1$D SPT states (as defined in Section~\ref{sec: definition of SPT phases}) protected by a finite Abelian symmetry satisfy Eq.~\eqref{eq: norm of strange correlator} for some choice of strange order parameter. 

\bibliographystyle{unsrtnat}
\bibliography{bibliography.bib}

\end{document}